\documentclass[3p,times]{elsarticle}
\PassOptionsToPackage{numbers,sort&compress}{natbib}
\usepackage[section]{placeins}
\usepackage{adjustbox}
\usepackage[T1]{fontenc}    
\usepackage{hyperref}       
\usepackage{url}            
\usepackage{booktabs}       
\usepackage{amsfonts}       
\usepackage{nicefrac}       
\usepackage{microtype}      
\usepackage{xcolor}         
\usepackage{algorithm}
\usepackage{algpseudocode}
\usepackage[para,online,flushleft]
{threeparttable}
\usepackage{pifont}
\usepackage{pgfplots}
\pgfplotsset{compat=1.18,
    /pgfplots/ybar legend/.style={
        /pgfplots/legend image code/.code={%
            \draw[##1,/tikz/.cd,bar width=6pt,yshift=-0.2em,bar shift=0pt]
            plot coordinates {(0cm,0.8em)};},
    },
}
\usepgfplotslibrary{groupplots}
\usepgfplotslibrary{statistics}
\usepackage{tikz}
\usepackage{pgfplots}
\usepackage{tikzviolinplots}
\usepackage{csvsimple}
\usepackage{overpic}
\usetikzlibrary{shapes, arrows.meta, positioning}
\usetikzlibrary{calc, positioning}

\usepackage{multirow}
\usepackage{graphicx}
\usetikzlibrary{
    shapes.geometric,
    arrows.meta,
    positioning,
    fit,
    calc,
    matrix,          
    decorations.pathreplacing
}
\usepackage{amsmath}
\usepackage{amsthm}
\usepackage[textfont=it,labelfont=bf]{caption} 
\usepackage{subcaption}
\usepackage{enumitem}
\usepackage{array}
\usepackage{wrapfig}
\usepackage{colortbl}
\usepackage{tabularray}
\usepackage{graphbox}
\usepackage[most]{tcolorbox}
\usepackage{algorithm}
\usepackage{algpseudocode}
\usepackage{subcaption}
\usepackage{mdframed}
\usepackage{shadethm}
\usepackage{thmtools}

\hypersetup{
    breaklinks=true,
    colorlinks,
    linkcolor={blue!80!black},
    citecolor={blue!80!black},
    urlcolor={blue!80!black},
    plainpages=true
}
\newcommand{\cref}[2]{\hyperref[#2]{#1~\ref*{#2}}}
\newcommand{\colref}[3]{\hyperref[#2]{#1~\ref*{#2}{#3}}}
\newcommand{\figref}[1]{\cref{Figure}{#1}}

\newcommand{\secref}[1]{\cref{Section}{#1}}

\newcommand{\tabref}[1]{\cref{Table}{#1}}
\newcommand{\algoref}[1]{\cref{Algorithm}{#1}}

\newcommand{\appendixref}[1]{\cref{}{#1}}

\declaretheoremstyle[%
  spaceabove=-6pt,%
  spacebelow=6pt,%
  headfont=\bfseries\itshape,%
  postheadspace=0.5em,%
  qed=\qedsymbol%
]{mystyle}

\theoremstyle{mystyle}

\definecolor{trueintercepted}{HTML}{FFFF00}
\definecolor{falseintercepted}{HTML}{F08080}
\definecolor{trueboundary}{HTML}{00007f}
\definecolor{notintercepted}{HTML}{008000}

\definecolor{uColor}{rgb}{0,0,0} 
\definecolor{gradColor}{rgb}{1.0,0.0,0.0} 
\definecolor{boundaryColor}{rgb}{0.6,0.2,0.8}

\journal{Computer Methods in Applied Mechanics and Engineering}

\usepackage{standalone}
\pgfdeclarehorizontalshading{cfdgradient}{100bp}{
  color(0bp)=(blue);
  color(50bp)=(white);
  color(100bp)=(red)
}

\begin{document}

\begin{frontmatter}
\title{Predicting time-dependent flow over complex geometries using operator networks}

\author[ISU]{Ali Rabeh}
\author[ISU]{Suresh Murugaiyan}
\author[ISU]{Adarsh Krishnamurthy\texorpdfstring{\corref{cor1}}{}}
\author[ISU]{Baskar Ganapathysubramanian\texorpdfstring{\corref{cor1}}{}}
\affiliation[ISU]{organization={Iowa State University}, 
            city={Ames},
            state={Iowa},
            country={USA}}
\cortext[cor1]{Corresponding Authors}

\begin{abstract}
Developing neural surrogates for unsteady flow that generalize across complex geometries and maintain accuracy over extended rollouts remains a critical challenge for accelerating flow simulations. We present a time-dependent, geometry-aware Deep Operator Network that predicts velocity fields for moderate-Re flows around parametric and non-parametric shapes. The model encodes geometry via a signed distance field (SDF) trunk and flow history via a CNN branch, and is trained using 841 high-fidelity simulations from the \href{https://baskargroup.bitbucket.io/}{FlowBench} dataset. The model attains $\sim 5\%$ relative $L_2$ single-step error and up to 1000$\times$ speedups over CFD on 262 test geometries not used for training. We provide physics-centric rollout diagnostics, including probe-based phase lag and Strouhal frequency analysis, to quantify long-horizon fidelity. These reveal accurate near-term transients but systematic error accumulation in fine-scale wakes, most pronounced for sharp-cornered geometries. We analyze failure modes and outline practical mitigation strategies including physics-informed regularization and diffusion-based refinement. Code, splits, and scripts are openly released \href{https://github.com/baskargroup/TimeDependent-DeepONet}{here} to support reproducibility and benchmarking.
\end{abstract}


\end{frontmatter}

\section{Introduction}
\label{Sec:Introduction}

Time-dependent flow simulations underpin tasks in shape optimization and flow control across aerospace, automotive, civil, and energy systems~\citep{he1997computational, mohammadi2009applied}. Yet high-fidelity CFD remains computationally intensive, especially for large design spaces or long horizons, and the need to mesh complex geometries further increases cost and engineering effort~\citep{li2024efficient, gou2024gpu}. A canonical stress test is unsteady, incompressible flow past immersed bodies: vortex formation, shedding frequency, and wake interactions must be captured accurately to avoid spurious loads and instabilities~\cite{forouzi2022review, wu2022modelling}. In practice, unsteady wakes around wings and vehicle bodies can induce oscillatory forces and fatigue~\citep{fuller2012unsteady, ericsson2001unsteady}; vortex shedding behind buildings and bridge piers can trigger hazardous vibrations~\citep{haldar2022state}; and wake interactions degrade turbine efficiency and lifetime~\citep{badshah2019coupled, lee2012atmospheric}. Resolving multiple shedding cycles at sufficiently fine spatial-temporal resolution often requires tens of thousands of time steps and hours of wall-clock time on HPC systems~\citep{Tali2024, franke1990numerical}, making large‐scale design sweeps or real-time inference infeasible. These challenges motivate the development of fast and reliable scientific machine learning (SciML) surrogates that retain physical fidelity while offering orders-of-magnitude speedups~\citep{rabeh2025benchmarking}.

Neural operators have emerged as a promising class of surrogates for PDEs by learning mappings between function spaces rather than individual solution instances~\citep{kovachki2023}. Among them, Deep Operator Networks (DeepONet) encode an input function (branch) and a query location (trunk) to regress field values~\citep{lu2020}, and have been applied across steady and transient physics in 2D/3D~\citep{rabeh20253d, wang2021learning, shadkhah2025mpfbench, li2023phase}. Beyond benchmark PDEs, DeepONet surrogates have been demonstrated in applied engineering settings including robust zero-shot generalization to unseen scenarios~\citep{kobayashi2024}, real-time prediction of complex spatio-temporal processes such as ocean wave dynamics~\citep{zhang2025deonwave}, and steel solidification modeling~\citep{koric2026seqdeeponet}. DeepONet variants have also been deployed in real-time workflows that fuse simulations with sparse/noisy monitoring data for field prediction in construction engineering~\citep{xu2024multifidelity}. Related efforts have extended DeepONet-style operators to industrial multiphysics problems in manufacturing, demonstrating feasibility under sequential loading~\citep{kushwaha2024addma}. Alternative operator families include spectral models (such as the Fourier Neural Operator, FNO) and transformer‐based operators that capture long-range spatio-temporal dependencies~\citep{li2023scalable, herde2024poseidon}. These approaches have demonstrated impressive speedups on canonical benchmarks (e.g., cylinder wakes) and even global weather surrogates~\citep{dai2023fourier, pathak2022fourcastnet}. 

Despite rapid progress, stable long-horizon rollouts for unsteady flows over arbitrary shapes remain difficult. Autoregressive prediction can accumulate small step-wise errors into unphysical fields~\citep{lippe2023pde, majid2024mixture}; moreover, fidelity depends critically on how geometry and flow history are represented. Recent work has addressed these challenges separately. Geometry-aware models encode shapes via point clouds, meshes, or low-dimensional parametric descriptors~\citep{he2024, li2023geometry, karki2025direct, li2023fourier}. Point-cloud DeepONet and its variants inject surface information from point clouds or meshes into the trunk network, improving shape generalization in steady and quasi-steady flows~\citep{he2024, he2024sequential}. Recent operator advances also combine attention mechanisms with geometry encoders to enable predictions on arbitrary 2D/3D geometries from surface point clouds, avoiding reliance on structured grids~\citep{liu2026ginot}. The Geometry-Informed Neural Operator (GINO) utilizes graph-based kernels to propagate signals across geometric graphs, offering resolution-invariant shape conditioning~\citep{li2023geometry}. These geometry-aware methods advance shape representation but lack explicit mechanisms for leveraging recent spatio-temporal evolution, which is crucial for capturing unsteady wake dynamics.

Complementary approaches target temporal stability. The Temporal Neural Operator (TNO) aggregates prior solution fields through a dedicated temporal branch, improving accuracy on time-dependent PDEs~\citep{diab2025temporal}. PDE-Refiner applies diffusion-based denoising to correct autoregressive predictions, stabilizing rollouts without modifying the base predictor~\citep{lippe2023pde}. Mixture operators blend multiple temporal pathways to learn complementary dynamics~\citep{majid2024mixture}. A complementary stability direction is to learn time-derivative operators and couple them with numerical time integration at inference~\citep{nayak2025tideeponet}. While these methods improve rollout stability, they provide limited or no geometry conditioning, restricting their effectiveness on diverse, non-parametric shapes. Achieving robustness across diverse geometries \emph{and} stability across long rollouts requires addressing both challenges simultaneously.

We address this gap by coupling geometry and temporal modeling within a unified architecture. Our approach builds on the Geometric Deep Operator Network~\citep{he2024} by retaining its SDF-based trunk for geometry encoding and introducing a CNN-based history encoder in the branch network to model temporal evolution. The SDF provides dense, resolution-agnostic implicit representation that is straightforward to mask and differentiate, offering practical advantages over point-cloud~\citep{he2024, he2024sequential} and graph-kernel schemes~\citep{li2023geometry}. Unlike purely temporal stabilizers~\citep{diab2025temporal, lippe2023pde, majid2024mixture}, our design explicitly couples geometry and history, targeting the dual sensitivity of unsteady wakes to both boundary shape and flow evolution. We train and evaluate on three FlowBench~\citep{Tali2024} shape families: (1) smooth NURBS, (2) irregular spherical-harmonic shapes, and (3) non-parametric SkelNetOn contours~\citep{demir2019skelneton}, studying both single-step accuracy and autoregressive rollouts with emphasis on cross-shape generalization.

Beyond standard error metrics, we adopt physics-based diagnostics tailored to unsteady wakes: (a) phase error at wake probes (time and frequency domain) to assess shedding frequency and lag, and (b) divergence norms to quantify incompressibility consistency over rollouts. These diagnostics, together with shape-conditioned analyses comparing sharp-cornered versus smooth geometries, illuminate failure modes and suggest practical remedies. Our contributions include:
\begin{itemize}
    \item A time-dependent, geometry- and history-aware DeepONet that couples SDF-based implicit geometry with CNN history encoding for unsteady flows over parametric and non-parametric shapes.
    \item A systematic study of history length and shape variability effects on single-step and rollout accuracy across three FlowBench families.
    \item Physics-centric rollout diagnostics (probe phase error; divergence norms) that expose long-horizon drift and relate it to geometric sharpness.
    \item Practical guidance on sequence length selection and geometric complexity effects that inform robust surrogate design for geometry-rich unsteady CFD.
\end{itemize}
Our evaluation emphasizes two aspects often overlooked in neural operator research: generalization across diverse geometric families and physical consistency in long-horizon rollouts. We assess performance not only through standard error metrics but also via wake-specific diagnostics that quantify temporal coherence and incompressibility violations.

We benchmark our method against MeshGraphNet~\citep{pfaff2020learning}, a message-passing neural network for learning physical dynamics on meshes and graphs. MeshGraphNet has demonstrated strong accuracy for fluid flow problems and has been used as a reference graph-based surrogate model~\citep{schmackcker2024generalization, physicsnemo2025}. In this work, we adapt MeshGraphNet to our problem and use it as a representative graph neural network baseline for comparison against operator-based models. Our MeshGraphNet implementation is based on the NVIDIA PhysicsNeMo repository~\footnote{\url{https://docs.nvidia.com/physicsnemo/25.11/physicsnemo/examples/cfd/vortex_shedding_mgn}}.

The rest of the paper is organized as follows. \secref{sec:data-model} details datasets and our time-dependent Geometric DeepONet. \secref{sec:results} reports quantitative/qualitative results and physics diagnostics. \secref{sec:conclusion} summarizes findings, limitations, and future directions.

\section{Dataset and Model Architecture} \label{sec:data-model}

We first describe our neural operator architecture, then detail the dataset used for training and evaluation.

\subsection{Time-Dependent Geometric DeepONet}

\begin{figure}[!t]
    \centering
    \input{Figures/overview_architecture.tex} \\[1em]
    \begin{subfigure}[b]{0.99\linewidth}
  \resizebox{\textwidth}{!}{%
    \begin{tikzpicture}[
        scale=0.8, transform shape,
        font=\scriptsize, >=Stealth,
        node distance=3mm and 4mm,
        arrow/.style={->, line width=0.6pt, shorten >=1pt, shorten <=1pt},
        input/.style={rectangle,draw,thick,fill=blue!10,
                      minimum width=5mm,minimum height=4mm,align=center},
        conv1/.style={rectangle,draw,thick,fill=orange!60,
                      minimum width=5mm,minimum height=4mm,align=center,font=\tiny\bfseries},
        conv3/.style={rectangle,draw,thick,fill=orange!45,
                      minimum width=5mm,minimum height=4mm,align=center,font=\tiny\bfseries},
        conv5/.style={rectangle,draw,thick,fill=orange!30,
                      minimum width=5mm,minimum height=4mm,align=center,font=\tiny\bfseries},
        pool/.style={rectangle,draw,thick,fill=green!50,
                      minimum width=5mm,minimum height=4mm,align=center,font=\tiny\bfseries},
        concat/.style={circle,draw,thick,fill=yellow!50,minimum size=5mm,align=center},
        fusion/.style={rectangle,draw,thick,fill=orange!60,
                       minimum width=6mm,minimum height=4mm,align=center,font=\tiny\bfseries},
        flat/.style={rectangle,draw,thick,fill=gray!30,
                      minimum width=5mm,minimum height=4mm,align=center,font=\tiny}
      ]

      \node[input] (inp) {Input};
      \node[above=4mm of inp,font=\scriptsize]   {$[B,2 \times N_t,H,W]$};
      \node[conv1,right=8mm of inp,yshift=4mm]  (b1c1) {Conv 1×1};
      \node[pool, right=of b1c1]            (b1p1) {Pool 2×2};
      \node[above=1mm of b1p1,font=\scriptsize]   {$[B,c_1,H/2,W/2]$};
      \node[conv1,right=of b1p1]            (b1c2) {Conv 1×1};
      \node[pool, right=of b1c2]            (b1p2) {Pool 2×2};
      \node[above=1mm of b1p2,font=\scriptsize]   {$[B,c_2,H/4,W/4]$};
      \node[conv1,right=of b1p2]            (b1c3) {Conv 1×1};
      \node[pool, right=of b1c3]            (b1p3) {Pool 2×2};
      \node[above=1mm of b1p3,font=\scriptsize]   {$[B,c_3,H/8,W/8]$};
      \node[conv3,right=8mm of inp]             (b2c1) {Conv 3×3};
      \node[pool, right=of b2c1]            (b2p1) {Pool 2×2};
      \node[conv3,right=of b2p1]            (b2c2) {Conv 3×3};
      \node[pool, right=of b2c2]            (b2p2) {Pool 2×2};
      \node[conv3,right=of b2p2]            (b2c3) {Conv 3×3};
      \node[pool, right=of b2c3]            (b2p3) {Pool 2×2};
      \node[conv5,right=8mm of inp,yshift=-4mm] (b3c1) {Conv 5×5};
      \node[pool, right=of b3c1]            (b3p1) {Pool 2×2};
      \node[conv5,right=of b3p1]            (b3c2) {Conv 5×5};
      \node[pool, right=of b3c2]            (b3p2) {Pool 2×2};
      \node[conv5,right=of b3p2]            (b3c3) {Conv 5×5};
      \node[pool, right=of b3c3]            (b3p3) {Pool 2×2};
      \draw[arrow] (inp.east) to[bend left=12]  (b1c1.west);
      \draw[arrow] (inp.east) --                 (b2c1.west);
      \draw[arrow] (inp.east) to[bend right=12] (b3c1.west);
      \draw[arrow] (b1c1)--(b1p1);
      \draw[arrow] (b1p1)--(b1c2);
      \draw[arrow] (b1c2)--(b1p2);
      \draw[arrow] (b1p2)--(b1c3);
      \draw[arrow] (b1c3)--(b1p3);
      \draw[arrow] (b2c1)--(b2p1);
      \draw[arrow] (b2p1)--(b2c2);
      \draw[arrow] (b2c2)--(b2p2);
      \draw[arrow] (b2p2)--(b2c3);
      \draw[arrow] (b2c3)--(b2p3);
      \draw[arrow] (b3c1)--(b3p1);
      \draw[arrow] (b3p1)--(b3c2);
      \draw[arrow] (b3c2)--(b3p2);
      \draw[arrow] (b3p2)--(b3c3);
      \draw[arrow] (b3c3)--(b3p3);
      \coordinate (mid) at ($(b1p3.east)!0.5!(b3p3.east)$);
      \node[concat] (cat) at ($(mid)+(15mm,0)$) {\Large$\oplus$};
      \draw[arrow] (b1p3.east) -- ++(2mm,0) |- (cat);
      \draw[arrow] (b2p3.east) -- (cat);
      \draw[arrow] (b3p3.east) -- ++(2mm,0) |- (cat);
      \node[above=0mm of cat,font=\scriptsize]   {$[B,3 \times c_3,H/8,W/8]$};
      \node[fusion,right=of cat] (f1) {Conv 1×1};
      \draw[arrow] (cat.east) -- (f1.west);
      \node[pool, right=of f1] (fp1) {Pool 2×2};
      \draw[arrow] (f1.east) -- (fp1.west);
      \node[above=1mm of fp1,font=\scriptsize]   {$[B,fc_1,H/16,W/16]$}; 
      \node[fusion,right=of fp1] (f2) {Conv 1×1};
      \draw[arrow] (fp1.east) -- (f2.west);
      \node[pool, right=of f2] (fp2) {Pool 2×2};
      \draw[arrow] (f2.east) -- (fp2.west);
      \node[above=1mm of fp2,font=\scriptsize]   {$[B,fc_2,H/32,W/32]$};
      \node[flat,right=8mm of fp2] (flatnode) {Flattened};
      \draw[arrow] (fp2.east) -- (flatnode.west);

    \end{tikzpicture}%
  }
\end{subfigure}
    \caption{Time-dependent Geometric DeepONet architecture: the branch network in Stage-1 processes $N_t$ velocity frames through parallel convolutional streams (Inception‐style CNN) fed into an MLP, while the trunk network encodes spatial $(x,y,\mathrm{SDF})$ via another MLP. These are fused element‐wise, passed through a Stage‐2 branch MLP (ReLU) and trunk MLP (sine), and finally contracted to predict the next‐step velocity field.}
    \label{fig:model-architecture}
\end{figure}

\figref{fig:model-architecture} illustrates our time-dependent Geometric DeepONet, which consists of two parallel networks -- branch and trunk -- and a two‐stage fusion process. The branch network extracts multi-scale features from a sequence of $N_t$ past velocity fields (we denote this input sequence length by $s=N_t$) by applying three parallel convolutional streams ($1\times1$, $3\times3$, and $5\times5$ kernels), each followed by $2\times2$ max‐pooling, to reduce the spatial resolution by a factor of 32. These feature maps are concatenated, flattened, and fed into a three‐layer MLP (Stage 1) to produce a global latent vector of dimension $m$. Simultaneously, the trunk network (Stage 1) processes each spatial query point by taking its coordinates $(x,y)$ and corresponding SDF value through a three‐layer MLP, yielding a local feature vector of dimension $m$ at each grid point. We fuse the branch and trunk outputs via an element‐wise product, thereby combining temporal and geometric information. Although the CNN history encoder operates on a Cartesian grid, the formulation can handle arbitrary outer-domain geometries by embedding the domain in a bounding box and using a binary occupancy mask to restrict loss evaluation, consistent with immersed-boundary neural formulations~\citep{Khara2024}.

In Stage 2, the fused tensor is split into two paths. The branch path first computes a spatial average over all grid points and processes the resulting vectors with a three‐layer MLP (Stage 2) using ReLU activations. The trunk path retains the full fused tensor (without averaging) and feeds it into another three‐layer MLP (Stage 2) using sine activations, generating per‐point outputs. A final dot‐product contraction along the latent modes between the branch and trunk outputs yields the predicted velocity components $(u,v)$ at each spatial location. For a concise summary of the entire data‐flow, see \algoref{alg:geo-deeponet}.

\begin{algorithm}[b!]
\caption{Time‐Dependent Geometric DeepONet}
\label{alg:geo-deeponet}
\begin{algorithmic}[1]
\Require Past $N_t$ velocity frames $\{u^{t-N_t},\dots,u^{t-1}\}$, SDF grid
\State \textbf{Branch encoding:}
\State \quad Stack past frames into $[B,C_{\rm out}\,N_t,H,W]$
\State \quad Apply three parallel conv streams (kernels $1\times1$, $3\times3$, $5\times5$), each with $2\times2$ max‐pool
\State \quad Fuse via $1\times1$ convs + pooling, then flatten
\State \quad MLP $\to$ global latent vector $\in\mathbb{R}^m$
\State \textbf{Trunk encoding:}
\State \quad For each query point $(x,y)$, read SDF value $\to (x,y,\mathrm{SDF})$
\State \quad MLP $\to$ local feature vector $\in\mathbb{R}^m$
\State \textbf{Stage 1 fusion:} element‐wise product $\to [B,P,m]$
\State \textbf{Stage 2 encoding:}
\State \quad Branch path: spatially average fused tensor, then MLP $\to [B,m\times C_{\rm out}]$
\State \quad Trunk path: apply MLP to each fused feature $\to [B,P,m\times C_{\rm out}]$
\State \textbf{Final fusion:} dot‐product over $m$ modes $\to [B,P,C_{\rm out}]$
\State \textbf{Loss computation:}  
\State \quad $\displaystyle \mathcal{L} = \mathrm{MSE}\bigl((u,v), (u_{\mathrm{gt}},v_{\mathrm{gt}})\bigr)$ over all points with $\mathrm{SDF}>0$
\end{algorithmic}
\end{algorithm}

\subsubsection*{Training Configuration}

We trained our model, which has 1.6 million parameters, using the Adam optimizer with a learning rate of $10^{-3}$, a batch size of 16, and 1000 epochs on a single A100 GPU. Hyperparameters (learning rate, batch size, and network width and depth) were carefully tuned via a structured grid search over multiple candidate values, selecting the configuration that minimized validation loss. Full architectural details, including layer dimensions and hyperparameters, are provided in~\appendixref{appendix:arch-details}. The training and validation loss curves for four different input sequence lengths \(s=1\) through \(s=16\) for our \emph{Time-Dependent Geometric-DeepONet} is reported in \figref{fig:train-val-loss} in~\appendixref{subsec:loss-plots}. Training these models took approximately 12 days.

\subsection{FlowBench Dataset}

\begin{figure}[!t]
    \centering
    \begin{tabular}{@{}cc@{}}
        \includegraphics[width=0.48\textwidth,trim=0 280 0 280,clip]{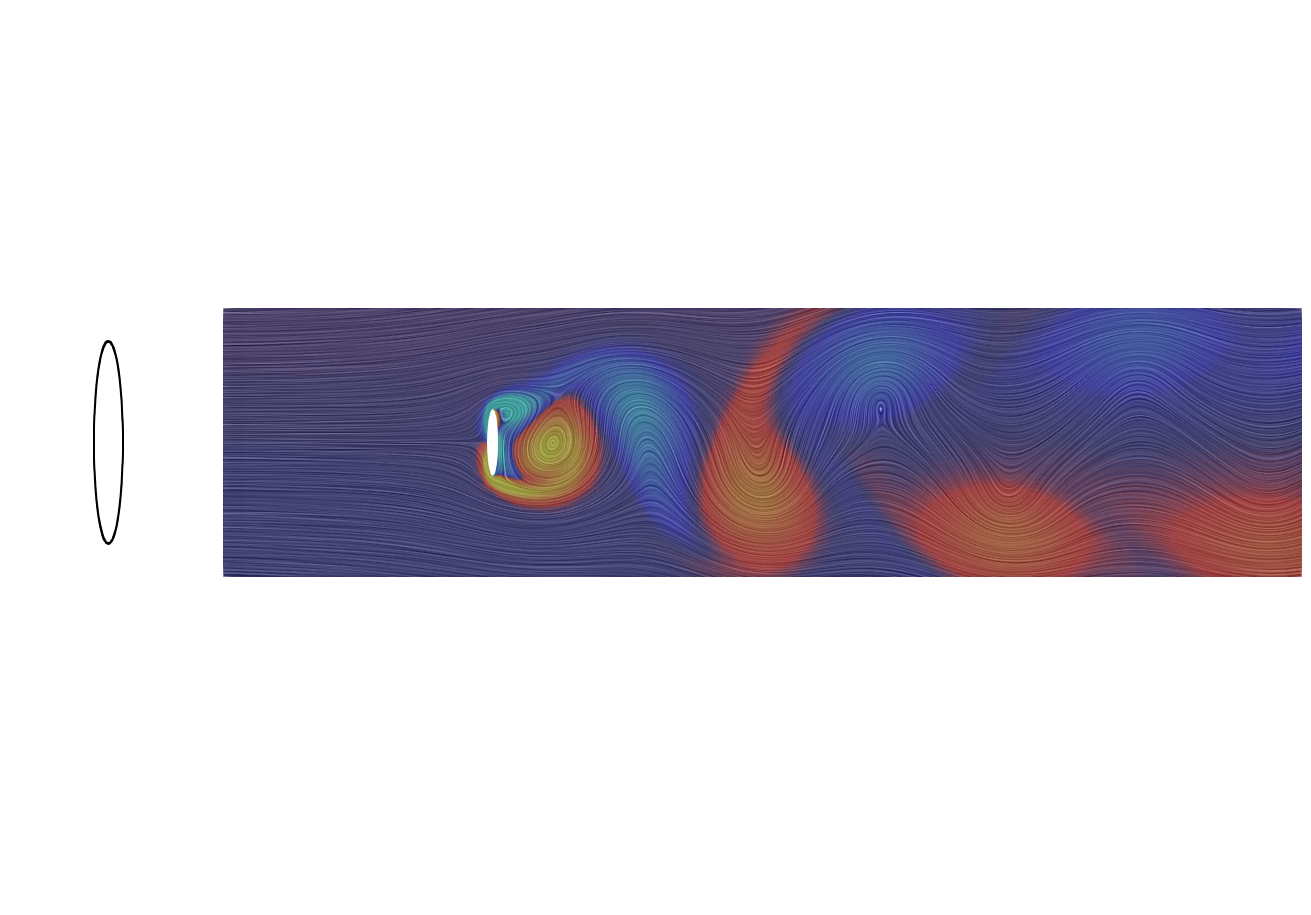} &
        \includegraphics[width=0.48\textwidth,trim=0 280 0 280,clip]{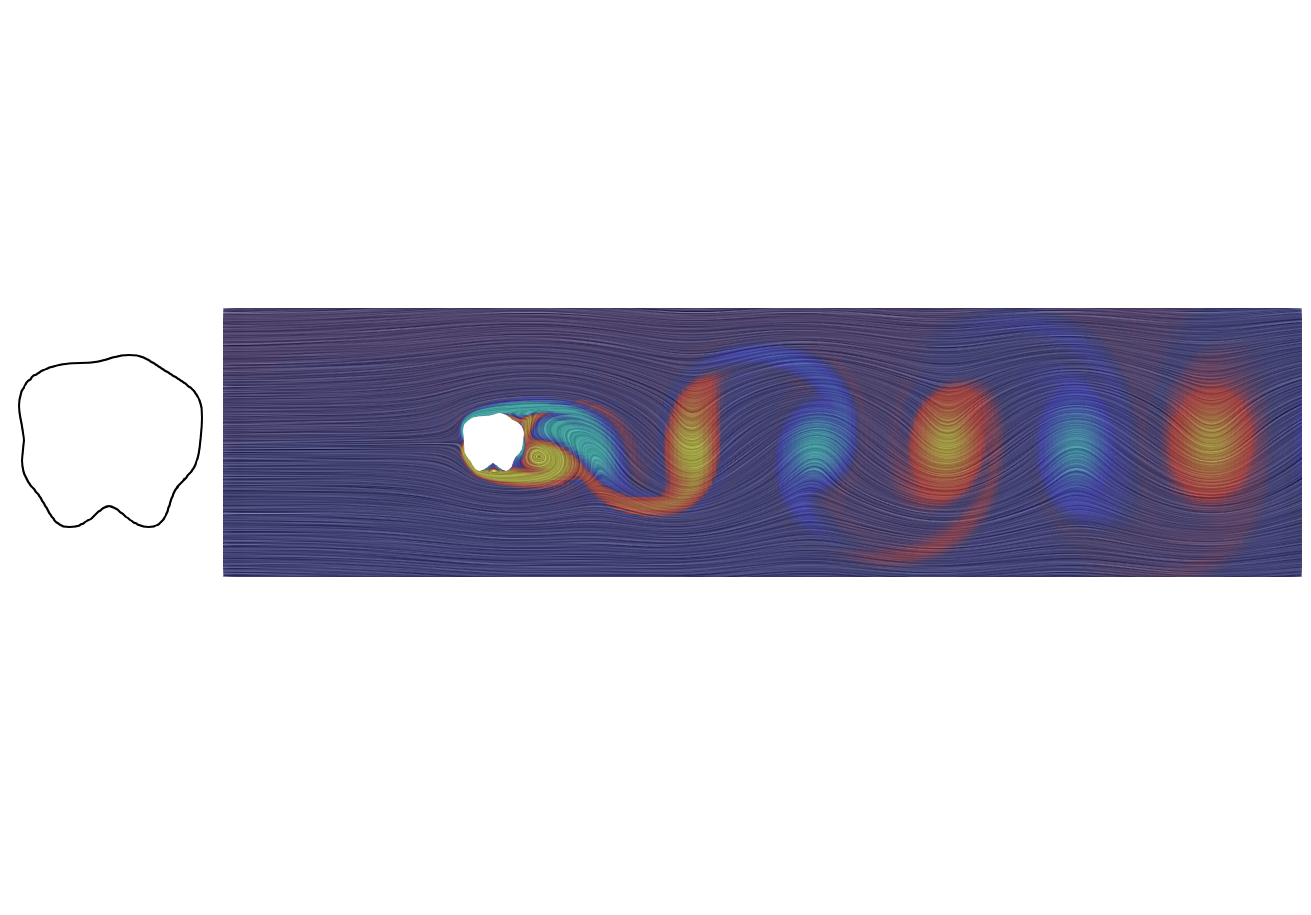} \\[-1pt] 
        \includegraphics[width=0.48\textwidth,trim=0 280 0 280,clip]{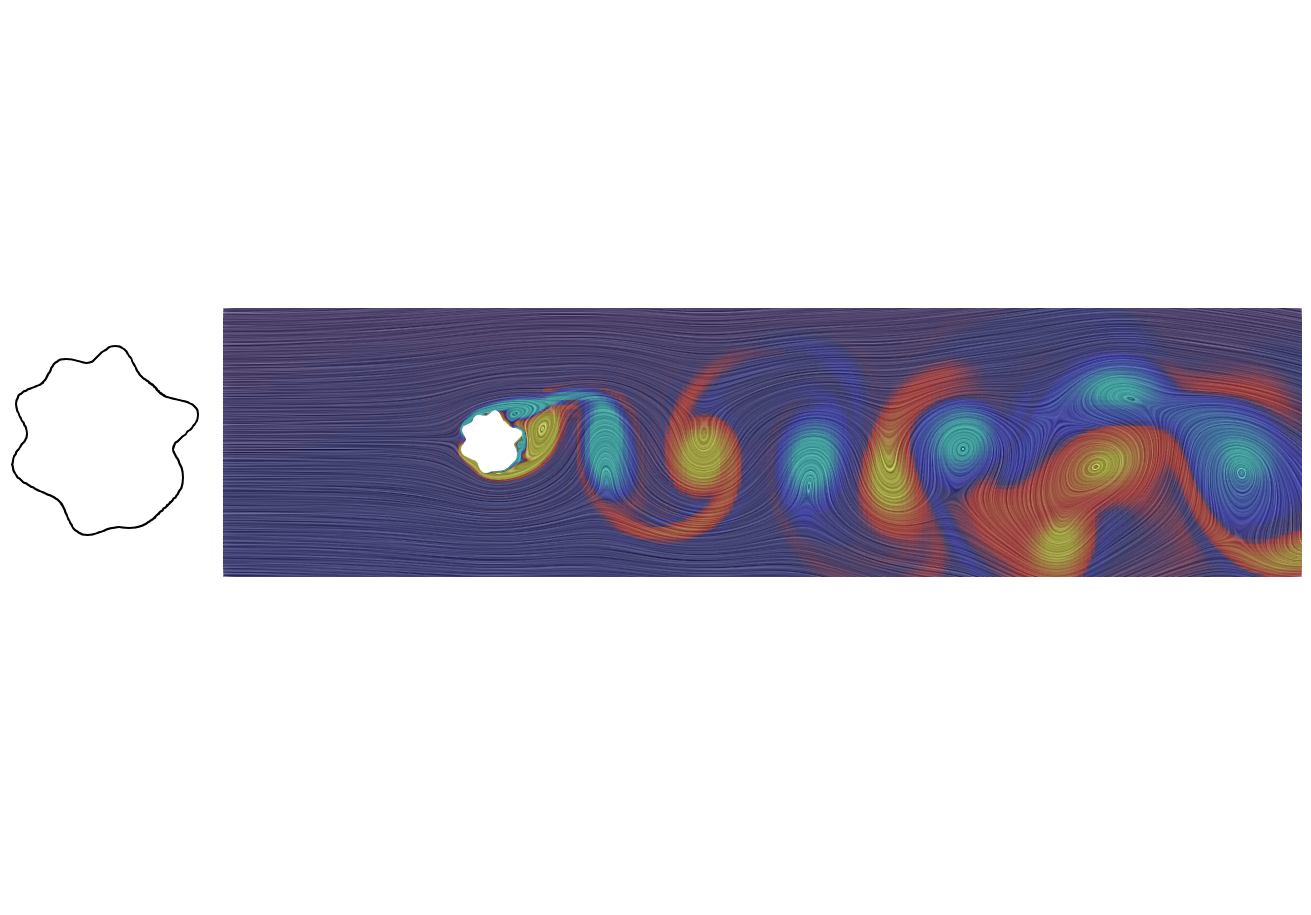} &
        \includegraphics[width=0.48\textwidth,trim=0 280 0 280,clip]{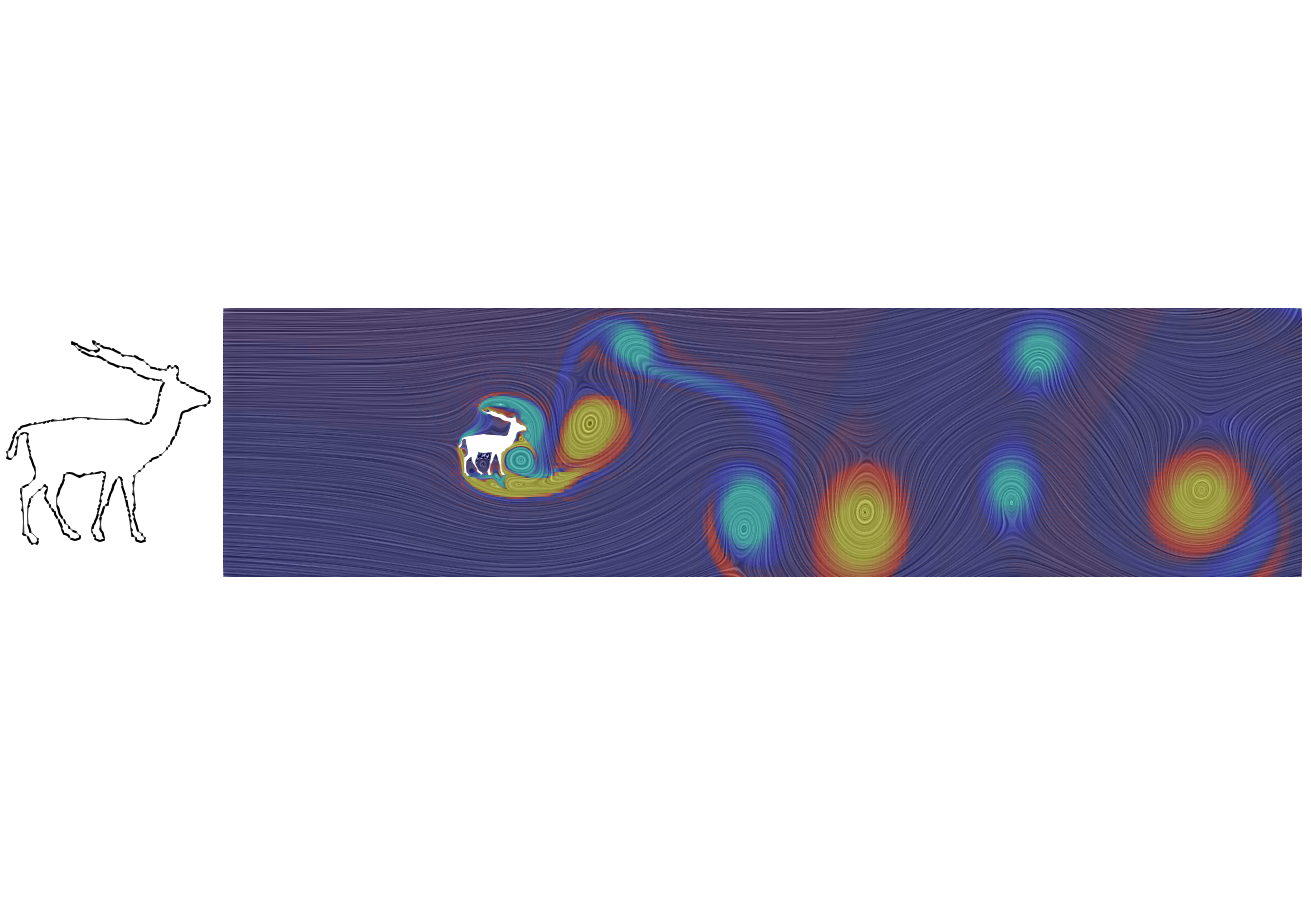} \\[-1pt]
    \end{tabular}
    \caption{Representative snapshots from the FlowBench FPO dataset, illustrating vortex‐shedding behind four representative shapes.}
    \label{fig:dataset-snapshots}
\end{figure}

We use the flow past an object (FPO) dataset from the publicly available FlowBench dataset~\citep{Tali2024}, hosted on  \href{https://huggingface.co/datasets/BGLab/FlowBench/tree/main/FPO_NS_2D_1024x256}{Hugging Face}. This AI‐ready dataset comprises 1,103 high-fidelity 2D simulations of unsteady, incompressible flows past complex shapes on a $1024 \times 256$ grid with 242 temporal snapshots per case. To balance accuracy and efficiency, we uniformly subsample every fourth frame, yielding 60 timesteps per case, while still capturing key vortex-shedding dynamics. Simulation data are generated using a rigorously validated Navier-Stokes solver using the shifted boundary method to enforce boundary conditions on complex geometries~\citep{main2018shifted,yang2024optimal}. Benchmark comparisons show velocity profiles, Strouhal numbers, drag \(C_D \) and lift \(C_L \) coefficients in good agreement with results in the literature~\citet{yang2024simulating}. Examples of flow around different geometries are shown in \figref{fig:dataset-snapshots}.

The FPO dataset is provided as NumPy compressed (\texttt{.npz}) files. We provide two \texttt{.npz} files: one for the inputs, suffixed with the marker \texttt{"\_X.npz"}, and one for the outputs, suffixed with the marker \texttt{"\_Y.npz"}. Each of these \texttt{.npz} files contains a 4D \texttt{NumPy} tensor structured as $[\mathbf{number\_of\_channels}][\mathbf{timesteps}][\mathbf{resolution\_x}][\mathbf{resolution\_y}]
$.

In our workflow, we omit the Reynolds channel at inference, letting the model infer flow conditions from the velocity history, and use only the SDF as input. Similarly, we predict only the velocity components \((u,v)\). since pressure can be recovered post hoc via a Poisson solve. We evaluate model performance by randomly splitting the 1,103 cases into 841 training and 262 test samples. We further divide the training set into an 80\%/20\% random shuffle to form training and validation subsets. The held-out 262-case test set is used exclusively for final evaluation.

\section{Results} \label{sec:results}

We evaluate the time-dependent Geometric DeepONet on the 262-case held-out test set spanning three shape families. Our analysis proceeds in three stages. First, we examine the effect of input sequence length on prediction accuracy to justify our choice of $s=1$ for subsequent experiments. Second, we assess single-step prediction accuracy, where ground-truth inputs are provided at each timestep, to establish the model's capacity for immediate-future prediction across diverse geometries. Third, we evaluate autoregressive rollout performance, where predictions are fed back as inputs, to quantify error accumulation and long-horizon stability.

\paragraph{Metrics}
We employ both standard error metrics and physics-based diagnostics, including Strouhal number analysis and phase lag measurements, to provide comprehensive insight into the surrogate's capabilities and failure modes.
We report two metrics over the test set at each timestep \(t\):

\[L_2(t) =\frac{\sqrt{\sum_{i=1}^{P}\bigl(u^{i}_{\rm pred}(t)-u^{i}_{\rm gt}(t)\bigr)^2}}{\sqrt{\sum_{i=1}^{P}\bigl(u^{i}_{\rm gt}(t)\bigr)^2}},
\]
and
\[ L_\infty(t) = \max_{1\le i\le P}\,\bigl|u^{i}_{\rm pred}(t)-u^{i}_{\rm gt}(t)\bigr|,
\]
where \(P\) is the number of spatial grid points and \(u^{i}\) denotes the velocity component (either \(u\) or \(v\)) at point \(i\).  

\subsection{Effect of Sequence Length on Prediction Accuracy} \label{subsec:sequence-length}

\figref{Fig:input-seq-length} compares the time evolution of relative \(L_{2}\) and \(L_{\infty}\) errors for both single step and autoregressive rollout predictions as the input sequence length \(s\) is varied from 1 to 16. In the single‐step setting (blue curves), all values of \(s\) yield virtually identical, low error from \(t=0\) onward, demonstrating that no additional past context beyond the immediately preceding field is affecting the short‐term accuracy. Under rollouts (red), longer sequences yield slightly lower error for the first few timesteps, an expected benefit of having more initial ground truth frames. But beyond \(t\approx20\), the error trajectories for \(s=1\), 4, 8, and 16 are similar. In other words, larger \(s\) only delays error growth by a handful of steps, without improving long‐term fidelity. Because using \(s>1\) requires that many more ground‐truth inputs (increasing data loading and memory demands) yet offers no lasting accuracy advantage, we select \(s=1\) for all subsequent experiments. This choice minimizes input requirements while preserving both single step and rollout performance.

\begin{figure}[!t]
  \centering
  \begin{subfigure}[t]{0.46\linewidth}
    \centering
    \begin{tikzpicture}
      \begin{axis}[
        width=0.98\textwidth, 
        height=0.6\textwidth,
        xmin=0, xmax=60,
        ymin=0, ymax=1,
        xlabel={time},
        ylabel={Relative $L_2$ Error},
        scaled y ticks  = true,
        scaled y ticks  = base 10:2,        
        legend pos=north east
        ]
        \addplot table[
          col sep=comma,
          x=timestep,
          y=rel_Overall
        ] {Figures/1in/single_step/errors/single_step_errors.csv};
        \addplot table[
          col sep=comma,
          x=timestep,
          y=rel_Overall
        ] {Figures/1in/rollout/errors/rollout_errors.csv};
        \legend{single-step,rollout}
      \end{axis}
    \end{tikzpicture}
    \caption{$L_2$ error using input sequence s=1.}
    \label{fig:sub-13a}
    \end{subfigure}%
    \begin{subfigure}[t]{0.46\linewidth}
    \centering
    \begin{tikzpicture}
      \begin{axis}[
        width=0.98\textwidth, 
        height=0.6\textwidth,
        xmin=0, xmax=60,
        ymin=0, ymax=3,
        xlabel={time},
        ylabel={$L_\infty$ Error},
        scaled y ticks  = true,
        scaled y ticks  = base 10:2,        
        legend pos=north east
        ]
        \addplot table[
          col sep=comma,
          x=timestep,
          y=linf_Overall
        ] {Figures/1in/single_step/errors/single_step_errors.csv};
        \addplot table[
          col sep=comma,
          x=timestep,
          y=linf_Overall
        ] {Figures/1in/rollout/errors/rollout_errors.csv};
        \legend{single-step,rollout}
      \end{axis}
    \end{tikzpicture}
    \caption{$L_\infty$ error using input sequence s=1.}
    \label{fig:sub-13b}
    \end{subfigure}%
    \hfill
    \begin{subfigure}[t]{0.46\linewidth}
    \centering
    \begin{tikzpicture}
      \begin{axis}[
        width=0.98\textwidth, 
        height=0.6\textwidth,
        xmin=3, xmax=60,
        ymin=0, ymax=1,
        xlabel={time},
        ylabel={Relative $L_2$ Error},
        scaled y ticks  = true,
        scaled y ticks  = base 10:2,        
        legend pos=north east
        ]
        \addplot table[
          col sep=comma,
          x=timestep,
          y=rel_Overall
        ] {Figures/4in/single_step/errors/single_step_errors.csv};
        \addplot table[
          col sep=comma,
          x=timestep,
          y=rel_Overall
        ] {Figures/4in/rollout/errors/rollout_errors.csv};
        \legend{single-step,rollout}
      \end{axis}
    \end{tikzpicture}
    \caption{$L_2$ error using input sequence s=4.}
    \label{fig:sub-13c}
    \end{subfigure}%
    \begin{subfigure}[t]{0.46\linewidth}
    \centering
    \begin{tikzpicture}
      \begin{axis}[
        width=0.98\textwidth, 
        height=0.6\textwidth,
        xmin=3, xmax=60,
        ymin=0, ymax=3,
        xlabel={time},
        ylabel={$L_\infty$ Error},
        scaled y ticks  = true,
        scaled y ticks  = base 10:2,        
        legend pos=north east
        ]
        \addplot table[
          col sep=comma,
          x=timestep,
          y=linf_Overall
        ] {Figures/4in/single_step/errors/single_step_errors.csv};
        \addplot table[
          col sep=comma,
          x=timestep,
          y=linf_Overall
        ] {Figures/4in/rollout/errors/rollout_errors.csv};
        \legend{single-step,rollout}
      \end{axis}
    \end{tikzpicture}
    \caption{$L_\infty$ error using input sequence s=4.}
    \label{fig:sub-13d}
    \end{subfigure}%
    \hfill
    \begin{subfigure}[t]{0.46\linewidth}
    \centering
    \begin{tikzpicture}
      \begin{axis}[
        width=0.98\textwidth, 
        height=0.6\textwidth,
        xmin=7, xmax=60,
        ymin=0, ymax=1,
        xlabel={time},
        ylabel={Relative $L_2$ Error},
        scaled y ticks  = true,
        scaled y ticks  = base 10:2,        
        legend pos=north east
        ]
        \addplot table[
          col sep=comma,
          x=timestep,
          y=rel_Overall
        ] {Figures/8in/single_step/errors/single_step_errors.csv};
        \addplot table[
          col sep=comma,
          x=timestep,
          y=rel_Overall
        ] {Figures/8in/rollout/errors/rollout_errors.csv};
        \legend{single-step,rollout}
      \end{axis}
    \end{tikzpicture}
    \caption{$L_2$ error using input sequence s=8.}
    \label{fig:sub-13e}
  \end{subfigure}%
    \begin{subfigure}[t]{0.46\linewidth}
    \centering
    \begin{tikzpicture}
      \begin{axis}[
        width=0.98\textwidth, 
        height=0.6\textwidth,
        xmin=7, xmax=60,
        ymin=0, ymax=3,
        xlabel={time},
        ylabel={$L_\infty$ Error},
        scaled y ticks  = true,
        scaled y ticks  = base 10:2,        
        legend pos=north east
        ]
        \addplot table[
          col sep=comma,
          x=timestep,
          y=linf_Overall
        ] {Figures/8in/single_step/errors/single_step_errors.csv};
        \addplot table[
          col sep=comma,
          x=timestep,
          y=linf_Overall
        ] {Figures/8in/rollout/errors/rollout_errors.csv};
        \legend{single-step,rollout}
      \end{axis}
    \end{tikzpicture}
    \caption{$L_\infty$ error using input sequence s=8.}
    \label{fig:sub-13f}
    \end{subfigure}%
    \hfill
    \begin{subfigure}[t]{0.46\linewidth}
    \centering
    \begin{tikzpicture}
      \begin{axis}[
        width=0.98\textwidth, 
        height=0.6\textwidth,
        xmin=15, xmax=60,
        ymin=0, ymax=1,
        xlabel={time},
        ylabel={Relative $L_2$ Error},
        scaled y ticks  = true,
        scaled y ticks  = base 10:2,        
        legend pos=north east
        ]
        \addplot table[
          col sep=comma,
          x=timestep,
          y=rel_Overall
        ] {Figures/16in/single_step/errors/single_step_errors.csv};
        \addplot table[
          col sep=comma,
          x=timestep,
          y=rel_Overall
        ] {Figures/16in/rollout/errors/rollout_errors.csv};
        \legend{single-step,rollout}
      \end{axis}
    \end{tikzpicture}
    \caption{$L_2$ error using input sequence s=16.}
    \label{fig:sub-13g}
    \end{subfigure}%
    \begin{subfigure}[t]{0.46\linewidth}
    \centering
    \begin{tikzpicture}
      \begin{axis}[
        width=0.98\textwidth, 
        height=0.6\textwidth,
        xmin=15, xmax=60,
        ymin=0, ymax=3,
        xlabel={time},
        ylabel={$L_\infty$ Error},
        scaled y ticks  = true,
        scaled y ticks  = base 10:2,        
        legend pos=north east
        ]
        \addplot table[
          col sep=comma,
          x=timestep,
          y=linf_Overall
        ] {Figures/16in/single_step/errors/single_step_errors.csv};
        \addplot table[
          col sep=comma,
          x=timestep,
          y=linf_Overall
        ] {Figures/16in/rollout/errors/rollout_errors.csv};
        \legend{single-step,rollout}
      \end{axis}
    \end{tikzpicture}
    \caption{$L_\infty$ error using input sequence s=16.}
    \label{fig:sub-13h}
  \end{subfigure}%
  \hfill
\caption{Time‐evolution of single‐step versus rollout prediction errors for varying input sequence lengths. Panels (a) and (b) plot the relative \(L_2\) and \(L_\infty\) errors over time using an input sequence of length \(s=1\). Panels (c) and (d) show the same metrics for \(s=4\); panels (e) and (f) for \(s=8\); and panels (g) and (h) for \(s=16\).}
\label{Fig:input-seq-length}
\end{figure}

\subsection{Single Step prediction} \label{subsec:single-step}

\begin{figure}[!t]
  \centering
\begin{subfigure}[t]{0.48\textwidth}
    \centering
    \begin{tikzpicture}
      \begin{axis}[
        name=relplot,
        width=0.98\textwidth, 
        height=0.6\textwidth,
        xmin=0, xmax=60,
        ymin=0, ymax=0.08,
        xlabel={time},
        ylabel={Relative $L_2$ Error},
        scaled y ticks  = true,
        scaled y ticks  = base 10:2,
        ]
        \addplot table[
          col sep=comma,
          x=timestep,
          y=rel_Overall
        ] {Figures/1in/single_step/errors/single_step_errors.csv};
      \end{axis}
    \end{tikzpicture}
    \caption{Relative $L_2$ error over time.}
    \label{fig:sub-3a}
  \end{subfigure}%
  \begin{subfigure}[t]{0.48\textwidth}
    \centering
    \begin{tikzpicture}
      \begin{axis}[
        width=0.98\textwidth, 
        height=0.6\textwidth,
        at={(relplot.east)}, anchor=west, xshift=2cm,
        xmin=0, xmax=60,
        ymin=0, ymax=0.08,
        xlabel={time},
        ylabel={RMSE},
        scaled y ticks  = true,
        scaled y ticks  = base 10:2,        
        legend pos=north east
      ]
        \addplot table[
          col sep=comma,
          x=timestep,
          y=rmse_U
        ] {Figures/1in/single_step/errors/single_step_errors.csv};
        \addplot table[
          col sep=comma,
          x=timestep,
          y=rmse_V
        ] {Figures/1in/single_step/errors/single_step_errors.csv};
        \legend{$u$,$v$}
      \end{axis}
    \end{tikzpicture}
    \caption{RMSE of $u$ and $v$.}
    \label{fig:sub-3b}
  \end{subfigure}%
  \hfill

  \vspace{1em}

  \begin{subfigure}[t]{\textwidth}
    \centering
    \begin{tikzpicture}
      \matrix [matrix of nodes,
               nodes={inner sep=0, anchor=south west},
               column sep=1pt, row sep=1pt] {
        \node{\subcaptionbox{$u$ GT, $t = 30$}[0.31\textwidth]{%
          \includegraphics[width=0.3\textwidth]{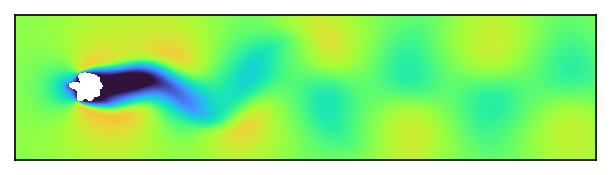}}}; &
        \node{\subcaptionbox{$u$ GT, $t = 45$}[0.31\textwidth]{%
          \includegraphics[width=0.3\textwidth]{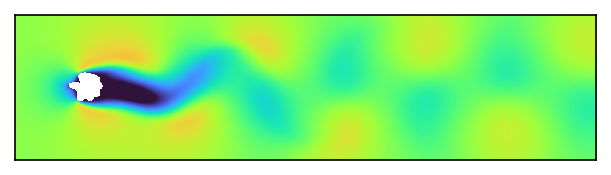}}}; &
        \node{\subcaptionbox{$u$ GT, $t = 59$}[0.31\textwidth]{%
          \includegraphics[width=0.3\textwidth]{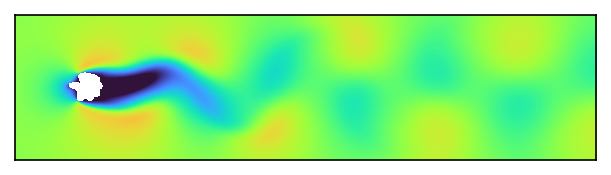}}}; &
        \node{\subcaptionbox*{}[0.037\textwidth]{\includegraphics[width=0.032\textwidth]{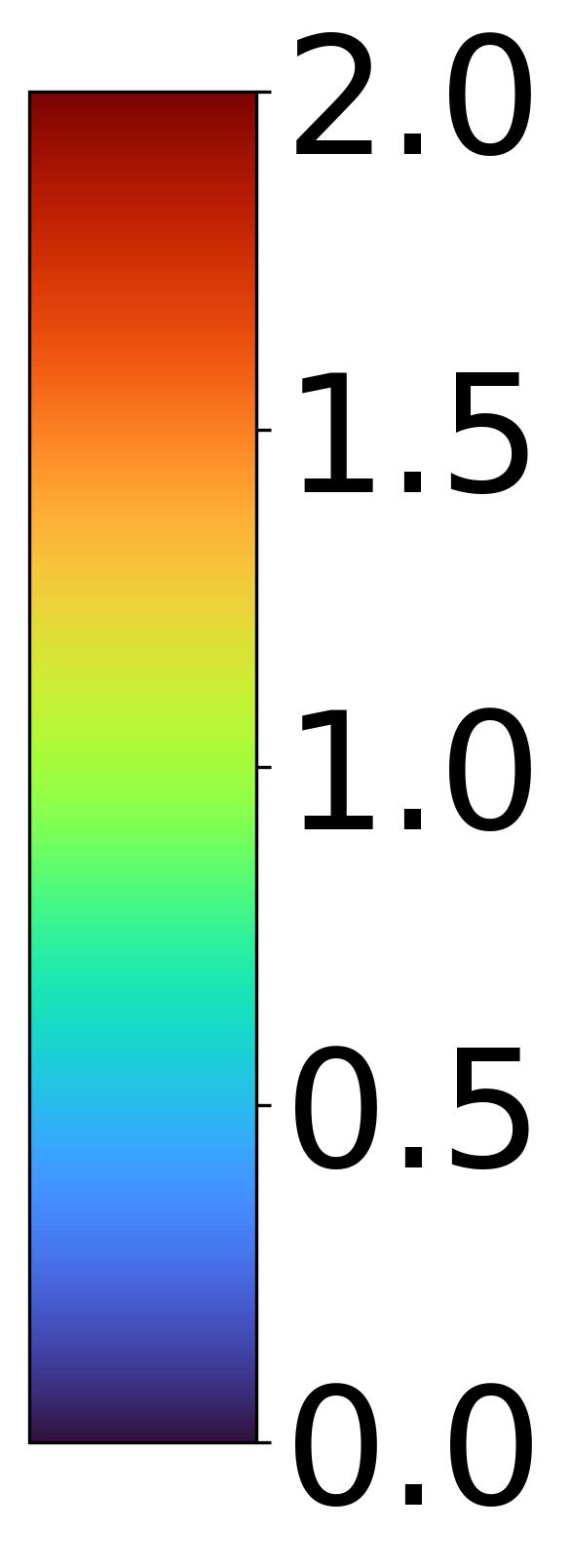}}}; \\
        \node{\subcaptionbox{$u$ Pred, $t = 30$}[0.31\textwidth]{%
          \includegraphics[width=0.3\textwidth]{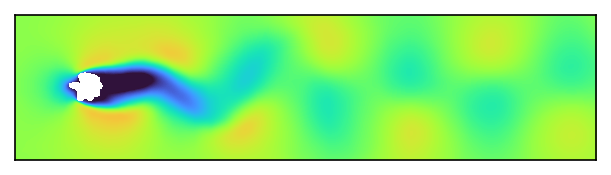}}}; &
        \node{\subcaptionbox{$u$ Pred, $t = 45$}[0.31\textwidth]{%
          \includegraphics[width=0.3\textwidth]{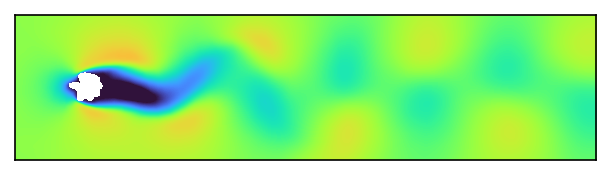}}}; &
        \node{\subcaptionbox{$u$ Pred, $t = 59$}[0.31\textwidth]{%
          \includegraphics[width=0.3\textwidth]{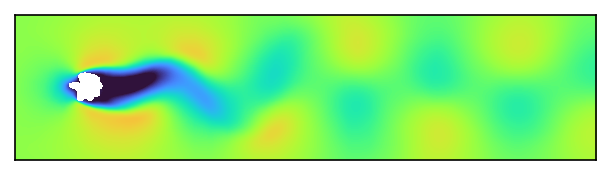}}}; &
        \node{\subcaptionbox*{}[0.037\textwidth] {\includegraphics[width=0.032\textwidth]{Figures/1in/single_step/sample200/u_colorbar.png}}}; \\
      };
    \end{tikzpicture}
    \label{fig:sub-b}
  \end{subfigure}
  \vspace{1em}
  \begin{subfigure}[t]{\textwidth}
    \centering
    \begin{tikzpicture}
      \matrix [matrix of nodes,
               nodes={inner sep=0, anchor=south west},
               column sep=1pt, row sep=1pt] {
        \node{\subcaptionbox{$v$ GT, $t = 30$}[0.31\textwidth]{%
          \includegraphics[width=0.3\textwidth]{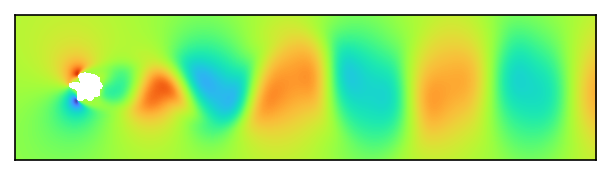}}}; &
        \node{\subcaptionbox{$v$ GT, $t = 45$}[0.31\textwidth]{%
          \includegraphics[width=0.3\textwidth]{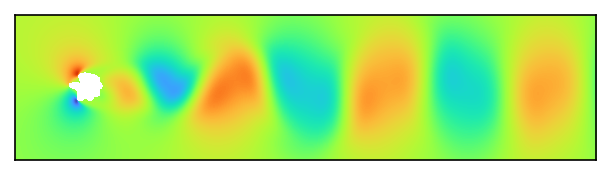}}}; &
        \node{\subcaptionbox{$v$ GT, $t = 59$}[0.31\textwidth]{%
          \includegraphics[width=0.3\textwidth]{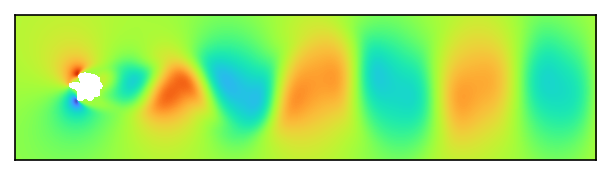}}}; &
        \node{\subcaptionbox*{}[0.04\textwidth] {\includegraphics[width=0.04\textwidth]{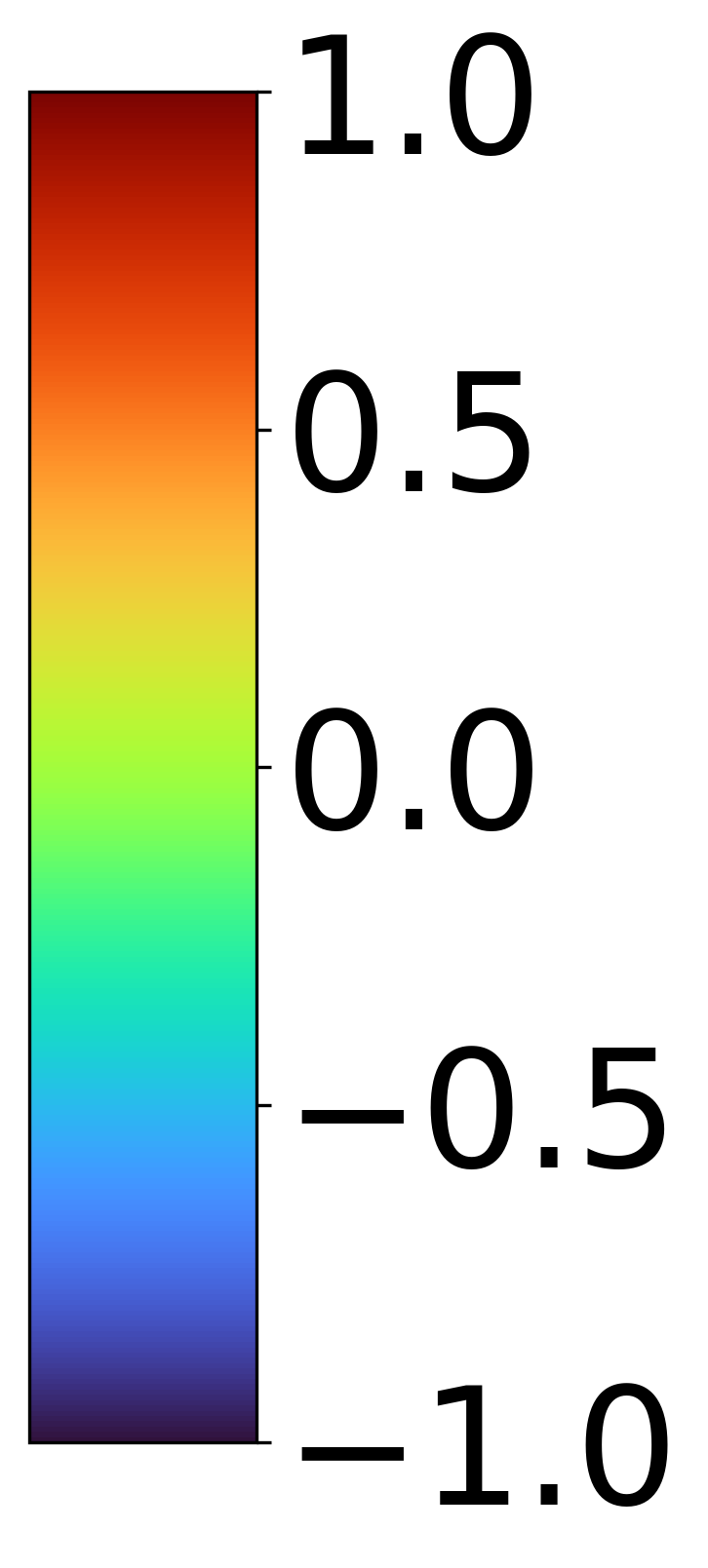}}}; \\
        \node{\subcaptionbox{$v$ Pred, $t = 30$}[0.31\textwidth]{%
          \includegraphics[width=0.3\textwidth]{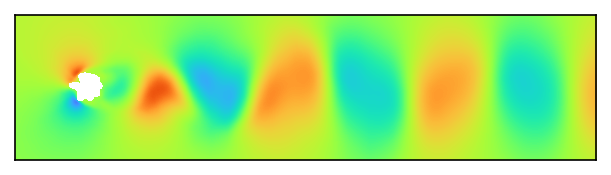}}}; &
        \node{\subcaptionbox{$v$ Pred, $t = 45$}[0.31\textwidth]{%
          \includegraphics[width=0.3\textwidth]{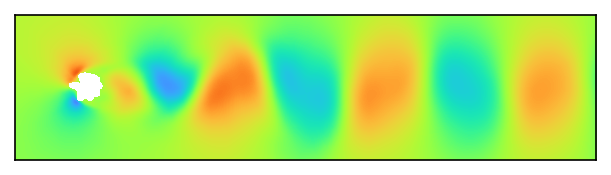}}}; &
        \node{\subcaptionbox{$v$ Pred, $t = 59$}[0.31\textwidth]{%
          \includegraphics[width=0.3\textwidth]{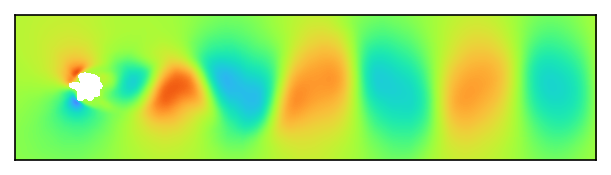}}}; &
        \node{\subcaptionbox*{}[0.04\textwidth] {\includegraphics[width=0.04\textwidth]{Figures/1in/single_step/sample200/v_colorbar.png}}}; \\
      };
    \end{tikzpicture}
    \label{fig:sub-c}
  \end{subfigure}

  \caption{Single‐step prediction of flow velocity for an example geometry. (a) Relative $L_2$ error over time. (b) RMSE of $u$ and $v$ over time. (c-e) Ground‐truth $u$ at $t = 30, 45, 59$. (f-h) Predicted $u$ at $t = 30, 45, 59$. (i-k) Ground‐truth $v$ at $t = 30, 45, 59$. (l-n) Predicted $v$ at $t = 30, 45, 59$. Colorbars for $u$ are shown in (e) and (h), and for $v$ in (k) and (n).}
  
  \label{Fig:single-step-sample200}
\end{figure}

\paragraph{Single‐Step Prediction Accuracy}
As shown in \figref{Fig:single-step-sample200}(a), the relative \(L_{2}\) error remains effectively constant at approximately 5\% over all 60 timesteps, reflecting the model’s one step evaluation where ground‐truth inputs are provided at each step.  Correspondingly, the RMSE for both \(u\) and \(v\) components holds steady at about 0.035 (see \figref{Fig:single-step-sample200}(b)), demonstrating that the network delivers uniform accuracy throughout the time sequence. This stable error profile confirms the model’s capacity to accurately predict the immediate future state when supplied with true past frames. \figref{Fig:single-step-sample200}(c)-(n) shows a comparison of ground truth and prediction velocity components fields (u, v) for a single geometry at three timesteps (\(t=30\), \(t=45\), \(t=59\)). The comparison shows good agreement, reflecting the model's ability to accurately predict the immediate future state.

\begin{figure}[!t]
  \centering
    \begin{tikzpicture}
      \matrix [matrix of nodes,
               nodes={inner sep=0, anchor=south west},
               column sep=1pt, row sep=1pt] {
        \node{\subcaptionbox*{}[0.08\textwidth]{%
          \includegraphics[width=0.08\textwidth]{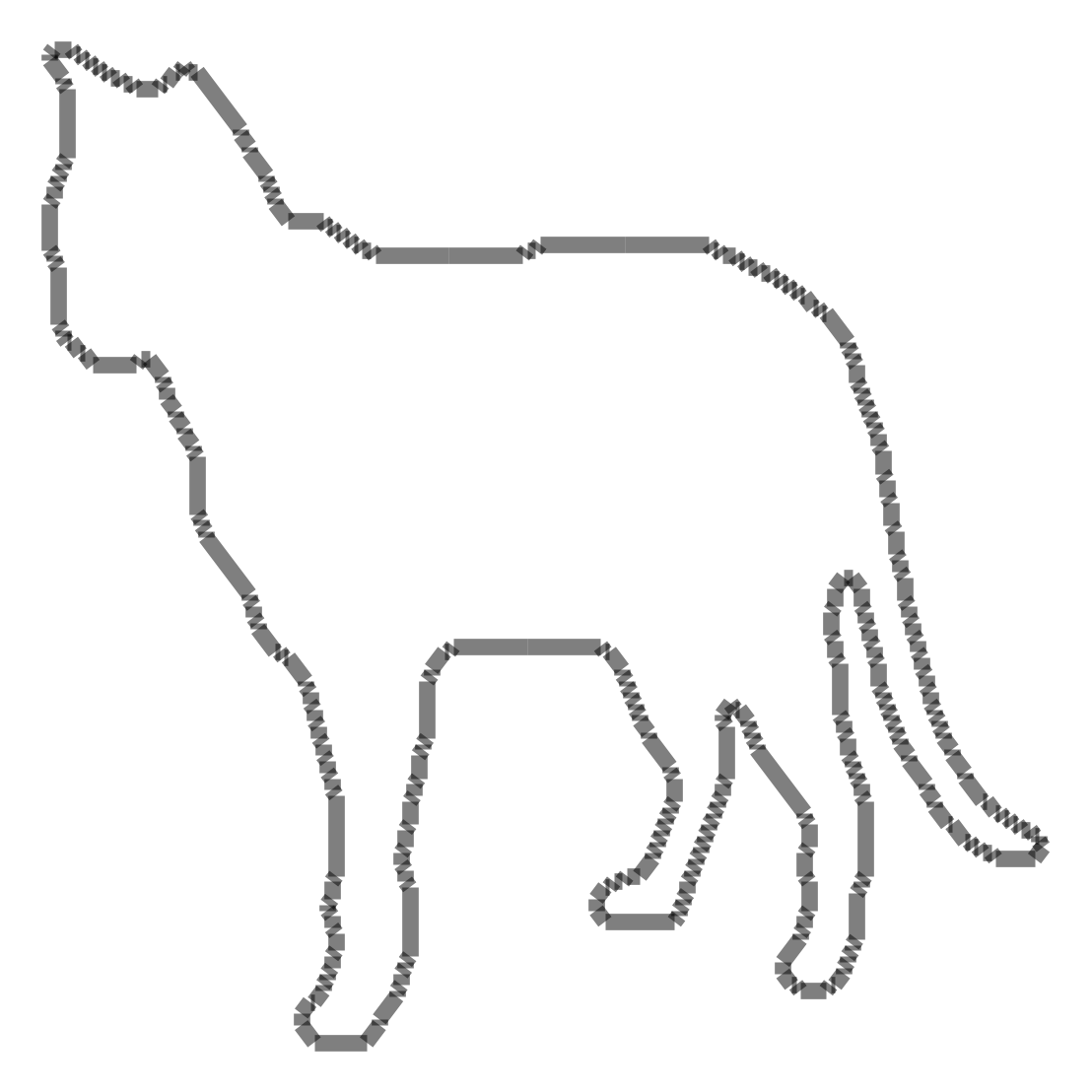}}}; &
        \node{\subcaptionbox{$u$ GT}[0.32\textwidth]{%
          \includegraphics[width=0.32\textwidth]{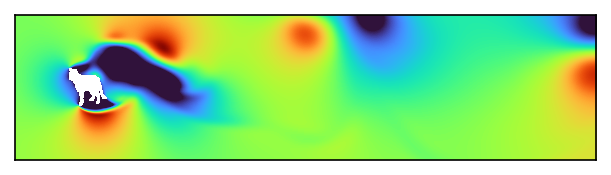}}}; &
        \node{\subcaptionbox{$u$ Pred}[0.32\textwidth]{%
          \includegraphics[width=0.32\textwidth]{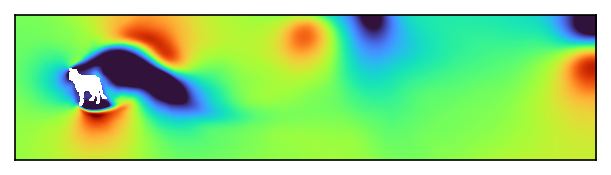}}}; &
        \node{\subcaptionbox*{}[0.037\textwidth]{\includegraphics[width=0.032\textwidth]{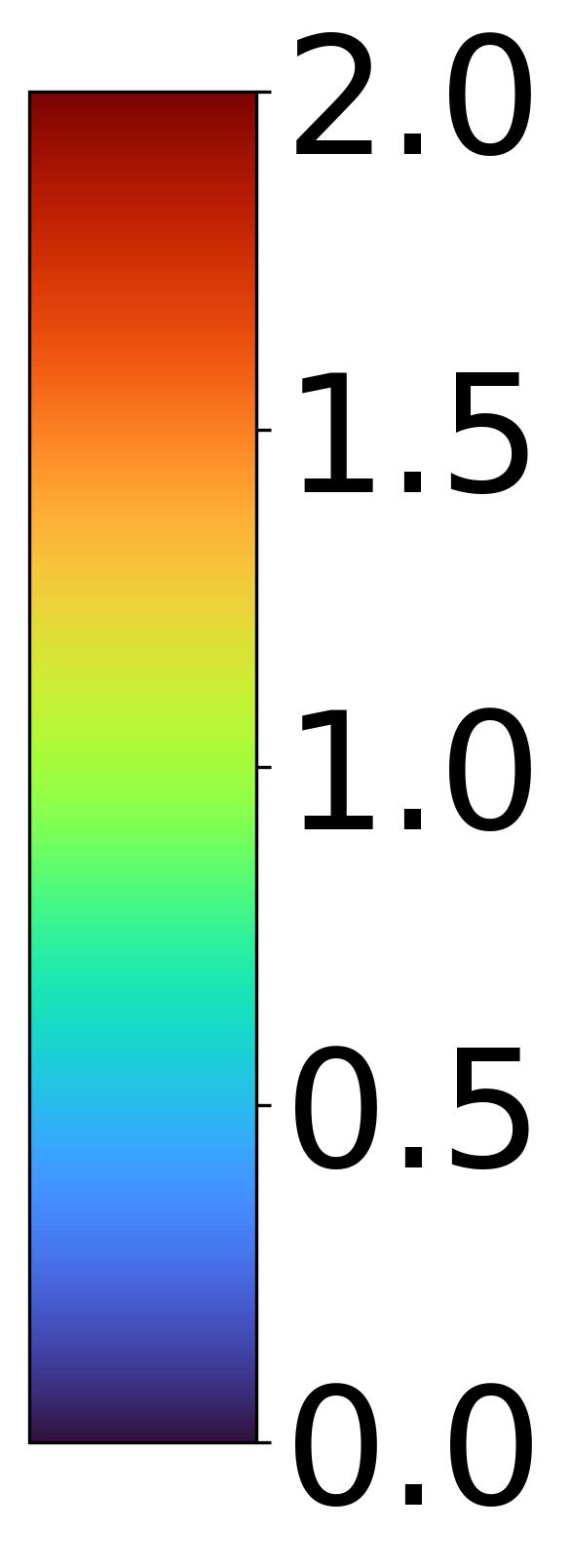}}}; \\
        \node{\subcaptionbox*{}[0.08\textwidth]{%
          \includegraphics[width=0.08\textwidth]{Figures/geometries/sample_0.png}}}; &
        \node{\subcaptionbox{$v$ GT}[0.32\textwidth]{%
          \includegraphics[width=0.32\textwidth]{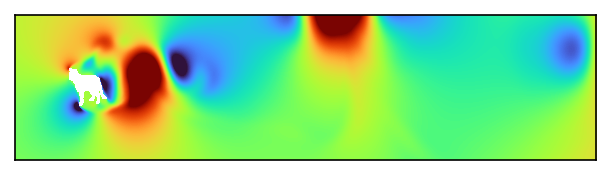}}}; &
        \node{\subcaptionbox{$v$ Pred}[0.32\textwidth]{%
          \includegraphics[width=0.32\textwidth]{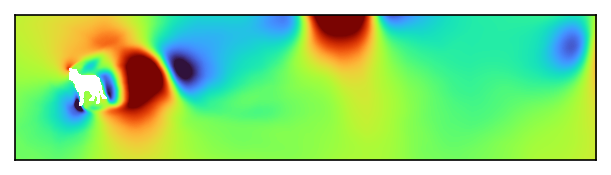}}}; &
        \node{\subcaptionbox*{}[0.04\textwidth]{\includegraphics[width=0.04\textwidth]{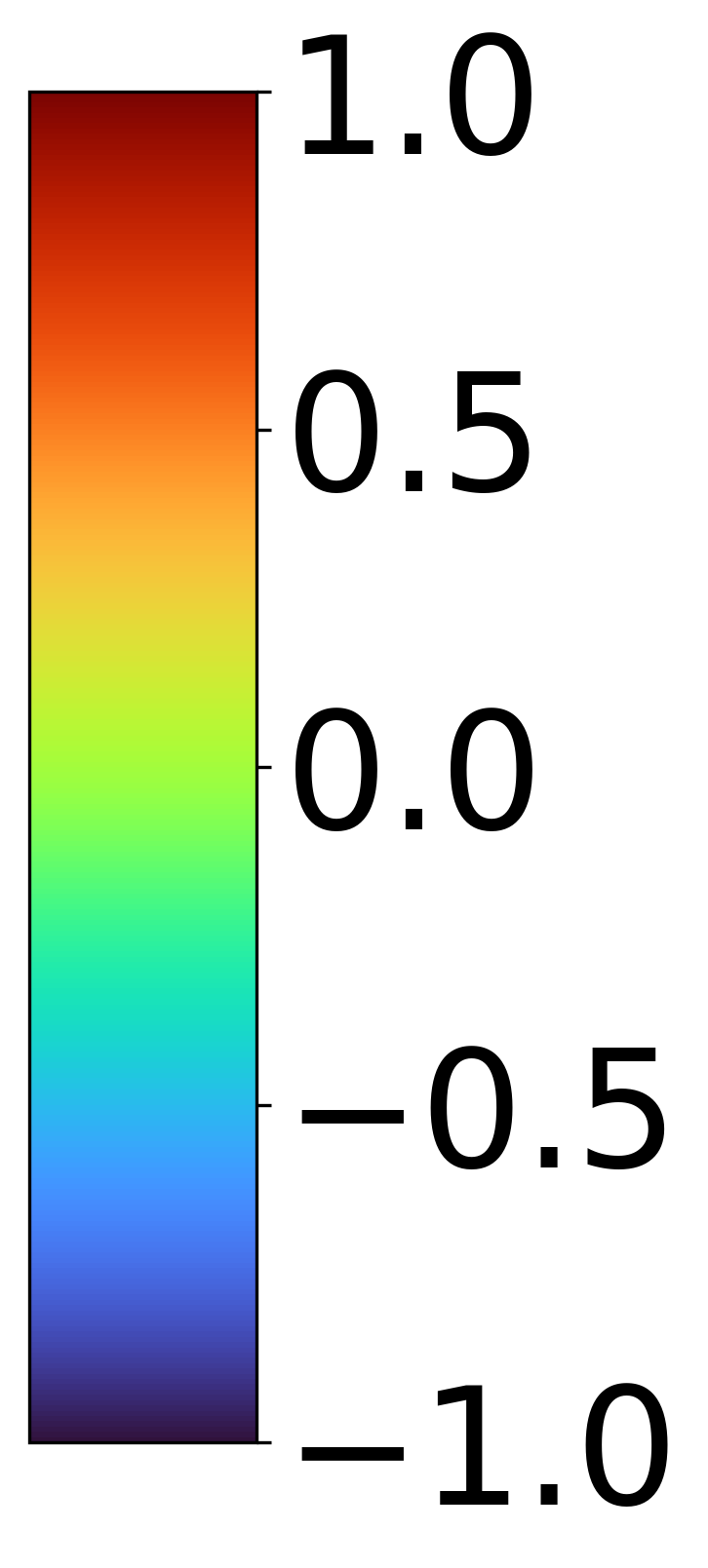}}}; \\
        \node{\subcaptionbox*{}[0.08\textwidth]{%
          \includegraphics[width=0.08\textwidth]{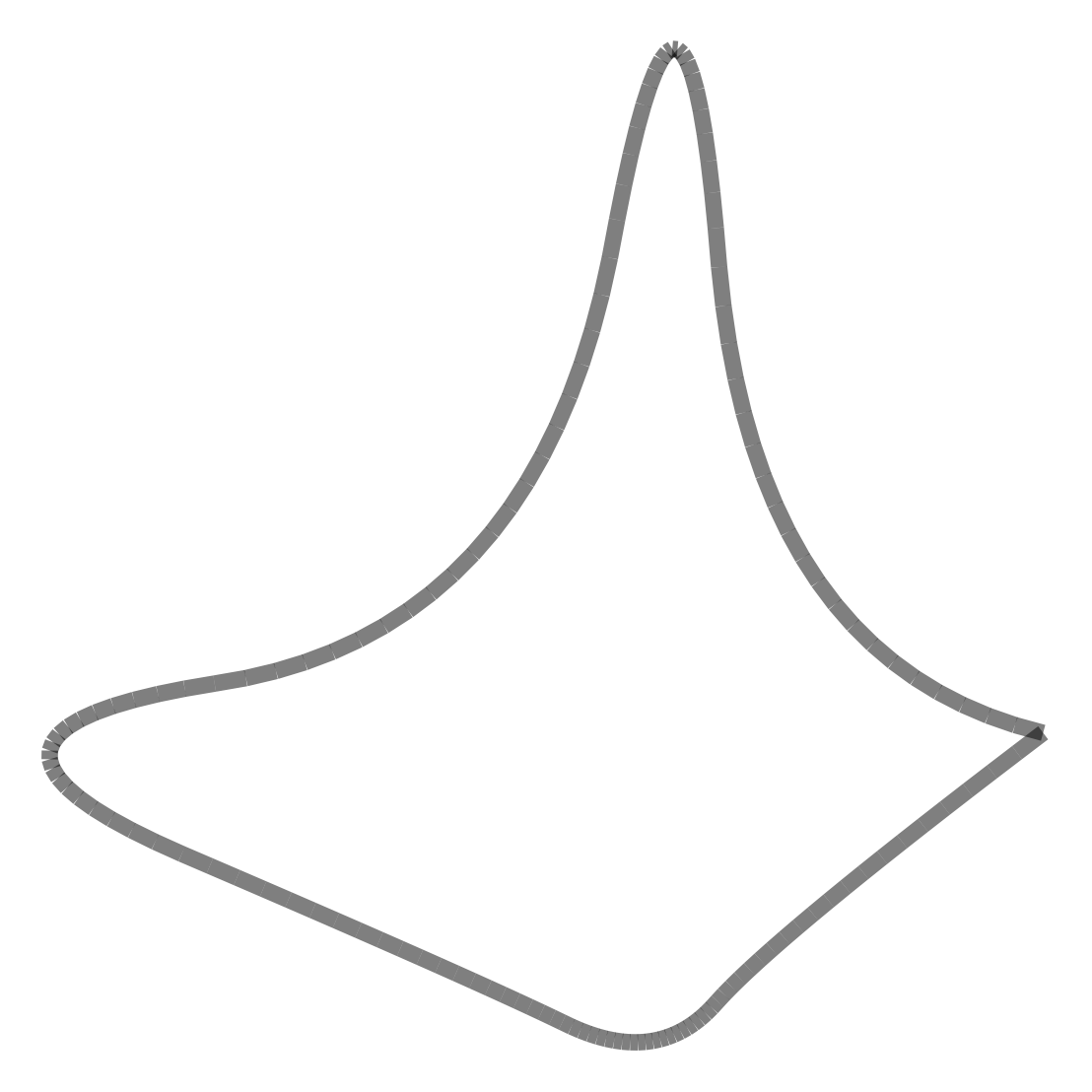}}}; &
        \node{\subcaptionbox{$u$ GT}[0.32\textwidth]{%
          \includegraphics[width=0.32\textwidth]{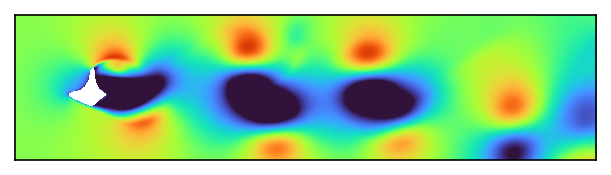}}}; &
        \node{\subcaptionbox{$u$ Pred}[0.32\textwidth]{%
          \includegraphics[width=0.32\textwidth]{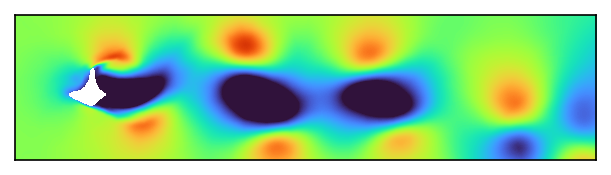}}}; &
        \node{\subcaptionbox*{}[0.037\textwidth]{\includegraphics[width=0.032\textwidth]{Figures/1in/single_step/t=30/u_colorbar.png}}}; \\
        \node{\subcaptionbox*{}[0.08\textwidth]{%
          \includegraphics[width=0.08\textwidth]{Figures/geometries/sample_50.png}}}; &
        \node{\subcaptionbox{$v$ GT}[0.32\textwidth]{%
          \includegraphics[width=0.32\textwidth]{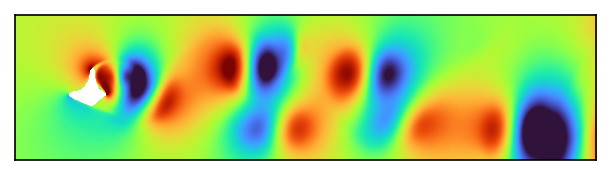}}}; &
        \node{\subcaptionbox{$v$ Pred}[0.32\textwidth]{%
          \includegraphics[width=0.32\textwidth]{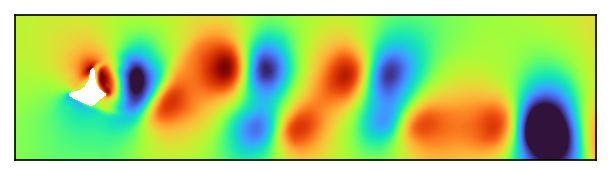}}}; &
        \node{\subcaptionbox*{}[0.04\textwidth]{\includegraphics[width=0.04\textwidth]{Figures/1in/single_step/t=30/v_colorbar.png}}}; \\
        \node{\subcaptionbox*{}[0.08\textwidth]{%
          \includegraphics[width=0.08\textwidth]{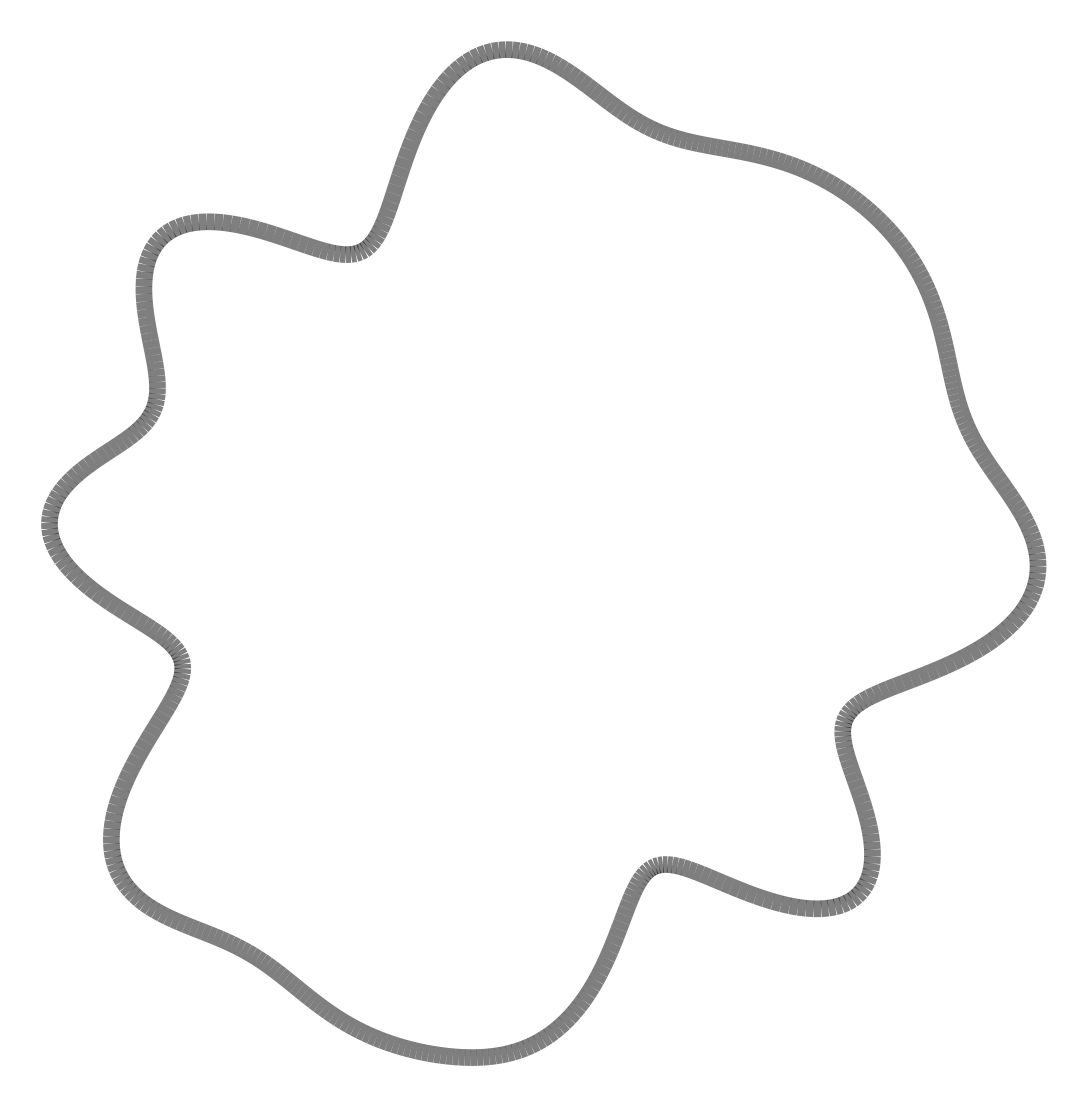}}}; &
        \node{\subcaptionbox{$u$ GT}[0.32\textwidth]{%
          \includegraphics[width=0.32\textwidth]{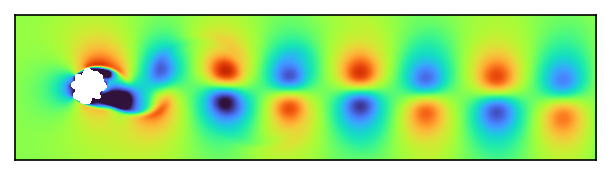}}}; &
        \node{\subcaptionbox{$u$ Pred}[0.32\textwidth]{%
          \includegraphics[width=0.32\textwidth]{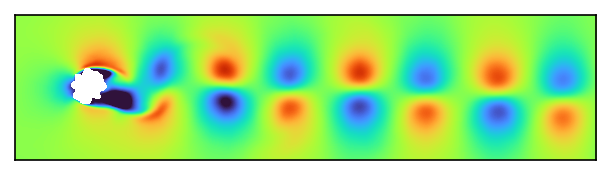}}}; &
        \node{\subcaptionbox*{}[0.037\textwidth]{\includegraphics[width=0.032\textwidth]{Figures/1in/single_step/t=30/u_colorbar.png}}}; \\
        \node{\subcaptionbox*{}[0.08\textwidth]{%
          \includegraphics[width=0.08\textwidth]{Figures/geometries/sample_150.png}}}; &
        \node{\subcaptionbox{$v$ GT}[0.32\textwidth]{%
          \includegraphics[width=0.32\textwidth]{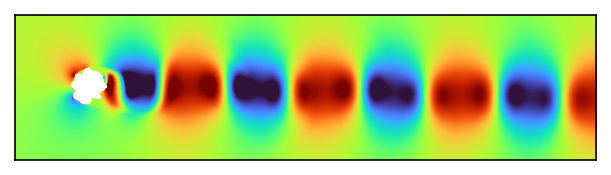}}}; &
        \node{\subcaptionbox{$v$ Pred}[0.32\textwidth]{%
          \includegraphics[width=0.32\textwidth]{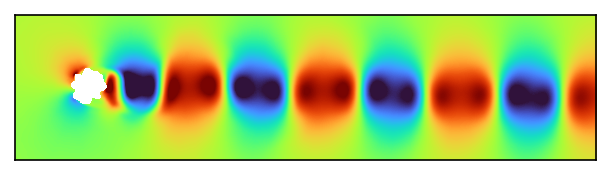}}}; &
        \node{\subcaptionbox*{}[0.04\textwidth]{\includegraphics[width=0.04\textwidth]{Figures/1in/single_step/t=30/v_colorbar.png}}}; \\
        \node{\subcaptionbox*{}[0.08\textwidth]{%
          \includegraphics[width=0.08\textwidth]{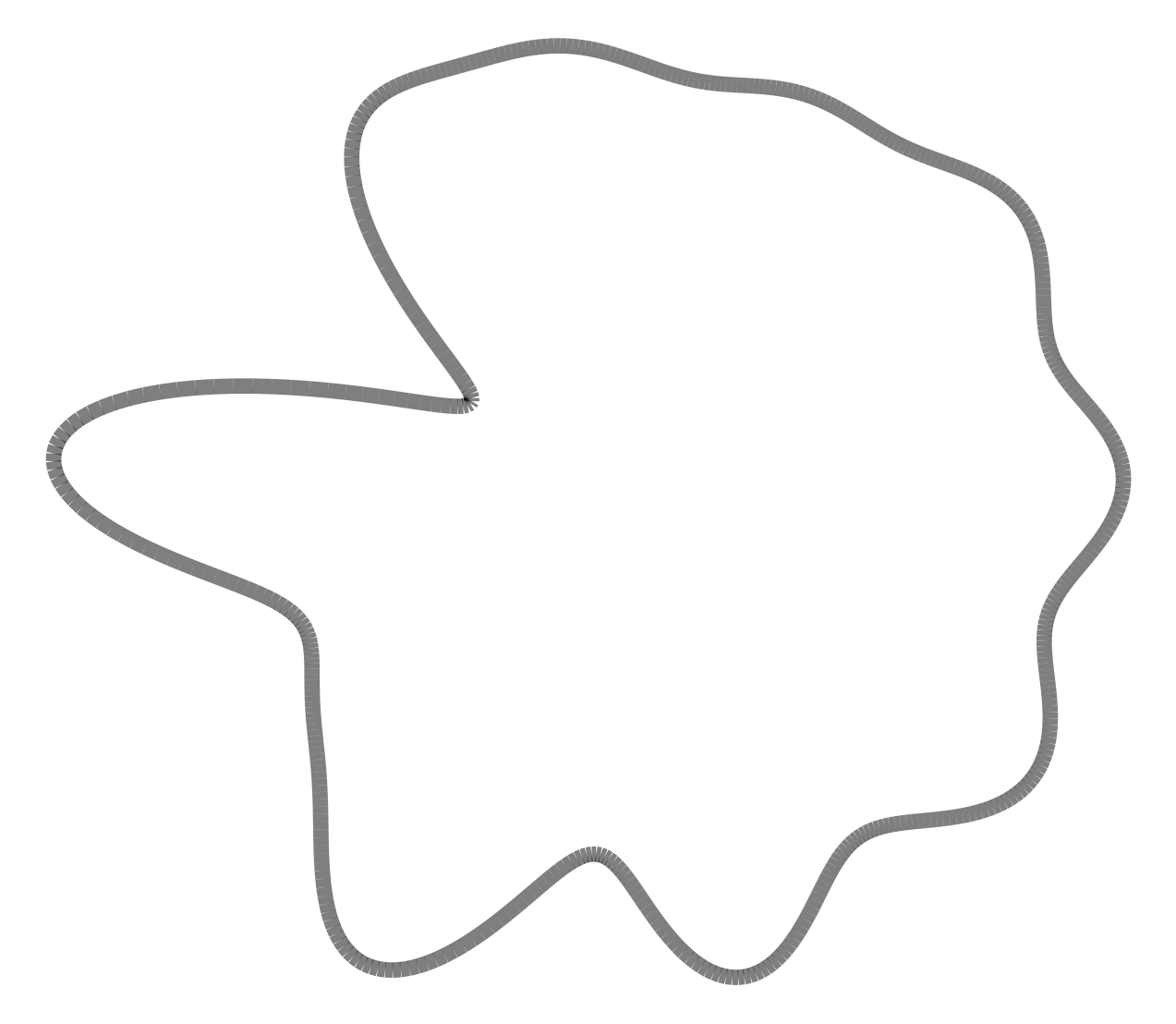}}}; &
        \node{\subcaptionbox{$u$ GT}[0.32\textwidth]{%
          \includegraphics[width=0.32\textwidth]{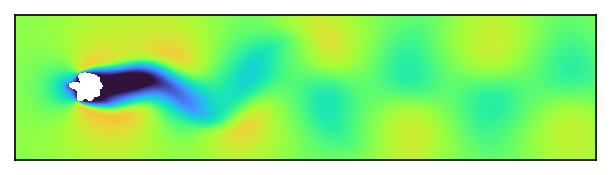}}}; &
        \node{\subcaptionbox{$u$ Pred}[0.32\textwidth]{%
          \includegraphics[width=0.32\textwidth]{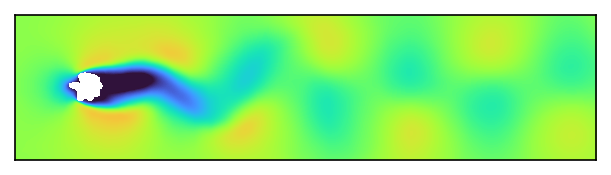}}}; &
        \node{\subcaptionbox*{}[0.037\textwidth]{\includegraphics[width=0.032\textwidth]{Figures/1in/single_step/t=30/u_colorbar.png}}}; \\
        \node{\subcaptionbox*{}[0.08\textwidth]{%
          \includegraphics[width=0.08\textwidth]{Figures/geometries/sample_200.png}}}; &
        \node{\subcaptionbox{$v$ GT}[0.32\textwidth]{%
          \includegraphics[width=0.32\textwidth]{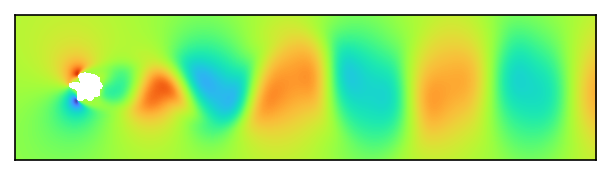}}}; &
        \node{\subcaptionbox{$v$ Pred}[0.32\textwidth]{%
          \includegraphics[width=0.32\textwidth]{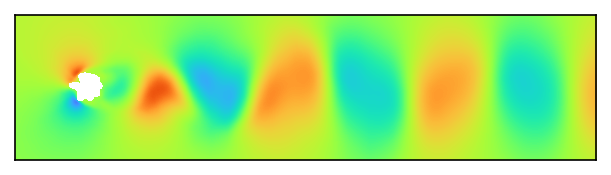}}}; &
        \node{\subcaptionbox*{}[0.04\textwidth]{\includegraphics[width=0.04\textwidth]{Figures/1in/single_step/t=30/v_colorbar.png}}}; \\        
      };
    \end{tikzpicture}
\caption{Single‐step predictions for four example geometries at $t = 30$. Each pair of rows corresponds to one shape: the top row shows the $u$‐component and the bottom row the $v$‐component.}
\label{Fig:single-step-shapes_t=30}
\end{figure}

\paragraph{Robust Geometric Generalization} \figref{Fig:single-step-shapes_t=30} compares predictions at \(t=30\) for four markedly different geometries ranging from smooth, symmetric NURBS shapes to highly irregular harmonic perturbations and non‐parametric skeleton outlines.  Despite the pronounced variations in wake dynamics induced by sharp corners and symmetry of the geometries, the predicted \(u\) and \(v\) fields remain in excellent agreement with CFD ground truth across all cases. Also, the number and spacing of shed vortices in the wake match between ground truth and prediction, indicating accurate capture of vortex-shedding frequency. These results underscore the surrogate’s ability to adapt to complex boundary geometries and capture the corresponding flow patterns, which can vary dramatically depending on local curvature and feature sharpness.

\begin{figure}[!t]
  \centering
  \begin{tikzpicture}
    \begin{groupplot}[
      group style={
        group size=5 by 4,
        horizontal sep=10mm,      
        vertical sep=76pt,
        },
        table/col sep=comma,       
        scale only axis,           
        x label style={yshift=2pt},
        every axis y label/.append style={
    at={(axis description cs:0,0.5)}, 
    anchor=center,                    
    xshift=8pt,                      
    yshift=20pt,
  },
    ]
      \nextgroupplot[
        width=0.1\textwidth,
        height=0.13\textwidth,
        axis x line=none,
        axis y line=none,
        ticks=none,
        enlarge x limits=false,
        enlarge y limits=false
      ]
      \addplot graphics [xmin=0,xmax=1,ymin=0,ymax=1]
        {Figures/geometries/sample_50.png};

      \nextgroupplot[
        width=0.15\linewidth,
        height=0.13\textwidth,
        xlabel={time},
        ylabel={$u$},
        title={$x=1D$},
        legend style={
          at={(0.5,-0.5)},anchor=north,
          legend columns=2,font=\scriptsize
        },
        clip marker paths=true
      ]
      \addplot table[x=time,y=u_gt2]{Figures/1in/single_step/plot_over_line/single_step_point_pred_s50.csv};
      \addlegendentry{GT}
      \addplot table[x=time,y=u_pr2]{Figures/1in/single_step/plot_over_line/single_step_point_pred_s50.csv};
      \addlegendentry{Pred};

      \nextgroupplot[
        width=0.15\linewidth,
        height=0.13\textwidth,
        xlabel={time},
        ylabel={$v$},
        title={$x=1D$},
        legend style={
          at={(0.5,-0.5)},anchor=north,
          legend columns=2,font=\scriptsize
        },
        clip marker paths=true
      ]
      \addplot table[x=time,y=v_gt2]{Figures/1in/single_step/plot_over_line/single_step_point_pred_s50.csv};
      \addlegendentry{GT}
      \addplot table[x=time,y=v_pr2]{Figures/1in/single_step/plot_over_line/single_step_point_pred_s50.csv};
      \addlegendentry{Pred};

      \nextgroupplot[
        width=0.15\linewidth,
        height=0.13\textwidth,
        xlabel={time},
        ylabel={$u$},
        title={$x=2D$},
        legend style={
          at={(0.5,-0.5)},anchor=north,
          legend columns=2,font=\scriptsize
        },
        clip marker paths=true
      ]
      \addplot table[x=time,y=u_gt3]{Figures/1in/single_step/plot_over_line/single_step_point_pred_s50.csv};
      \addlegendentry{GT}
      \addplot table[x=time,y=u_pr3]{Figures/1in/single_step/plot_over_line/single_step_point_pred_s50.csv};
      \addlegendentry{Pred};

      \nextgroupplot[
        width=0.15\linewidth,
        height=0.13\textwidth,
        xlabel={time},
        ylabel={$v$},
        title={$x=2D$},
        legend style={
          at={(0.5,-0.5)},anchor=north,
          legend columns=2,font=\scriptsize
        },
        clip marker paths=true
      ]
      \addplot table[x=time,y=v_gt3]{Figures/1in/single_step/plot_over_line/single_step_point_pred_s50.csv};
      \addlegendentry{GT}
      \addplot table[x=time,y=v_pr3]{Figures/1in/single_step/plot_over_line/single_step_point_pred_s50.csv};
      \addlegendentry{Pred};

      \nextgroupplot[
        width=0.1\textwidth,
        height=0.13\textwidth,
        axis x line=none,
        axis y line=none,
        ticks=none,
        enlarge x limits=false,
        enlarge y limits=false
      ]
      \addplot graphics [xmin=0,xmax=1,ymin=0,ymax=1]
        {Figures/geometries/sample_150.png};

      \nextgroupplot[
        width=0.15\linewidth,
        height=0.13\textwidth,
        xlabel={time},
        ylabel={$u$},
        title={$x=1D$},
        legend style={
          at={(0.5,-0.5)},anchor=north,
          legend columns=2,font=\scriptsize
        },
        clip marker paths=true
      ]
      \addplot table[x=time,y=u_gt2]{Figures/1in/single_step/plot_over_line/single_step_point_pred_s150.csv};
      \addlegendentry{GT}
      \addplot table[x=time,y=u_pr2]{Figures/1in/single_step/plot_over_line/single_step_point_pred_s150.csv};
      \addlegendentry{Pred};

      \nextgroupplot[
        width=0.15\linewidth,
        height=0.13\textwidth,
        xlabel={time},
        ylabel={$v$},
        title={$x=1D$},
        legend style={
          at={(0.5,-0.5)},anchor=north,
          legend columns=2,font=\scriptsize
        },
        clip marker paths=true
      ]
      \addplot table[x=time,y=v_gt2]{Figures/1in/single_step/plot_over_line/single_step_point_pred_s150.csv};
      \addlegendentry{GT}
      \addplot table[x=time,y=v_pr2]{Figures/1in/single_step/plot_over_line/single_step_point_pred_s150.csv};
      \addlegendentry{Pred};

      \nextgroupplot[
        width=0.15\linewidth,
        height=0.13\textwidth,
        xlabel={time},
        ylabel={$u$},
        title={$x=2D$},
        legend style={
          at={(0.5,-0.5)},anchor=north,
          legend columns=2,font=\scriptsize
        },
        clip marker paths=true
      ]
      \addplot table[x=time,y=u_gt3]{Figures/1in/single_step/plot_over_line/single_step_point_pred_s150.csv};
      \addlegendentry{GT}
      \addplot table[x=time,y=u_pr3]{Figures/1in/single_step/plot_over_line/single_step_point_pred_s150.csv};
      \addlegendentry{Pred};

      \nextgroupplot[
        width=0.15\linewidth,
        height=0.13\textwidth,
        xlabel={time},
        ylabel={$v$},
        title={$x=2D$},
        legend style={
          at={(0.5,-0.5)},anchor=north,
          legend columns=2,font=\scriptsize
        },
        clip marker paths=true
      ]
      \addplot table[x=time,y=v_gt3]{Figures/1in/single_step/plot_over_line/single_step_point_pred_s150.csv};
      \addlegendentry{GT}
      \addplot table[x=time,y=v_pr3]{Figures/1in/single_step/plot_over_line/single_step_point_pred_s150.csv};
      \addlegendentry{Pred};

      \nextgroupplot[
        width=0.1\textwidth,
        height=0.13\textwidth,
        axis x line=none,
        axis y line=none,
        ticks=none,
        enlarge x limits=false,
        enlarge y limits=false
      ]
      \addplot graphics [xmin=0,xmax=1,ymin=0,ymax=1]
        {Figures/geometries/sample_200.png};

      \nextgroupplot[
        width=0.15\linewidth,
        height=0.13\textwidth,
        xlabel={time},
        ylabel={$u$},
        title={$x=1D$},
        legend style={
          at={(0.5,-0.5)},anchor=north,
          legend columns=2,font=\scriptsize
        },
        clip marker paths=true
      ]
      \addplot table[x=time,y=u_gt2]{Figures/1in/single_step/plot_over_line/single_step_point_pred_s200.csv};
      \addlegendentry{GT}
      \addplot table[x=time,y=u_pr2]{Figures/1in/single_step/plot_over_line/single_step_point_pred_s200.csv};
      \addlegendentry{Pred};

      \nextgroupplot[
        width=0.15\linewidth,
        height=0.13\textwidth,
        xlabel={time},
        ylabel={$v$},
        title={$x=1D$},
        legend style={
          at={(0.5,-0.5)},anchor=north,
          legend columns=2,font=\scriptsize
        },
        clip marker paths=true
      ]
      \addplot table[x=time,y=v_gt2]{Figures/1in/single_step/plot_over_line/single_step_point_pred_s200.csv};
      \addlegendentry{GT}
      \addplot table[x=time,y=v_pr2]{Figures/1in/single_step/plot_over_line/single_step_point_pred_s200.csv};
      \addlegendentry{Pred};

      \nextgroupplot[
        width=0.15\linewidth,
        height=0.13\textwidth,
        xlabel={time},
        ylabel={$u$},
        title={$x=2D$},
        legend style={
          at={(0.5,-0.5)},anchor=north,
          legend columns=2,font=\scriptsize
        },
        clip marker paths=true
      ]
      \addplot table[x=time,y=u_gt3]{Figures/1in/single_step/plot_over_line/single_step_point_pred_s200.csv};
      \addlegendentry{GT}
      \addplot table[x=time,y=u_pr3]{Figures/1in/single_step/plot_over_line/single_step_point_pred_s200.csv};
      \addlegendentry{Pred};

      \nextgroupplot[
        width=0.15\linewidth,
        height=0.13\textwidth,
        xlabel={time},
        ylabel={$v$},
        title={$x=2D$},
        legend style={
          at={(0.5,-0.5)},anchor=north,
          legend columns=2,font=\scriptsize
        },
        clip marker paths=true
      ]
      \addplot table[x=time,y=v_gt3]{Figures/1in/single_step/plot_over_line/single_step_point_pred_s200.csv};
      \addlegendentry{GT}
      \addplot table[x=time,y=v_pr3]{Figures/1in/single_step/plot_over_line/single_step_point_pred_s200.csv};
      \addlegendentry{Pred};

      \nextgroupplot[
        width=0.1\textwidth,
        height=0.13\textwidth,
        axis x line=none,
        axis y line=none,
        ticks=none,
        enlarge x limits=false,
        enlarge y limits=false
      ]
      \addplot graphics [xmin=0,xmax=1,ymin=0,ymax=1]
        {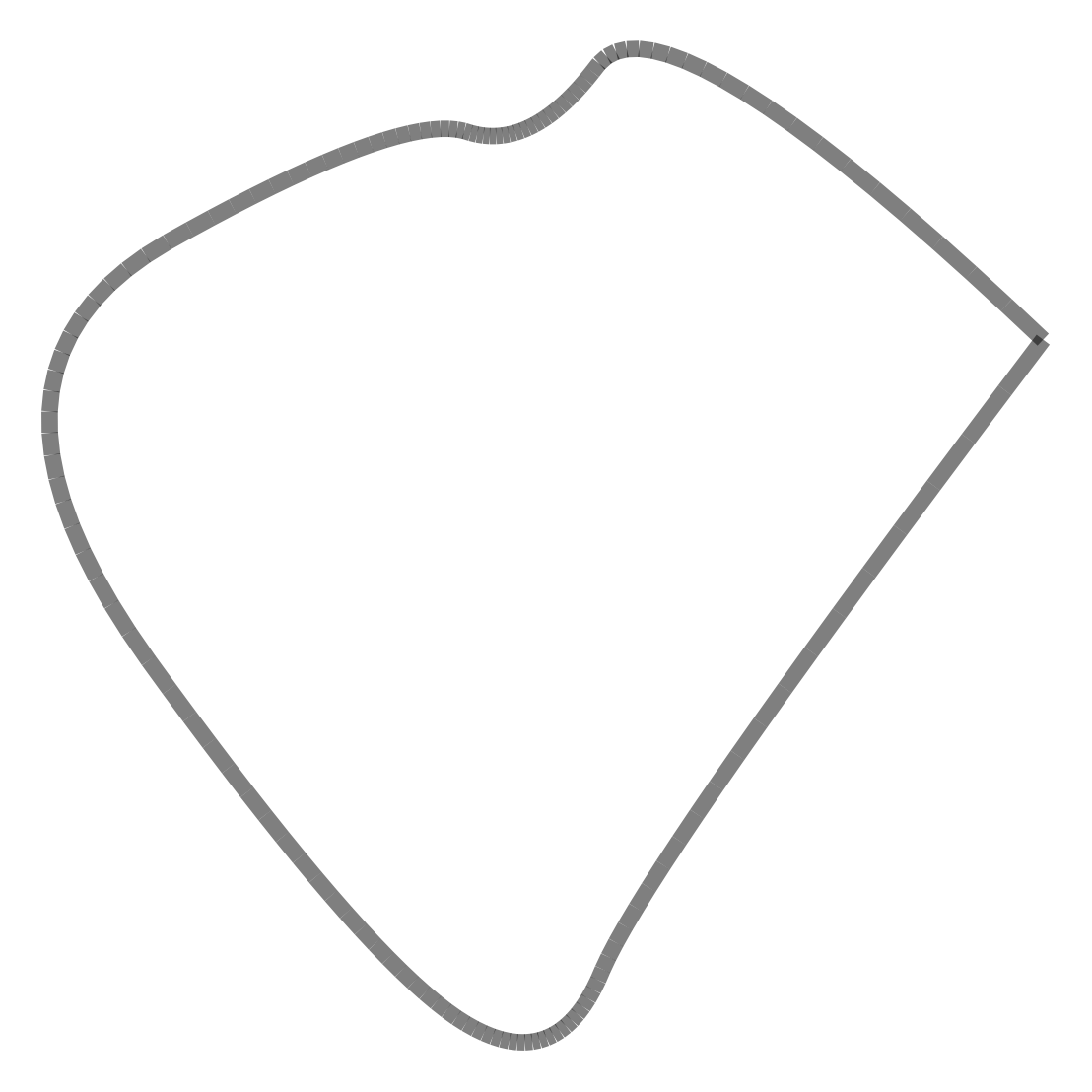};

      \nextgroupplot[
        width=0.15\linewidth,
        height=0.13\textwidth,
        xlabel={time},
        ylabel={$u$},
        title={$x=1D$},
        legend style={
          at={(0.5,-0.5)},anchor=north,
          legend columns=2,font=\scriptsize
        },
        clip marker paths=true
      ]
      \addplot table[x=time,y=u_gt2]{Figures/1in/single_step/plot_over_line/single_step_point_pred_s250.csv};
      \addlegendentry{GT}
      \addplot table[x=time,y=u_pr2]{Figures/1in/single_step/plot_over_line/single_step_point_pred_s250.csv};
      \addlegendentry{Pred};

      \nextgroupplot[
        width=0.15\linewidth,
        height=0.13\textwidth,
        xlabel={time},
        ylabel={$v$},
        title={$x=1D$},
        legend style={
          at={(0.5,-0.5)},anchor=north,
          legend columns=2,font=\scriptsize
        },
        clip marker paths=true
      ]
      \addplot table[x=time,y=v_gt2]{Figures/1in/single_step/plot_over_line/single_step_point_pred_s250.csv};
      \addlegendentry{GT}
      \addplot table[x=time,y=v_pr2]{Figures/1in/single_step/plot_over_line/single_step_point_pred_s250.csv};
      \addlegendentry{Pred};

      \nextgroupplot[
        width=0.15\linewidth,
        height=0.13\textwidth,
        xlabel={time},
        ylabel={$u$},
        title={$x=2D$},
        legend style={
          at={(0.5,-0.5)},anchor=north,
          legend columns=2,font=\scriptsize
        },
        clip marker paths=true
      ]
      \addplot table[x=time,y=u_gt3]{Figures/1in/single_step/plot_over_line/single_step_point_pred_s250.csv};
      \addlegendentry{GT}
      \addplot table[x=time,y=u_pr3]{Figures/1in/single_step/plot_over_line/single_step_point_pred_s250.csv};
      \addlegendentry{Pred};

      \nextgroupplot[
        width=0.15\linewidth,
        height=0.13\textwidth,
        xlabel={time},
        ylabel={$v$},
        title={$x=2D$},
        legend style={
          at={(0.5,-0.5)},anchor=north,
          legend columns=2,font=\scriptsize
        },
        clip marker paths=true
      ]
      \addplot table[x=time,y=v_gt3]{Figures/1in/single_step/plot_over_line/single_step_point_pred_s250.csv};
      \addlegendentry{GT}
      \addplot table[x=time,y=v_pr3]{Figures/1in/single_step/plot_over_line/single_step_point_pred_s250.csv};
      \addlegendentry{Pred};

    \end{groupplot}
    \end{tikzpicture}
    \caption{Single step time‐series of $u$ and $v$ at two points at downstream distance from geometry $x=1D$ and $x=2D$, where D is the geometry diameter. We show a collection of 4 shapes where each row corresponds to a single geometry.}
    \label{Fig:single-step-shapes_tseries}
\end{figure}

\paragraph{Pointwise Temporal Dynamics Near Boundaries}
In \figref{Fig:single-step-shapes_tseries}, we plot the time series of \(u\) and \(v\) at two downstream probes located at \(x=1D\) and \(x=2D\) (where \(D\) is the characteristic diameter of each shape).  These locations lie within the near‐geometry region, where viscous boundary‐layer effects dominate and flow transition depends sensitively on leading‐edge shape and curvature.  Despite the inherent difficulty of modeling highly transient, non‐sinusoidal signals in this boundary-layer, the predicted time series closely follow the ground truth in both phase and amplitude.  This agreement highlights the surrogate model’s ability to resolve fine‐scale, geometry‐driven unsteady phenomena at critical downstream positions.

\clearpage
\subsection{Rollout prediction} \label{subsec:rollout}

\begin{figure}[!t]
  \centering
\begin{subfigure}[t]{0.48\textwidth}
    \centering
    \begin{tikzpicture}
      \begin{axis}[
        name=relplot,
        width=0.98\textwidth, 
        height=0.6\textwidth,
        xmin=0, xmax=60,
        ymin=0, ymax=1,
        xlabel={time},
        ylabel={Relative $L_2$ Error},
        scaled y ticks  = true,
        scaled y ticks  = base 10:2,
        ]
        \addplot table[
          col sep=comma,
          x=timestep,
          y=rel_Overall
        ] {Figures/1in/rollout/errors/rollout_errors.csv};
      \end{axis}
    \end{tikzpicture}
    \caption{Relative $L_2$ error over time.}
  \end{subfigure}%
  \begin{subfigure}[t]{0.48\textwidth}
    \centering
    \begin{tikzpicture}
      \begin{axis}[
        width=0.98\textwidth, 
        height=0.6\textwidth,
        at={(relplot.east)}, anchor=west, xshift=2cm,
        xmin=0, xmax=60,
        ymin=0, ymax=1,
        xlabel={time},
        ylabel={RMSE},
        scaled y ticks  = true,
        scaled y ticks  = base 10:2,        
        legend pos=north east
      ]
        \addplot table[
          col sep=comma,
          x=timestep,
          y=rmse_U
        ] {Figures/1in/rollout/errors/rollout_errors.csv};
        \addplot table[
          col sep=comma,
          x=timestep,
          y=rmse_V
        ] {Figures/1in/rollout/errors/rollout_errors.csv};
        \legend{$u$,$v$}
      \end{axis}
    \end{tikzpicture}
    \caption{RMSE of $u$ and $v$.}
  \end{subfigure}%
  \hfill

  \vspace{1em}

  \begin{subfigure}[t]{\textwidth}
    \centering
    \begin{tikzpicture}
      \matrix [matrix of nodes,
               nodes={inner sep=0, anchor=south west},
               column sep=1pt, row sep=1pt] {
        \node{\subcaptionbox{$u$ GT, $t = 30$}[0.31\textwidth]{%
          \includegraphics[width=0.3\textwidth]{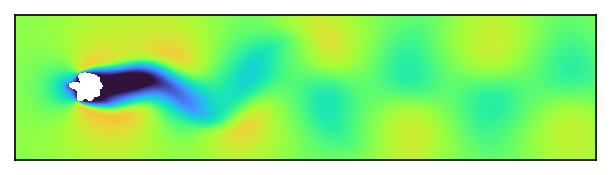}}}; &
        \node{\subcaptionbox{$u$ GT, $t = 45$}[0.31\textwidth]{%
          \includegraphics[width=0.3\textwidth]{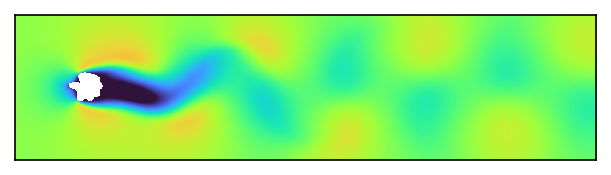}}}; &
        \node{\subcaptionbox{$u$ GT, $t = 59$}[0.31\textwidth]{%
          \includegraphics[width=0.3\textwidth]{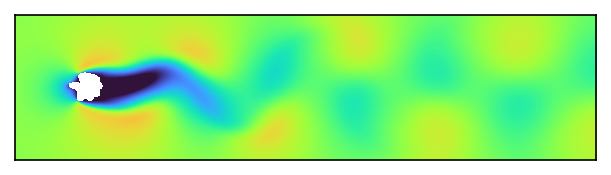}}}; &
        \node{\subcaptionbox*{}[0.037\textwidth]{\includegraphics[width=0.032\textwidth]{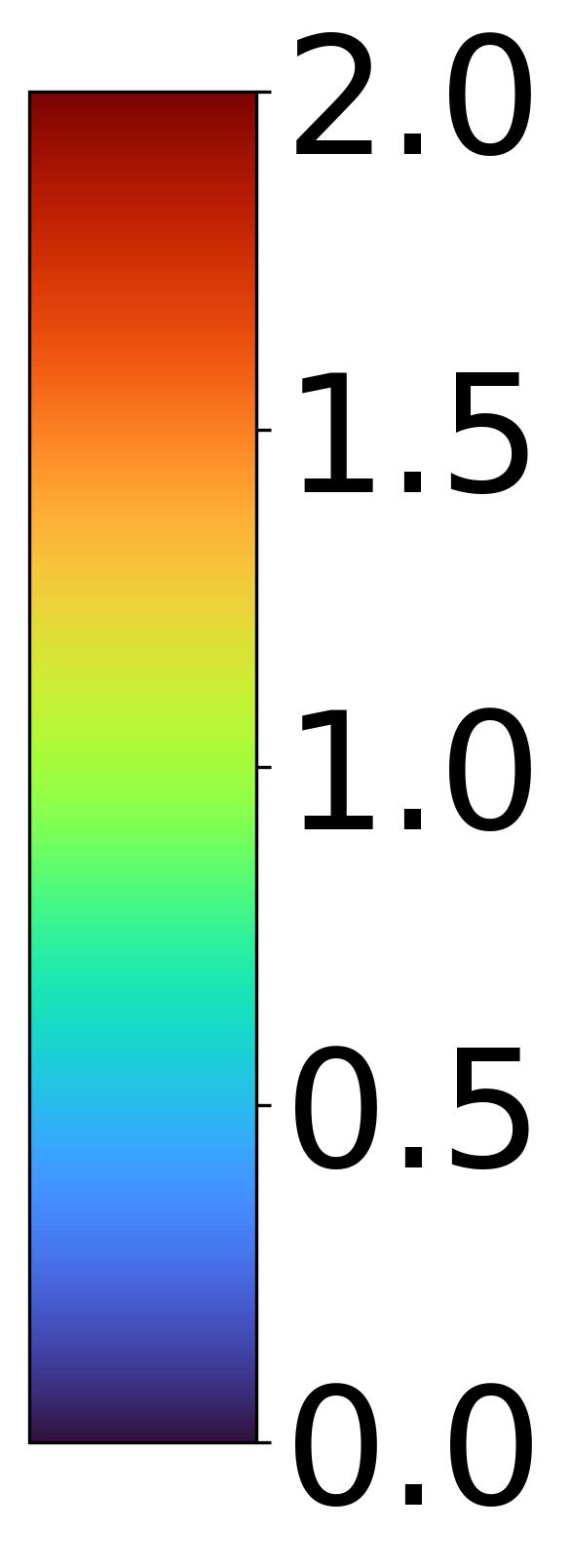}}}; \\
        \node{\subcaptionbox{$u$ Pred, $t = 30$}[0.31\textwidth]{%
          \includegraphics[width=0.3\textwidth]{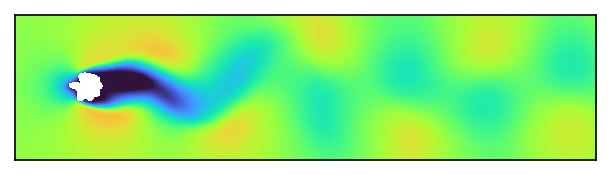}}}; &
        \node{\subcaptionbox{$u$ Pred, $t = 45$}[0.31\textwidth]{%
          \includegraphics[width=0.3\textwidth]{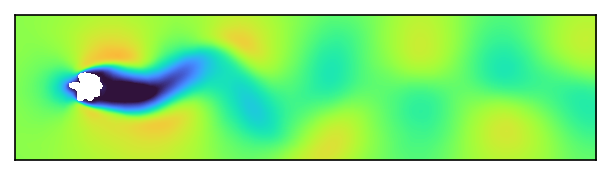}}}; &
        \node{\subcaptionbox{$u$ Pred, $t = 59$}[0.31\textwidth]{%
          \includegraphics[width=0.3\textwidth]{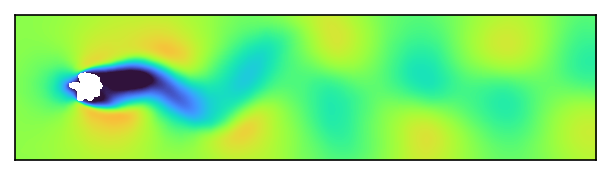}}}; &
        \node{\subcaptionbox*{}[0.037\textwidth] {\includegraphics[width=0.032\textwidth]{Figures/1in/rollout/sample200/u_colorbar.png}}}; \\
      };
    \end{tikzpicture}
  \end{subfigure}
  \vspace{1em}
  \begin{subfigure}[t]{\textwidth}
    \centering
    \begin{tikzpicture}
      \matrix [matrix of nodes,
               nodes={inner sep=0, anchor=south west},
               column sep=1pt, row sep=1pt] {
        \node{\subcaptionbox{$v$ GT, $t = 30$}[0.31\textwidth]{%
          \includegraphics[width=0.3\textwidth]{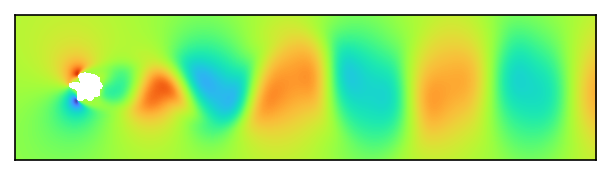}}}; &
        \node{\subcaptionbox{$v$ GT, $t = 45$}[0.31\textwidth]{%
          \includegraphics[width=0.3\textwidth]{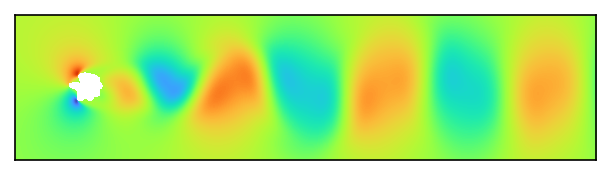}}}; &
        \node{\subcaptionbox{$v$ GT, $t = 59$}[0.31\textwidth]{%
          \includegraphics[width=0.3\textwidth]{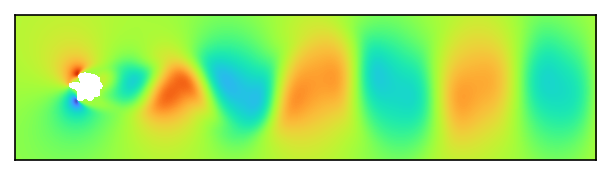}}}; &
        \node{\subcaptionbox*{}[0.04\textwidth] {\includegraphics[width=0.04\textwidth]{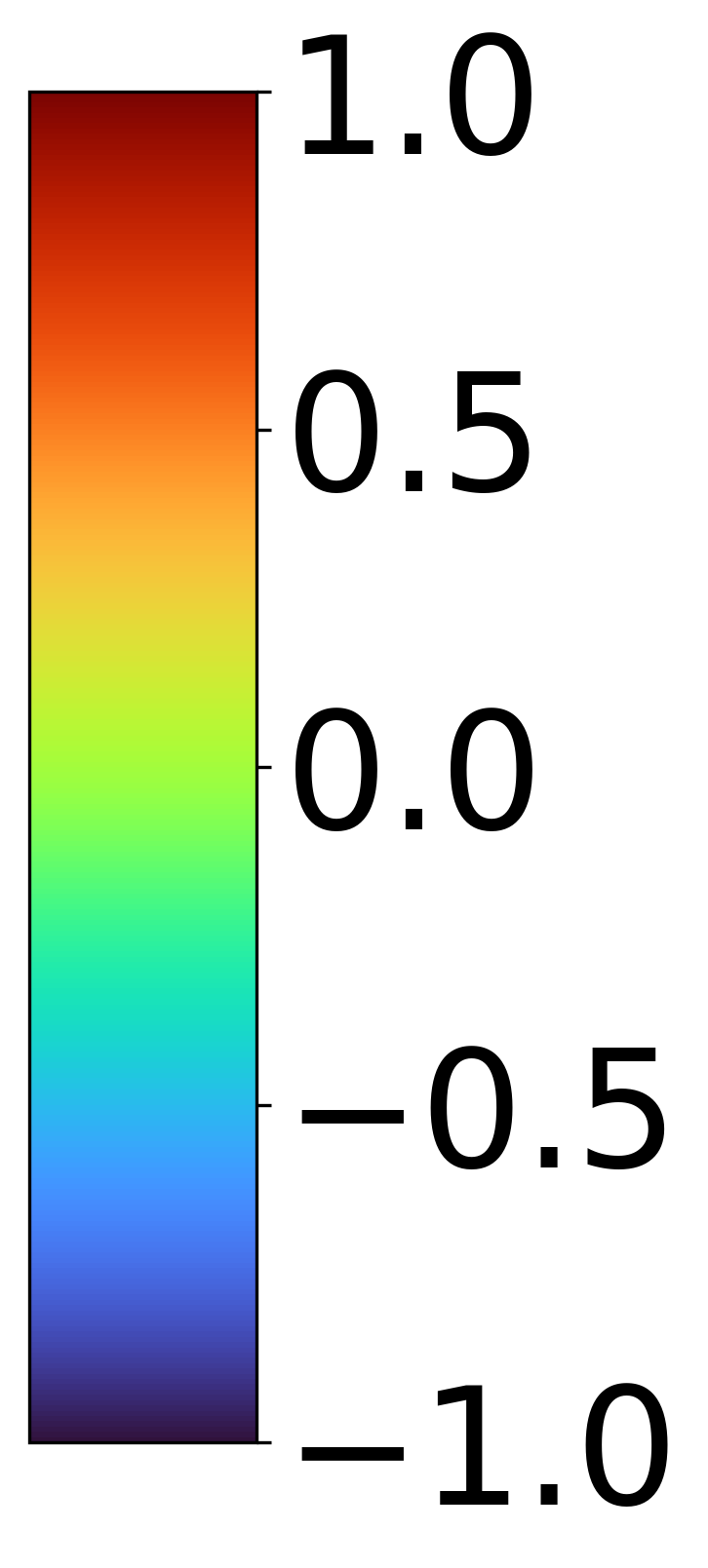}}}; \\
        \node{\subcaptionbox{$v$ Pred, $t = 30$}[0.31\textwidth]{%
          \includegraphics[width=0.3\textwidth]{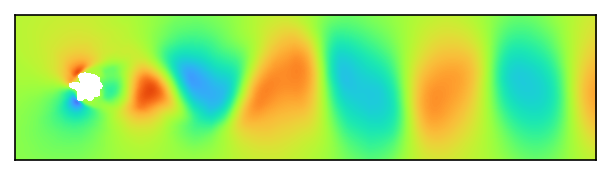}}}; &
        \node{\subcaptionbox{$v$ Pred, $t = 45$}[0.31\textwidth]{%
          \includegraphics[width=0.3\textwidth]{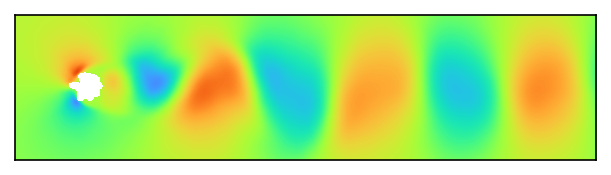}}}; &
        \node{\subcaptionbox{$v$ Pred, $t = 59$}[0.31\textwidth]{%
          \includegraphics[width=0.3\textwidth]{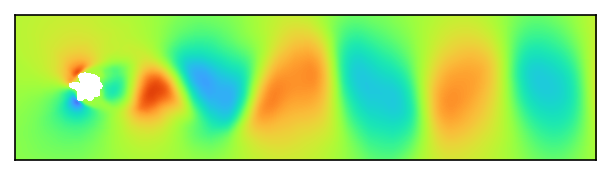}}}; &
        \node{\subcaptionbox*{}[0.04\textwidth] {\includegraphics[width=0.04\textwidth]{Figures/1in/rollout/sample200/v_colorbar.png}}}; \\
      };
    \end{tikzpicture}
  \end{subfigure}

  \caption{Rollout prediction of flow velocity for an example geometry. (a) Relative $L_2$ error over time. (b) RMSE of $u$ and $v$ over time. (c-e) Ground‐truth $u$ at $t = 30, 45, 59$. (f-h) Predicted $u$ at $t = 30, 45, 59$. (i-k) Ground‐truth $v$ at $t = 30, 45, 59$. (l-n) Predicted $v$ at $t = 30, 45, 59$. Colorbars for $u$ are shown in (e) and (h), and for $v$ in (k) and (n).}
  \label{Fig:rollout-sample200}
\end{figure}

\paragraph{Rollout Prediction Accuracy}
As shown in \figref{Fig:rollout-sample200}(a), the overall relative \(L_{2}\) error begins at approximately 5\% and grows to about 55\% by \(t=60\), indicating cumulative error accumulation when previous predictions are fed back into the model. This trend reflects the degradation of the model’s accuracy during rollouts, where each predicted field becomes the input for the next timestep. Correspondingly, the RMSE for both the \(u\) and \(v\) components increases monotonically (see \figref{Fig:rollout-sample200}(b)), rising from roughly 0.02 and 0.05 at \(t=0\) to 0.4 and 0.7 at \(t=60\) for \(u\) and \(v\), respectively. This shows that the surrogate struggles to maintain accuracy over extended time horizons.  \figref{Fig:rollout-sample200}(c)-(h) compare ground-truth and predicted velocity fields for a single geometry at \(t=30\), \(t=45\), and \(t=59\). At \(t=30\), the prediction closely matches the reference, with minor discrepancies confined near the solid boundary. By \(t=45\), wake vortices exhibit slight phase shifts and reduced amplitude. At \(t=59\), the predicted wake structure is noticeably different and vortex centers are displaced, highlighting the challenge of long-horizon rollouts.

\clearpage

\begin{figure}[!t]
  \centering
    \begin{tikzpicture}
      \matrix [matrix of nodes,
               nodes={inner sep=0, anchor=south west},
               column sep=1pt, row sep=1pt] {
        \node{\subcaptionbox*{}[0.08\textwidth]{%
          \includegraphics[width=0.08\textwidth]{Figures/geometries/sample_0.png}}}; &
        \node{\subcaptionbox{$u$ GT}[0.32\textwidth]{%
          \includegraphics[width=0.32\textwidth]{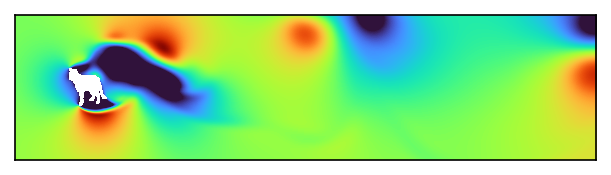}}}; &
        \node{\subcaptionbox{$u$ Pred}[0.32\textwidth]{%
          \includegraphics[width=0.32\textwidth]{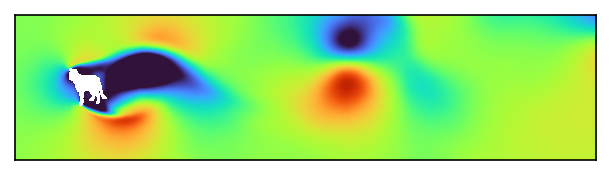}}}; &
        \node{\subcaptionbox*{}[0.037\textwidth]{\includegraphics[width=0.032\textwidth]{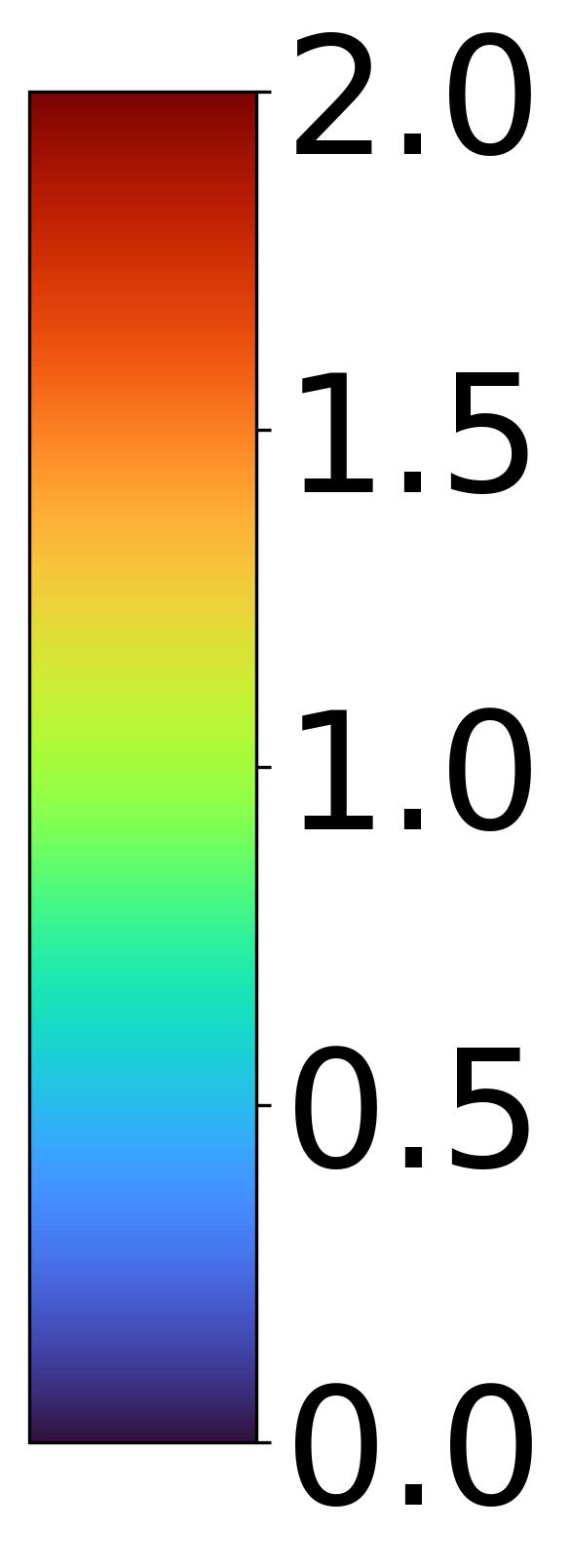}}}; \\
        \node{\subcaptionbox*{}[0.08\textwidth]{%
          \includegraphics[width=0.08\textwidth]{Figures/geometries/sample_0.png}}}; &
        \node{\subcaptionbox{$v$ GT}[0.32\textwidth]{%
          \includegraphics[width=0.32\textwidth]{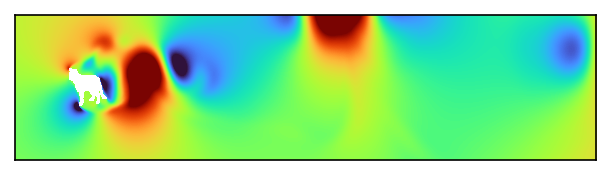}}}; &
        \node{\subcaptionbox{$v$ Pred}[0.32\textwidth]{%
          \includegraphics[width=0.32\textwidth]{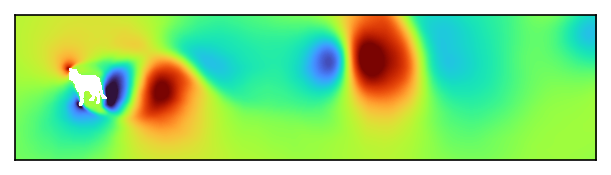}}}; &
        \node{\subcaptionbox*{}[0.04\textwidth]{\includegraphics[width=0.04\textwidth]{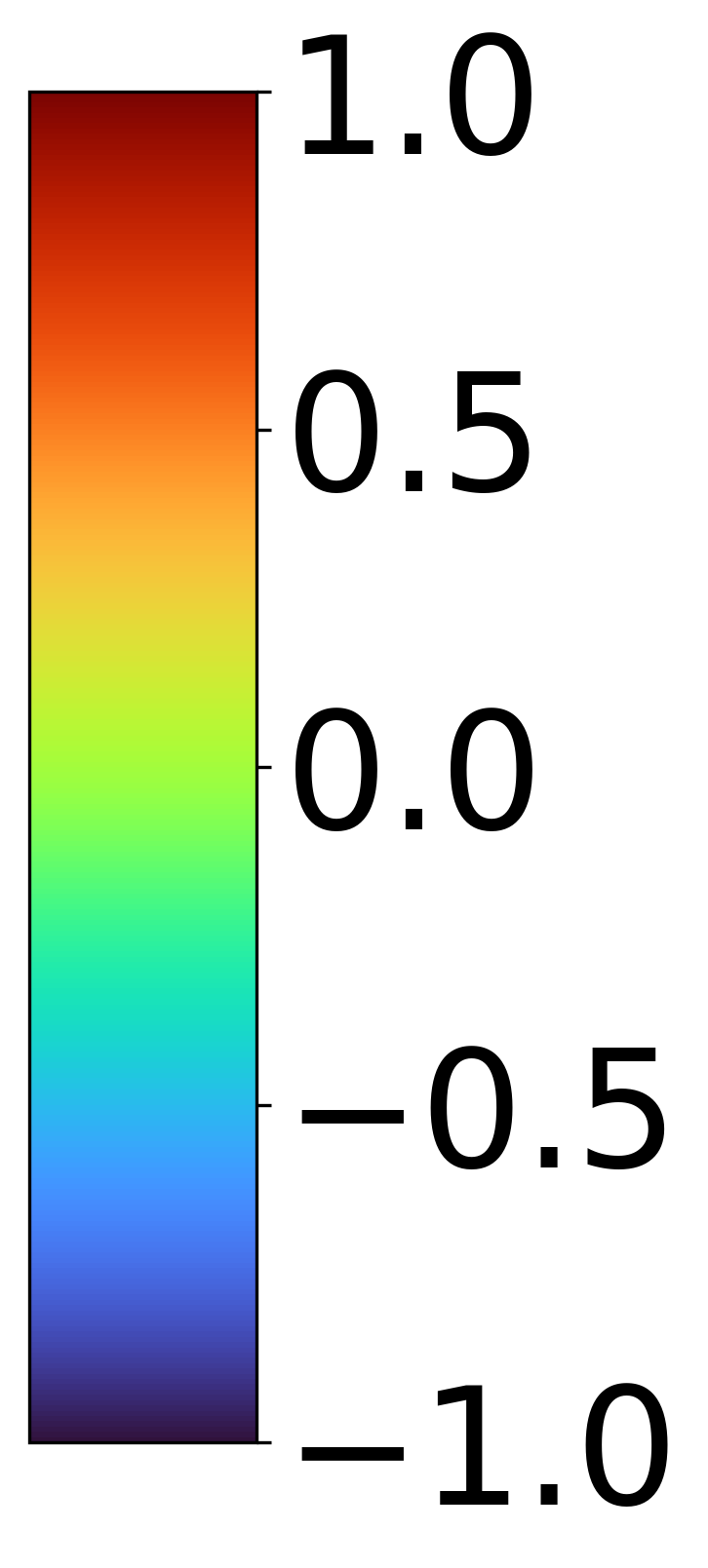}}}; \\
        \node{\subcaptionbox*{}[0.08\textwidth]{%
          \includegraphics[width=0.08\textwidth]{Figures/geometries/sample_50.png}}}; &
        \node{\subcaptionbox{$u$ GT}[0.32\textwidth]{%
          \includegraphics[width=0.32\textwidth]{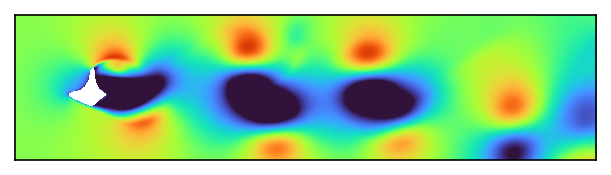}}}; &
        \node{\subcaptionbox{$u$ Pred}[0.32\textwidth]{%
          \includegraphics[width=0.32\textwidth]{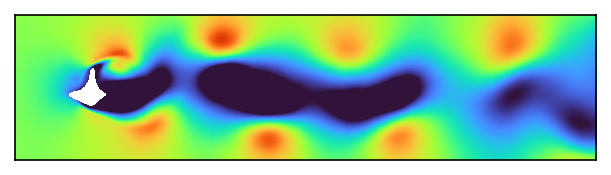}}}; &
        \node{\subcaptionbox*{}[0.037\textwidth]{\includegraphics[width=0.032\textwidth]{Figures/1in/rollout/t=30/u_colorbar.png}}}; \\
        \node{\subcaptionbox*{}[0.08\textwidth]{%
          \includegraphics[width=0.08\textwidth]{Figures/geometries/sample_50.png}}}; &
        \node{\subcaptionbox{$v$ GT}[0.32\textwidth]{%
          \includegraphics[width=0.32\textwidth]{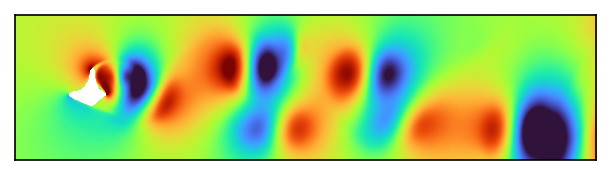}}}; &
        \node{\subcaptionbox{$v$ Pred}[0.32\textwidth]{%
          \includegraphics[width=0.32\textwidth]{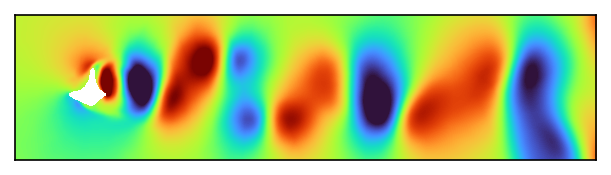}}}; &
        \node{\subcaptionbox*{}[0.04\textwidth]{\includegraphics[width=0.04\textwidth]{Figures/1in/rollout/t=30/v_colorbar.png}}}; \\
        \node{\subcaptionbox*{}[0.08\textwidth]{%
          \includegraphics[width=0.08\textwidth]{Figures/geometries/sample_150.png}}}; &
        \node{\subcaptionbox{$u$ GT}[0.32\textwidth]{%
          \includegraphics[width=0.32\textwidth]{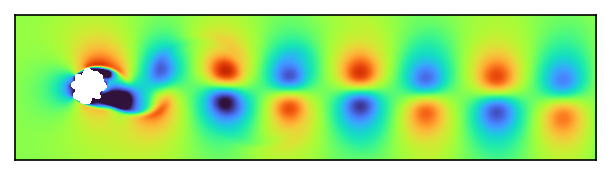}}}; &
        \node{\subcaptionbox{$u$ Pred}[0.32\textwidth]{%
          \includegraphics[width=0.32\textwidth]{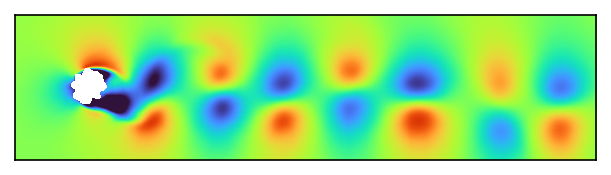}}}; &
        \node{\subcaptionbox*{}[0.037\textwidth]{\includegraphics[width=0.032\textwidth]{Figures/1in/rollout/t=30/u_colorbar.png}}}; \\
        \node{\subcaptionbox*{}[0.08\textwidth]{%
          \includegraphics[width=0.08\textwidth]{Figures/geometries/sample_150.png}}}; &
        \node{\subcaptionbox{$v$ GT}[0.32\textwidth]{%
          \includegraphics[width=0.32\textwidth]{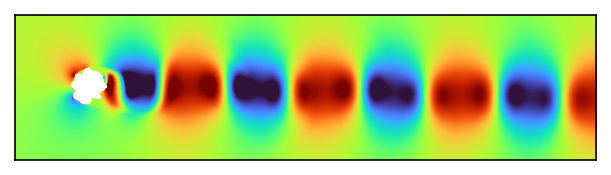}}}; &
        \node{\subcaptionbox{$v$ Pred}[0.32\textwidth]{%
          \includegraphics[width=0.32\textwidth]{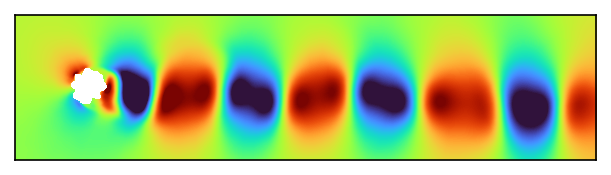}}}; &
        \node{\subcaptionbox*{}[0.04\textwidth]{\includegraphics[width=0.04\textwidth]{Figures/1in/rollout/t=30/v_colorbar.png}}}; \\
        \node{\subcaptionbox*{}[0.08\textwidth]{%
          \includegraphics[width=0.08\textwidth]{Figures/geometries/sample_200.png}}}; &
        \node{\subcaptionbox{$u$ GT}[0.32\textwidth]{%
          \includegraphics[width=0.32\textwidth]{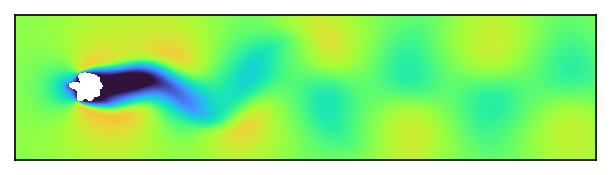}}}; &
        \node{\subcaptionbox{$u$ Pred}[0.32\textwidth]{%
          \includegraphics[width=0.32\textwidth]{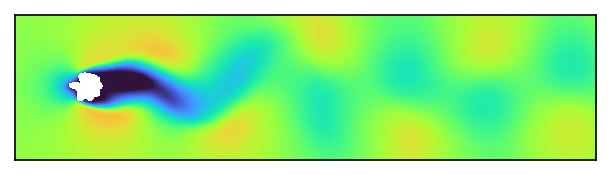}}}; &
        \node{\subcaptionbox*{}[0.037\textwidth]{\includegraphics[width=0.032\textwidth]{Figures/1in/rollout/t=30/u_colorbar.png}}}; \\
        \node{\subcaptionbox*{}[0.08\textwidth]{%
          \includegraphics[width=0.08\textwidth]{Figures/geometries/sample_200.png}}}; &
        \node{\subcaptionbox{$v$ GT}[0.32\textwidth]{%
          \includegraphics[width=0.32\textwidth]{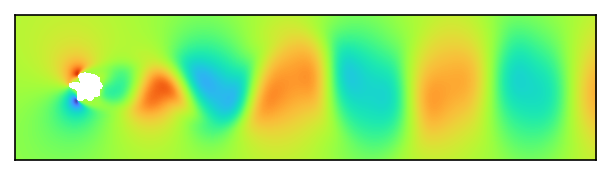}}}; &
        \node{\subcaptionbox{$v$ Pred}[0.32\textwidth]{%
          \includegraphics[width=0.32\textwidth]{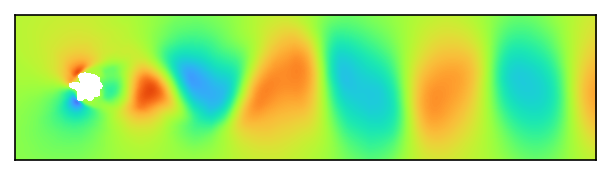}}}; &
        \node{\subcaptionbox*{}[0.041\textwidth]{\includegraphics[width=0.04\textwidth]{Figures/1in/rollout/t=30/v_colorbar.png}}}; \\        
      };
    \end{tikzpicture}
  \caption{Rollout predictions for four example geometries at $t = 30$. Each pair of rows corresponds to one shape: the top row shows the $u$‐component and the bottom row the $v$‐component.}
  \label{Fig:rollout-shapes_t=30}
\end{figure}

\paragraph{Robust Geometric Generalization During Rollouts} \figref{Fig:rollout-shapes_t=30} compares rollout predictions at \(t=30\) for four markedly different geometries ranging from smooth, symmetric NURBS shapes to highly irregular harmonic perturbations and non-parametric skeleton outlines. The first two shapes (panels (a)--(h)) show high deviation from the ground-truth data, particularly near sharp edges where the surrogate smooths peak velocities and displaces vortex cores. The last two shapes are in better agreement (panels (i)--(p)), with wake vortices correctly positioned and velocity amplitudes closely matching the CFD reference. These results indicate that while rollout accuracy degrades for geometries with pronounced corners, the surrogate maintains robust prediction quality for smoother boundaries, preserving key flow structures across diverse shape complexities.

\begin{figure}[!t]
  \centering
  \begin{subfigure}[t]{\textwidth}
    \centering
    \begin{tikzpicture}
      \matrix [matrix of nodes,
               nodes={inner sep=0, anchor=south west},
               column sep=1pt, row sep=1pt] {
        \node{\subcaptionbox{$u$ GT, $10$th percentile}[0.31\textwidth]{%
          \includegraphics[width=0.3\textwidth]{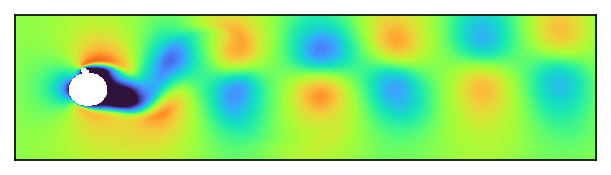}}}; &
        \node{\subcaptionbox{$u$ Predicted, $10$th percentile}[0.31\textwidth]{%
          \includegraphics[width=0.3\textwidth]{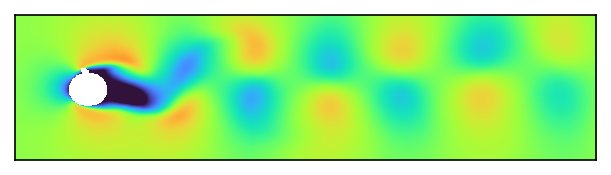}}}; &
        \node{\subcaptionbox*{}[0.037\textwidth]{\includegraphics[width=0.032\textwidth]{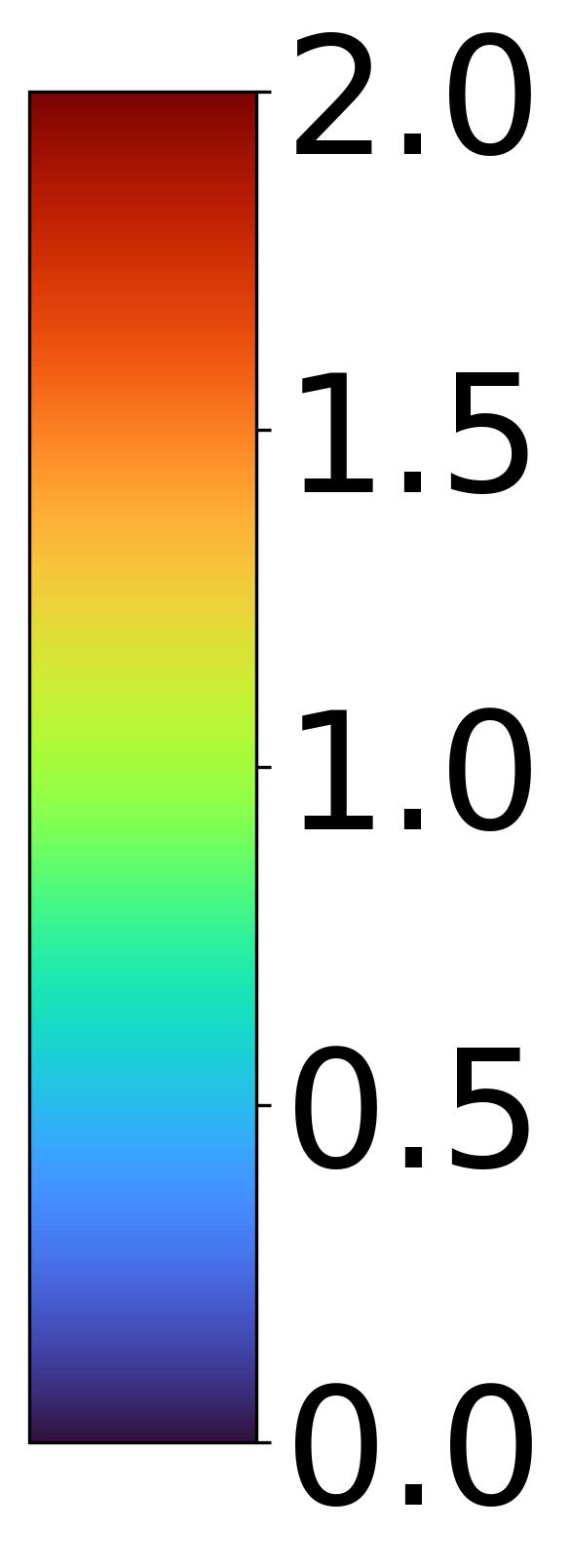}}}; &
        \node{\subcaptionbox{$u$ Error, $10$th percentile}[0.31\textwidth]{%
          \includegraphics[width=0.3\textwidth]{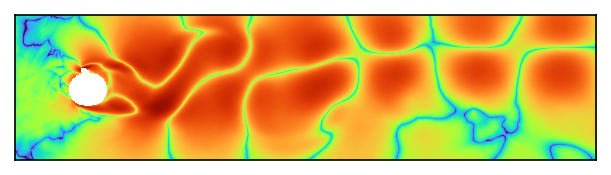}}}; &
        \node{\subcaptionbox*{}[0.037\textwidth]{\includegraphics[width=0.024\textwidth]{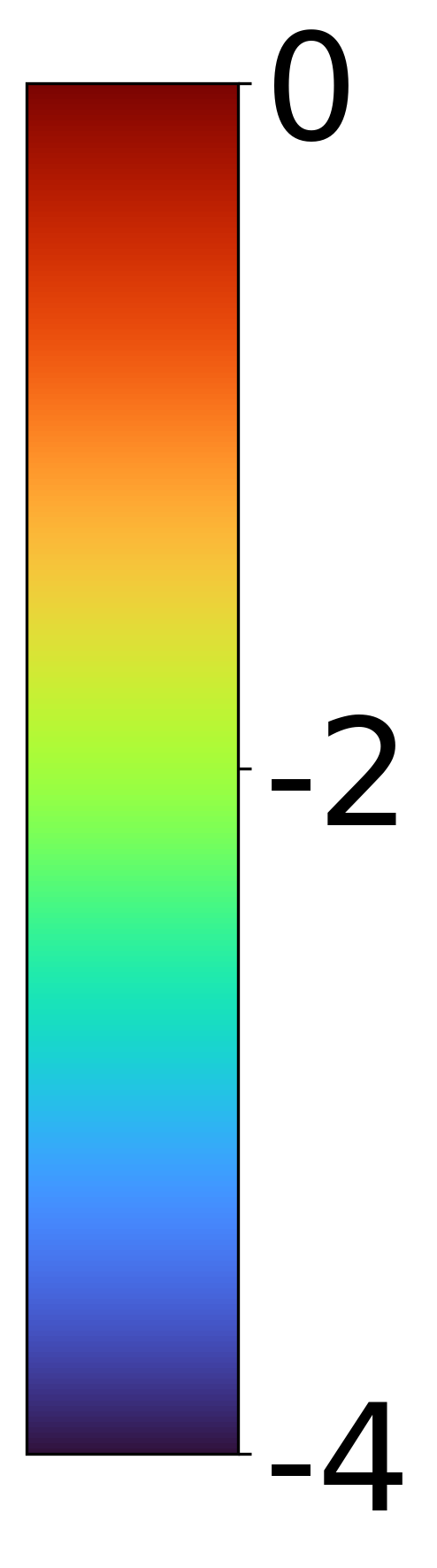}}}; \\
        \node{\subcaptionbox{$v$ GT, $10$th percentile}[0.31\textwidth]{%
          \includegraphics[width=0.3\textwidth]{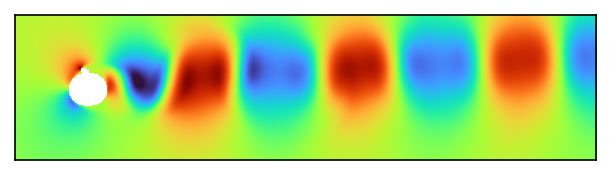}}}; &
        \node{\subcaptionbox{$v$ Predicted, $10$th percentile}[0.31\textwidth]{%
          \includegraphics[width=0.3\textwidth]{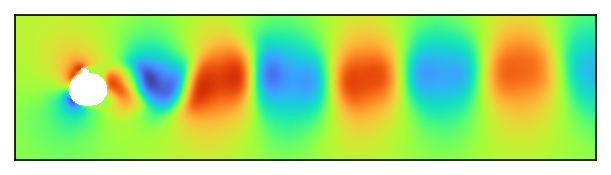}}}; &
        \node{\subcaptionbox*{}[0.037\textwidth]{\includegraphics[width=0.032\textwidth]{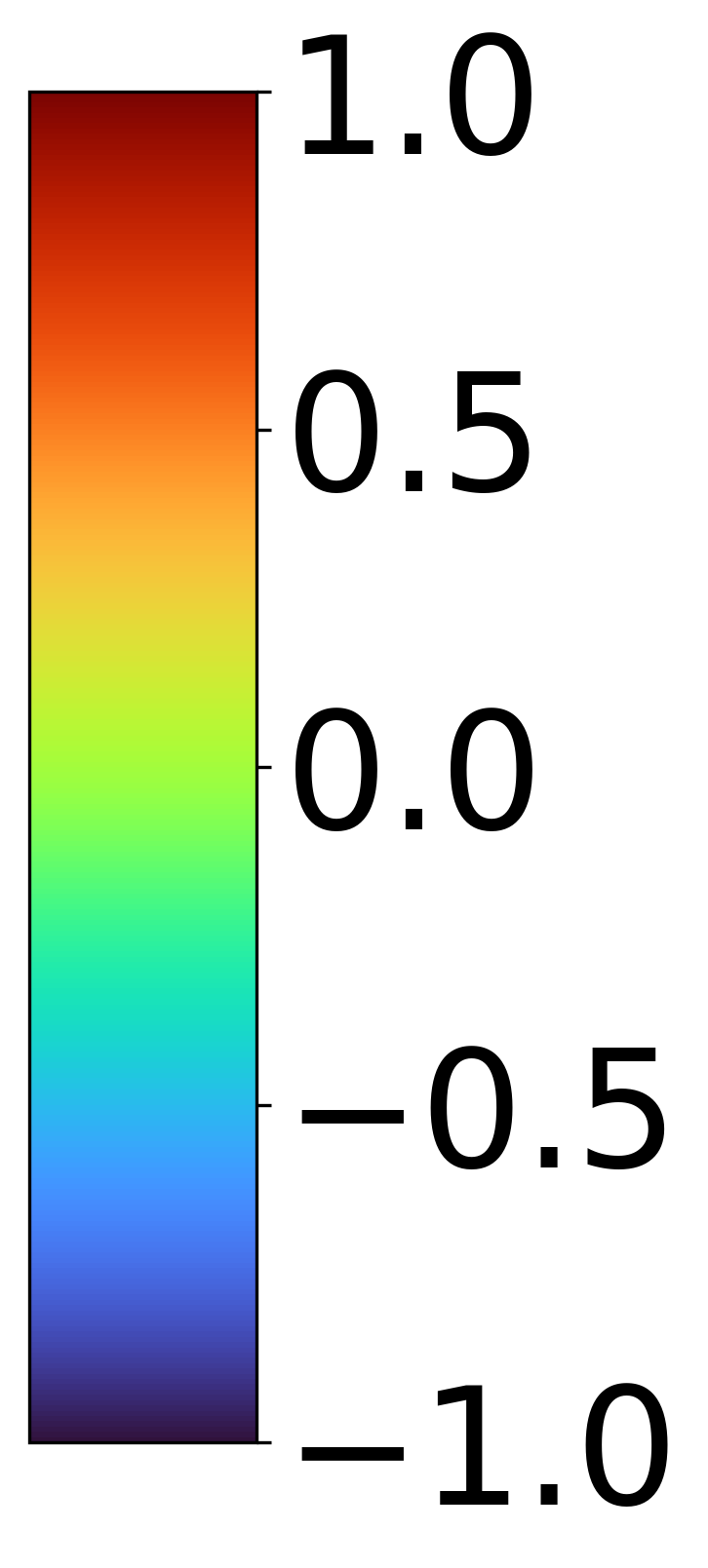}}}; &
        \node{\subcaptionbox{$v$ Error, $10$th percentile}[0.31\textwidth]{%
          \includegraphics[width=0.3\textwidth]{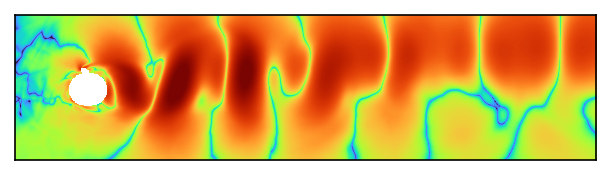}}}; &
        \node{\subcaptionbox*{}[0.037\textwidth]{\includegraphics[width=0.024\textwidth]{Figures/percentile/error_colorbar.png}}}; \\        
        \node{\subcaptionbox{$u$ GT, $50$th percentile}[0.31\textwidth]{%
          \includegraphics[width=0.3\textwidth]{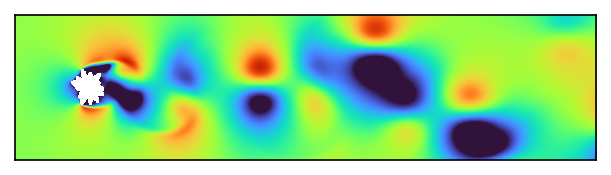}}}; &
        \node{\subcaptionbox{$u$ Predicted, $50$th percentile}[0.31\textwidth]{%
          \includegraphics[width=0.3\textwidth]{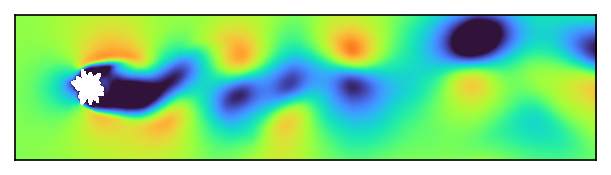}}}; &
        \node{\subcaptionbox*{}[0.037\textwidth]{\includegraphics[width=0.032\textwidth]{Figures/percentile/u_colorbar.png}}}; &
        \node{\subcaptionbox{$u$ Error, $50$th percentile}[0.31\textwidth]{%
          \includegraphics[width=0.3\textwidth]{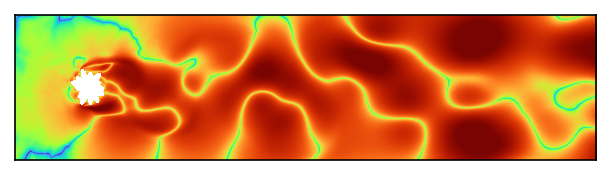}}}; &
        \node{\subcaptionbox*{}[0.037\textwidth]{\includegraphics[width=0.024\textwidth]{Figures/percentile/error_colorbar.png}}}; \\
        \node{\subcaptionbox{$v$ GT, $50$th percentile}[0.31\textwidth]{%
          \includegraphics[width=0.3\textwidth]{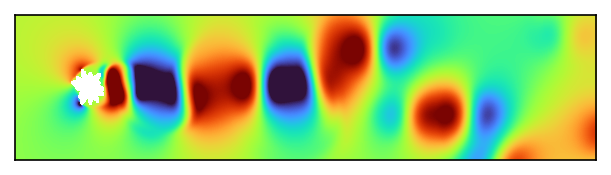}}}; &
        \node{\subcaptionbox{$v$ Predicted, $50$th percentile}[0.31\textwidth]{%
          \includegraphics[width=0.3\textwidth]{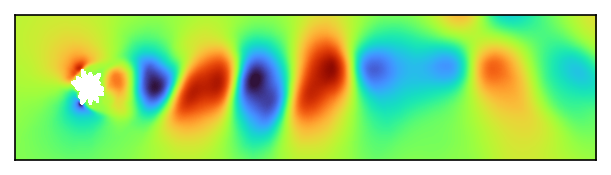}}}; &
        \node{\subcaptionbox*{}[0.037\textwidth]{\includegraphics[width=0.032\textwidth]{Figures/percentile/v_colorbar.png}}}; &
        \node{\subcaptionbox{$v$ Error, $50$th percentile}[0.31\textwidth]{%
          \includegraphics[width=0.3\textwidth]{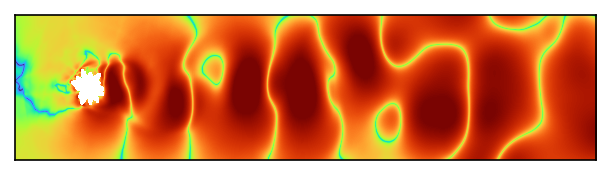}}}; &
        \node{\subcaptionbox*{}[0.037\textwidth]{\includegraphics[width=0.024\textwidth]{Figures/percentile/error_colorbar.png}}}; \\    
        \node{\subcaptionbox{$u$ GT, $90$th percentile}[0.31\textwidth]{%
          \includegraphics[width=0.3\textwidth]{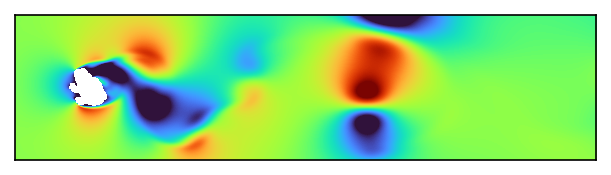}}}; &
        \node{\subcaptionbox{$u$ Predicted, $90$th percentile}[0.31\textwidth]{%
          \includegraphics[width=0.3\textwidth]{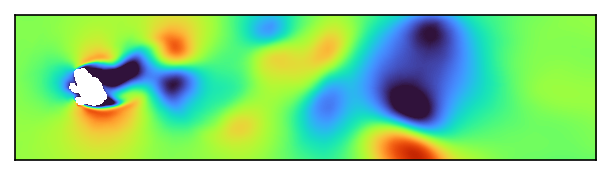}}}; &
        \node{\subcaptionbox*{}[0.037\textwidth]{\includegraphics[width=0.032\textwidth]{Figures/percentile/u_colorbar.png}}}; &
        \node{\subcaptionbox{$u$ Error, $90$th percentile}[0.31\textwidth]{%
          \includegraphics[width=0.3\textwidth]{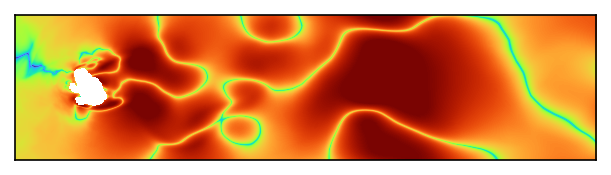}}}; &
        \node{\subcaptionbox*{}[0.037\textwidth]{\includegraphics[width=0.024\textwidth]{Figures/percentile/error_colorbar.png}}}; \\
        \node{\subcaptionbox{$v$ GT, $90$th percentile}[0.31\textwidth]{%
          \includegraphics[width=0.3\textwidth]{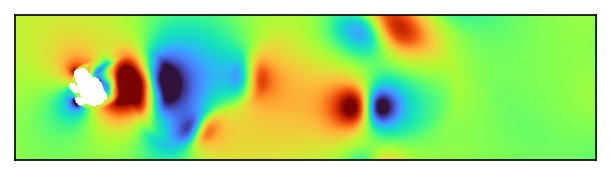}}}; &
        \node{\subcaptionbox{$v$ Predicted, $90$th percentile}[0.31\textwidth]{%
          \includegraphics[width=0.3\textwidth]{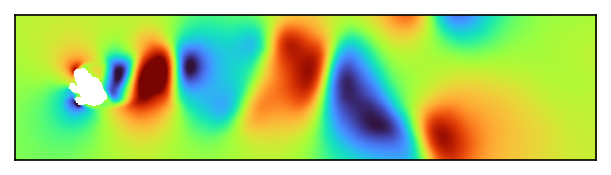}}}; &
        \node{\subcaptionbox*{}[0.037\textwidth]{\includegraphics[width=0.032\textwidth]{Figures/percentile/v_colorbar.png}}}; &
        \node{\subcaptionbox{$v$ Error, $90$th percentile}[0.31\textwidth]{%
          \includegraphics[width=0.3\textwidth]{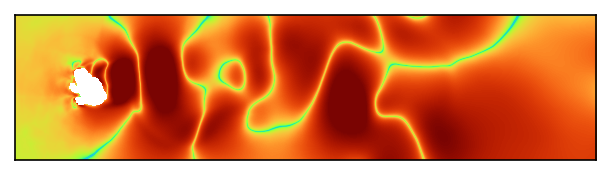}}}; &
        \node{\subcaptionbox*{}[0.037\textwidth]{\includegraphics[width=0.024\textwidth]{Figures/percentile/error_colorbar.png}}}; \\           
      };
    \end{tikzpicture}
    \label{fig:sub-6d}
  \end{subfigure}

  \caption{Rollout predictions at the final timestep ($t = 60$) for representative test cases selected at the $10$th, $50$th, and $90$th percentiles error. Shown are the ground-truth (GT), predicted, and absolute error fields for the velocity components $u$ and $v$. The absolute error fields are plotted on a logarithmic scale, showing increased error accumulation for higher-percentile geometries.}
  \label{Fig:rollout-percentile}
\end{figure}

\paragraph{Representative Rollout Behavior Across Error Percentiles}
\figref{Fig:rollout-percentile} visualizes rollout predictions at the final timestep (\(t = 60\)) for test geometries selected at the \(10\)th, \(50\)th, and \(90\)th percentiles of the rollout error. In addition to the ground-truth and predicted velocity fields, pointwise absolute error fields are reported on a logarithmic scale to show error accumulation. The low-percentile (\(10\)th) case corresponds to a smoother geometry resulting in a more symmetric flow about the centerline. In contrast, the high-percentile cases (\(50\)th, \(90\)th) contain sharp corners and asymmetric boundary features, leading to substantially larger errors in the downstream wake. These results provide a quantitative comparison of easy and challenging geometries and illustrate that sharper geometric features systematically increase error accumulation over long prediction horizons.

\begin{figure}[!t]
  \centering
  \begin{tikzpicture}
    \begin{groupplot}[
      group style={
        group size=5 by 4,
        horizontal sep=10mm,      
        vertical sep=76pt,
        },
        table/col sep=comma,       
        scale only axis,           
        x label style={yshift=2pt},
        every axis y label/.append style={
    at={(axis description cs:0,0.5)}, 
    anchor=center,                    
    xshift=8pt,                      
    yshift=20pt,
  },
    ]
      \nextgroupplot[
        width=0.1\textwidth,
        height=0.13\textwidth,
        axis x line=none,
        axis y line=none,
        ticks=none,
        enlarge x limits=false,
        enlarge y limits=false
      ]
      \addplot graphics [xmin=0,xmax=1,ymin=0,ymax=1]
        {Figures/geometries/sample_50.png};

      \nextgroupplot[
        width=0.15\linewidth,
        height=0.13\textwidth,
        xlabel={time},
        ylabel={$u$},
        title={$x=1D$},
        legend style={
          at={(0.5,-0.5)},anchor=north,
          legend columns=2,font=\scriptsize
        },
        clip marker paths=true
      ]
      \addplot table[x=time,y=u_gt2]{Figures/1in/rollout/plot_over_line/rollout_point_pred_s50.csv};
      \addlegendentry{GT}
      \addplot table[x=time,y=u_pr2]{Figures/1in/rollout/plot_over_line/rollout_point_pred_s50.csv};
      \addlegendentry{Pred};

      \nextgroupplot[
        width=0.15\linewidth,
        height=0.13\textwidth,
        xlabel={time},
        ylabel={$v$},
        title={$x=1D$},
        legend style={
          at={(0.5,-0.5)},anchor=north,
          legend columns=2,font=\scriptsize
        },
        clip marker paths=true
      ]
      \addplot table[x=time,y=v_gt2]{Figures/1in/rollout/plot_over_line/rollout_point_pred_s50.csv};
      \addlegendentry{GT}
      \addplot table[x=time,y=v_pr2]{Figures/1in/rollout/plot_over_line/rollout_point_pred_s50.csv};
      \addlegendentry{Pred};

      \nextgroupplot[
        width=0.15\linewidth,
        height=0.13\textwidth,
        xlabel={time},
        ylabel={$u$},
        title={$x=2D$},
        legend style={
          at={(0.5,-0.5)},anchor=north,
          legend columns=2,font=\scriptsize
        },
        clip marker paths=true
      ]
      \addplot table[x=time,y=u_gt3]{Figures/1in/rollout/plot_over_line/rollout_point_pred_s50.csv};
      \addlegendentry{GT}
      \addplot table[x=time,y=u_pr3]{Figures/1in/rollout/plot_over_line/rollout_point_pred_s50.csv};
      \addlegendentry{Pred};

      \nextgroupplot[
        width=0.15\linewidth,
        height=0.13\textwidth,
        xlabel={time},
        ylabel={$v$},
        title={$x=2D$},
        legend style={
          at={(0.5,-0.5)},anchor=north,
          legend columns=2,font=\scriptsize
        },
        clip marker paths=true
      ]
      \addplot table[x=time,y=v_gt3]{Figures/1in/rollout/plot_over_line/rollout_point_pred_s50.csv};
      \addlegendentry{GT}
      \addplot table[x=time,y=v_pr3]{Figures/1in/rollout/plot_over_line/rollout_point_pred_s50.csv};
      \addlegendentry{Pred};

      \nextgroupplot[
        width=0.1\textwidth,
        height=0.13\textwidth,
        axis x line=none,
        axis y line=none,
        ticks=none,
        enlarge x limits=false,
        enlarge y limits=false
      ]
      \addplot graphics [xmin=0,xmax=1,ymin=0,ymax=1]
        {Figures/geometries/sample_150.png};

      \nextgroupplot[
        width=0.15\linewidth,
        height=0.13\textwidth,
        xlabel={time},
        ylabel={$u$},
        title={$x=1D$},
        legend style={
          at={(0.5,-0.5)},anchor=north,
          legend columns=2,font=\scriptsize
        },
        clip marker paths=true
      ]
      \addplot table[x=time,y=u_gt2]{Figures/1in/rollout/plot_over_line/rollout_point_pred_s150.csv};
      \addlegendentry{GT}
      \addplot table[x=time,y=u_pr2]{Figures/1in/rollout/plot_over_line/rollout_point_pred_s150.csv};
      \addlegendentry{Pred};

      \nextgroupplot[
        width=0.15\linewidth,
        height=0.13\textwidth,
        xlabel={time},
        ylabel={$v$},
        title={$x=1D$},
        legend style={
          at={(0.5,-0.5)},anchor=north,
          legend columns=2,font=\scriptsize
        },
        clip marker paths=true
      ]
      \addplot table[x=time,y=v_gt2]{Figures/1in/rollout/plot_over_line/rollout_point_pred_s150.csv};
      \addlegendentry{GT}
      \addplot table[x=time,y=v_pr2]{Figures/1in/rollout/plot_over_line/rollout_point_pred_s150.csv};
      \addlegendentry{Pred};

      \nextgroupplot[
        width=0.15\linewidth,
        height=0.13\textwidth,
        xlabel={time},
        ylabel={$u$},
        title={$x=2D$},
        legend style={
          at={(0.5,-0.5)},anchor=north,
          legend columns=2,font=\scriptsize
        },
        clip marker paths=true
      ]
      \addplot table[x=time,y=u_gt3]{Figures/1in/rollout/plot_over_line/rollout_point_pred_s150.csv};
      \addlegendentry{GT}
      \addplot table[x=time,y=u_pr3]{Figures/1in/rollout/plot_over_line/rollout_point_pred_s150.csv};
      \addlegendentry{Pred};

      \nextgroupplot[
        width=0.15\linewidth,
        height=0.13\textwidth,
        xlabel={time},
        ylabel={$v$},
        title={$x=2D$},
        legend style={
          at={(0.5,-0.5)},anchor=north,
          legend columns=2,font=\scriptsize
        },
        clip marker paths=true
      ]
      \addplot table[x=time,y=v_gt3]{Figures/1in/rollout/plot_over_line/rollout_point_pred_s150.csv};
      \addlegendentry{GT}
      \addplot table[x=time,y=v_pr3]{Figures/1in/rollout/plot_over_line/rollout_point_pred_s150.csv};
      \addlegendentry{Pred};

      \nextgroupplot[
        width=0.1\textwidth,
        height=0.13\textwidth,
        axis x line=none,
        axis y line=none,
        ticks=none,
        enlarge x limits=false,
        enlarge y limits=false
      ]
      \addplot graphics [xmin=0,xmax=1,ymin=0,ymax=1]
        {Figures/geometries/sample_200.png};

      \nextgroupplot[
        width=0.15\linewidth,
        height=0.13\textwidth,
        xlabel={time},
        ylabel={$u$},
        title={$x=1D$},
        legend style={
          at={(0.5,-0.5)},anchor=north,
          legend columns=2,font=\scriptsize
        },
        clip marker paths=true
      ]
      \addplot table[x=time,y=u_gt2]{Figures/1in/rollout/plot_over_line/rollout_point_pred_s200.csv};
      \addlegendentry{GT}
      \addplot table[x=time,y=u_pr2]{Figures/1in/rollout/plot_over_line/rollout_point_pred_s200.csv};
      \addlegendentry{Pred};

      \nextgroupplot[
        width=0.15\linewidth,
        height=0.13\textwidth,
        xlabel={time},
        ylabel={$v$},
        title={$x=1D$},
        legend style={
          at={(0.5,-0.5)},anchor=north,
          legend columns=2,font=\scriptsize
        },
        clip marker paths=true
      ]
      \addplot table[x=time,y=v_gt2]{Figures/1in/rollout/plot_over_line/rollout_point_pred_s200.csv};
      \addlegendentry{GT}
      \addplot table[x=time,y=v_pr2]{Figures/1in/rollout/plot_over_line/rollout_point_pred_s200.csv};
      \addlegendentry{Pred};

      \nextgroupplot[
        width=0.15\linewidth,
        height=0.13\textwidth,
        xlabel={time},
        ylabel={$u$},
        title={$x=2D$},
        legend style={
          at={(0.5,-0.5)},anchor=north,
          legend columns=2,font=\scriptsize
        },
        clip marker paths=true
      ]
      \addplot table[x=time,y=u_gt3]{Figures/1in/rollout/plot_over_line/rollout_point_pred_s200.csv};
      \addlegendentry{GT}
      \addplot table[x=time,y=u_pr3]{Figures/1in/rollout/plot_over_line/rollout_point_pred_s200.csv};
      \addlegendentry{Pred};

      \nextgroupplot[
        width=0.15\linewidth,
        height=0.13\textwidth,
        xlabel={time},
        ylabel={$v$},
        title={$x=2D$},
        legend style={
          at={(0.5,-0.5)},anchor=north,
          legend columns=2,font=\scriptsize
        },
        clip marker paths=true
      ]
      \addplot table[x=time,y=v_gt3]{Figures/1in/rollout/plot_over_line/rollout_point_pred_s200.csv};
      \addlegendentry{GT}
      \addplot table[x=time,y=v_pr3]{Figures/1in/rollout/plot_over_line/rollout_point_pred_s200.csv};
      \addlegendentry{Pred};

      \nextgroupplot[
        width=0.1\textwidth,
        height=0.13\textwidth,
        axis x line=none,
        axis y line=none,
        ticks=none,
        enlarge x limits=false,
        enlarge y limits=false
      ]
      \addplot graphics [xmin=0,xmax=1,ymin=0,ymax=1]
        {Figures/geometries/sample_250.png};

      \nextgroupplot[
        width=0.15\linewidth,
        height=0.13\textwidth,
        xlabel={time},
        ylabel={$u$},
        title={$x=1D$},
        legend style={
          at={(0.5,-0.5)},anchor=north,
          legend columns=2,font=\scriptsize
        },
        clip marker paths=true
      ]
      \addplot table[x=time,y=u_gt2]{Figures/1in/rollout/plot_over_line/rollout_point_pred_s250.csv};
      \addlegendentry{GT}
      \addplot table[x=time,y=u_pr2]{Figures/1in/rollout/plot_over_line/rollout_point_pred_s250.csv};
      \addlegendentry{Pred};

      \nextgroupplot[
        width=0.15\linewidth,
        height=0.13\textwidth,
        xlabel={time},
        ylabel={$v$},
        title={$x=1D$},
        legend style={
          at={(0.5,-0.5)},anchor=north,
          legend columns=2,font=\scriptsize
        },
        clip marker paths=true
      ]
      \addplot table[x=time,y=v_gt2]{Figures/1in/rollout/plot_over_line/rollout_point_pred_s250.csv};
      \addlegendentry{GT}
      \addplot table[x=time,y=v_pr2]{Figures/1in/rollout/plot_over_line/rollout_point_pred_s250.csv};
      \addlegendentry{Pred};

      \nextgroupplot[
        width=0.15\linewidth,
        height=0.13\textwidth,
        xlabel={time},
        ylabel={$u$},
        title={$x=2D$},
        legend style={
          at={(0.5,-0.5)},anchor=north,
          legend columns=2,font=\scriptsize
        },
        clip marker paths=true
      ]
      \addplot table[x=time,y=u_gt3]{Figures/1in/rollout/plot_over_line/rollout_point_pred_s250.csv};
      \addlegendentry{GT}
      \addplot table[x=time,y=u_pr3]{Figures/1in/rollout/plot_over_line/rollout_point_pred_s250.csv};
      \addlegendentry{Pred};

      \nextgroupplot[
        width=0.15\linewidth,
        height=0.13\textwidth,
        xlabel={time},
        ylabel={$v$},
        title={$x=2D$},
        legend style={
          at={(0.5,-0.5)},anchor=north,
          legend columns=2,font=\scriptsize
        },
        clip marker paths=true
      ]
      \addplot table[x=time,y=v_gt3]{Figures/1in/rollout/plot_over_line/rollout_point_pred_s250.csv};
      \addlegendentry{GT}
      \addplot table[x=time,y=v_pr3]{Figures/1in/rollout/plot_over_line/rollout_point_pred_s250.csv};
      \addlegendentry{Pred};

    \end{groupplot}
  \end{tikzpicture}
  \caption{Rollout time‐series of $u$ and $v$ at two points at downstream distance from geometry $x=1D$ and $x=2D$, where D is the geometry diameter. We show a collection of 4 shapes where each row correspond to a single geometry.}
  \label{Fig:rollout-shapes_tseries}
\end{figure}

\paragraph{Point Wise Temporal Dynamics During Rollouts} \figref{Fig:rollout-shapes_tseries} shows the rollout time series of \(u\) and \(v\) at two downstream probes located at \(x=1D\) and \(x=2D\) (where \(D\) is the characteristic diameter of each shape) for four representative geometries. These probes lie within the near--geometry boundary layer region, where unsteady viscous effects and wake development are most pronounced. The predicted signals maintain accurate phase alignment and amplitude matching with ground truth for the first 20--30 timesteps across all shapes but diverge gradually thereafter. The two shapes with sharp corners (rows one and two) exhibit larger deviations, characterized by phase lag and damped peak values, compared to the smoother geometries (row four), which demonstrate closer agreement through \(t=60\). These results demonstrate the surrogate’s ability to capture essential unsteady boundary layer phenomena under feedback, while highlighting systematic error growth during rollouts, which scales with boundary complexity.

\clearpage

\paragraph{Strouhal Number and Phase Lag}
\figref{Fig:strouhal-phase-lag-1in} evaluates how well the surrogate preserves the periodic wake dynamics for a single-timestep history (\(s=1\)). For each test geometry, we record the vertical velocity \(v(x,t)\) on the wake centerline at four probes (\(x/D = 1,2,3,4\)) and define the Strouhal number as the dominant non-dimensional frequency of \(v(x,t)\). The left column compares predicted versus ground-truth Strouhal numbers at all probes: the points form a tight cloud around the \(y = x\) line with only a few high- and low-frequency outliers, and the number of these extremes decreases further downstream as the wake becomes less sensitive to local geometric details. 

\begin{figure}[!t]
  \centering

  \begin{subfigure}[t]{0.48\textwidth}
    \centering
    \begin{tikzpicture}
      \begin{axis}[
        width=0.98\textwidth,
        height=0.6\textwidth,
        xmin=0, xmax=0.42,
        ymin=0, ymax=0.42,
        xlabel={GT $St$},
        ylabel={Pred.\ $St$},
        title={$x = 1D$},
      ]
        \addplot[
          only marks,
          mark=*,
          mark size=1pt
        ] table[
          col sep=comma,
          x=St_gt_x_2.5,
          y=St_pred_x_2.5
        ] {Figures/Strouhal/1in_rollout_strouhal.csv};

        \addplot[dashed] coordinates {(0,0) (0.4,0.4)};
      \end{axis}
    \end{tikzpicture}
  \end{subfigure}%
  \begin{subfigure}[t]{0.48\textwidth}
    \centering
    \begin{tikzpicture}
      \begin{axis}[
        width=0.98\textwidth,
        height=0.6\textwidth,
        xmin=0, xmax=265,
        ymin=-3.14, ymax=3.14,
        xlabel={Sample index},
        ylabel={Phase lag},
        title={$x = 1D$}
      ]
        \addplot[
          only marks,
          mark=*,
          mark size=1pt
        ] table[
          col sep=comma,
          x=sample_id,
          y=phase_lag_x_2.5
        ] {Figures/Strouhal/1in_rollout_phase_lag.csv};
      \end{axis}
    \end{tikzpicture}
  \end{subfigure}%
  \hfill

  \begin{subfigure}[t]{0.48\textwidth}
    \centering
    \begin{tikzpicture}
      \begin{axis}[
        width=0.98\textwidth,
        height=0.6\textwidth,
        xmin=0, xmax=0.42,
        ymin=0, ymax=0.42,
        xlabel={GT $St$},
        ylabel={Pred.\ $St$},
        title={$x = 2D$},
      ]
        \addplot[
          only marks,
          mark=*,
          mark size=1pt
        ] table[
          col sep=comma,
          x=St_gt_x_3.5,
          y=St_pred_x_3.5
        ] {Figures/Strouhal/1in_rollout_strouhal.csv};

        \addplot[dashed] coordinates {(0,0) (0.4,0.4)};
      \end{axis}
    \end{tikzpicture}
  \end{subfigure}%
  \begin{subfigure}[t]{0.48\textwidth}
    \centering
    \begin{tikzpicture}
      \begin{axis}[
        width=0.98\textwidth,
        height=0.6\textwidth,
        xmin=0, xmax=265,
        ymin=-3.14, ymax=3.14,
        xlabel={Sample index},
        ylabel={Phase lag},
        title={$x = 2D$},
      ]
        \addplot[
          only marks,
          mark=*,
          mark size=1pt
        ] table[
          col sep=comma,
          x=sample_id,
          y=phase_lag_x_3.5
        ] {Figures/Strouhal/1in_rollout_phase_lag.csv};
      \end{axis}
    \end{tikzpicture}
  \end{subfigure}%
  \hfill

  \begin{subfigure}[t]{0.48\textwidth}
    \centering
    \begin{tikzpicture}
      \begin{axis}[
        width=0.98\textwidth,
        height=0.6\textwidth,
        xmin=0, xmax=0.42,
        ymin=0, ymax=0.42,
        xlabel={GT $St$},
        ylabel={Pred.\ $St$},
        title={$x = 3D$},
      ]
        \addplot[
          only marks,
          mark=*,
          mark size=1pt
        ] table[
          col sep=comma,
          x=St_gt_x_4.5,
          y=St_pred_x_4.5
        ] {Figures/Strouhal/1in_rollout_strouhal.csv};

        \addplot[dashed] coordinates {(0,0) (0.4,0.4)};
      \end{axis}
    \end{tikzpicture}
  \end{subfigure}%
  \begin{subfigure}[t]{0.48\textwidth}
    \centering
    \begin{tikzpicture}
      \begin{axis}[
        width=0.98\textwidth,
        height=0.6\textwidth,
        xmin=0, xmax=265,
        ymin=-3.14, ymax=3.14,
        xlabel={Sample index},
        ylabel={Phase lag},
        title={$x = 3D$},
      ]
        \addplot[
          only marks,
          mark=*,
          mark size=1pt
        ] table[
          col sep=comma,
          x=sample_id,
          y=phase_lag_x_4.5
        ] {Figures/Strouhal/1in_rollout_phase_lag.csv};
      \end{axis}
    \end{tikzpicture}
  \end{subfigure}%
  \hfill

  \begin{subfigure}[t]{0.48\textwidth}
    \centering
    \begin{tikzpicture}
      \begin{axis}[
        width=0.98\textwidth,
        height=0.6\textwidth,
        xmin=0, xmax=0.42,
        ymin=0, ymax=0.42,
        xlabel={GT $St$},
        ylabel={Pred.\ $St$},
        title={$x = 4D$},
      ]
        \addplot[
          only marks,
          mark=*,
          mark size=1pt
        ] table[
          col sep=comma,
          x=St_gt_x_5.5,
          y=St_pred_x_5.5
        ] {Figures/Strouhal/1in_rollout_strouhal.csv};

        \addplot[dashed] coordinates {(0,0) (0.4,0.4)};
      \end{axis}
    \end{tikzpicture}
  \end{subfigure}%
  \begin{subfigure}[t]{0.48\textwidth}
    \centering
    \begin{tikzpicture}
      \begin{axis}[
        width=0.98\textwidth,
        height=0.6\textwidth,
        xmin=0, xmax=265,
        ymin=-3.14, ymax=3.14,
        xlabel={Sample index},
        ylabel={Phase lag},
        title={$x = 4D$},
      ]
        \addplot[
          only marks,
          mark=*,
          mark size=1pt
        ] table[
          col sep=comma,
          x=sample_id,
          y=phase_lag_x_5.5
        ] {Figures/Strouhal/1in_rollout_phase_lag.csv};
      \end{axis}
    \end{tikzpicture}
  \end{subfigure}%
  \hfill

  \caption{Strouhal number and phase‐lag (in radians) for the case of \(s=1\) at four downstream probe locations. Left column: predicted versus ground‐truth Strouhal number with a dashed $y=x$ reference line. Right column: phase lag (prediction minus ground truth) for test samples. Each row corresponds to a different streamwise position $x=1D$, $x=2D$, $x=3D$, and $x=4D$, where D is the geometry diameter.}
  \label{Fig:strouhal-phase-lag-1in}
\end{figure}

\begin{table}[t!]
    \centering
    \caption{Relative $L_2$ and $L_\infty$ errors in the predicted Strouhal number, and mean / maximum phase lag between predicted and ground-truth wake signals, for different input sequence lengths $s$ and downstream probe locations $x/D$. Bold values indicate the best-performing sequence length for each metric across all probe locations $(x/D)$.}

    \label{tab:strouhal_phase_all}
    \begin{tabular}{cccccc}
        \toprule
        $x/D$ & $s$ 
        & $\mathrm{relative}\,L_2(\mathrm{Strouhal})$ 
        & $L_\infty(\mathrm{Strouhal})$ 
        & mean phase lag [rad] 
        & max phase lag [rad] \\
        \midrule
        \multirow{4}{*}{1} 
        & 1  & 0.233 & 0.563 & 0.337 & 3.004 \\
        & 4  & 0.228 & 0.546 & 0.353 & 2.876 \\
        & 8  & 0.210 & 0.521 & \textbf{0.128} & 2.886 \\
        & 16 & \textbf{0.190} & \textbf{0.418} & 0.168 & \textbf{2.666} \\
        \midrule
        \multirow{4}{*}{2} 
        & 1  & 0.209 & 0.589 & 0.267 & 2.807 \\
        & 4  & 0.210 & 0.575 & 0.365 & 2.832 \\
        & 8  & 0.195 & 0.557 & 0.140 & \textbf{2.399} \\
        & 16 & \textbf{0.182} & \textbf{0.525} & \textbf{0.033} & 2.981 \\
        \midrule
        \multirow{4}{*}{3} 
        & 1  & 0.196 & 0.578 & 0.274 & 3.099 \\
        & 4  & 0.213 & 0.558 & 0.314 & 3.009 \\
        & 8  & 0.193 & 0.581 & 0.160 & 3.011 \\
        & 16 & \textbf{0.178} & \textbf{0.513} & \textbf{0.049} & \textbf{2.617} \\
        \midrule
        \multirow{4}{*}{4} 
        & 1  & 0.190 & 0.565 & 0.241 & \textbf{3.028} \\
        & 4  & 0.210 & 0.510 & 0.326 & 3.033 \\
        & 8  & 0.211 & 0.565 & 0.184 & 3.124 \\
        & 16 & \textbf{0.171} & \textbf{0.495} & \textbf{0.091} & 3.101 \\
        \bottomrule
    \end{tabular}
\end{table}

To quantify phase coherence, we estimate at each probe a phase offset \(\phi(x)\) by finding the time delay \(\tau(x)\) that maximizes the cross-correlation between predicted and ground-truth signals at the dominant shedding frequency, and then converting this delay into \(\phi(x) = -2\pi f_{\text{shed}}(x)\,\tau(x)\) (right column). Phase lags are concentrated near zero, with fluctuations and no systematic tendency to lead or lag. In addition to these distributions, ~\tabref{tab:strouhal_phase_all} summarizes error statistics for Strouhal number and phase lag across downstream probes and sequence lengths. For a single-timestep input (\(s=1\)), the relative \(L_2\) error in Strouhal number lies between \(0.19\) and \(0.23\) with \(L_\infty < 0.59\) at all locations. The mean phase lag is about \(0.3\)~rad, with outliers approaching \(3\)~rad that correspond to the most challenging, sharp-cornered geometries. As the history length increases (\(s=4,8,16\)), both metrics generally improve: the best relative \(L_2\) error in Strouhal number decreases to \(\approx 0.17\) and \(L_\infty\) drops below \(\approx 0.52\) at most probes, while the mean phase lag is reduced to \(< 0.15\)~rad for \(s=8\) and below \(0.1\)~rad at several probes for \(s=16\). The maximum phase lag remains \(\mathcal{O}(3)\)~rad due to a small number of difficult cases, but these outliers do not dominate the statistics. Overall, the model captures the dominant shedding frequency and its phase with minimal temporal context, and longer input sequences further improve both frequency and phase predictions.

\paragraph{Error Amplification at Sharp Corners}
We find that sharp corners are especially prone to error accumulation. First, the SDF at our $1024 \times 256$ grid only approximates sharp corners in a pixelated way, smoothing out true corner geometry. Second, the CNN encoder downsamples spatial resolution by $32 \times$, which further blurs small‐scale vortical structures that originate at those corners. During rollouts, these initial insufficient encodings at sharp edges propagate downstream and amplify, leading to the larger errors observed for high curvature shapes.

\clearpage

\begin{figure}[!t]
  \centering
  \begin{subfigure}[t]{0.48\textwidth}
    \centering
    \includegraphics[width=0.98\textwidth]{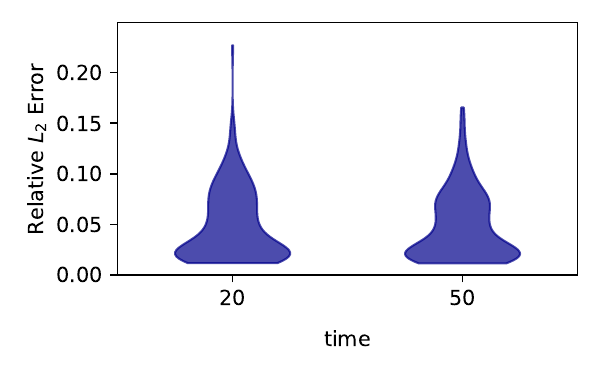}
    \caption{Single‐step relative $L_2$ error at $t = 20$ and $t = 50$.}
    \label{fig:ss-violin-rel20-50}
  \end{subfigure}%
  \hfill
    \begin{subfigure}[t]{0.48\textwidth}
    \centering
    \includegraphics[width=0.98\textwidth]{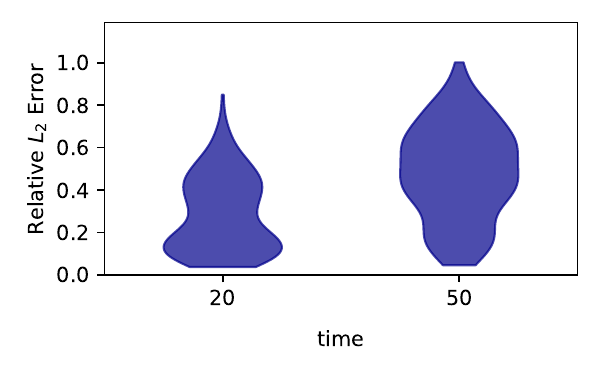}
    \caption{Rollout relative $L_2$ error at $t = 20$ and $t = 50$.}
    \label{fig:rollout-violin-rel20-50}
  \end{subfigure}%
  \caption{Violin‐style density estimates of relative $L_2$ error at $t = 20$ and $t = 50$ for (a) single‐step predictions and (b) autoregressive rollouts.}
  \label{fig:violin_errors}
\end{figure}

\begin{figure}[!t]
    \centering
    \begin{subfigure}[t]{0.48\textwidth}
    \centering
    \begin{tikzpicture}
      \begin{axis}[
        width=0.98\textwidth, 
        height=0.6\textwidth,
        xmin=0, xmax=60,
        ymin=0, ymax=0.1,
        xlabel={time},
        ylabel={Relative $L_2$ Error},
        scaled y ticks  = true,
        scaled y ticks  = base 10:2,        
        legend pos=north east
      ]
        \addplot table[
          col sep=comma,
          x=timestep,
          y=rel_Overall
        ] {Figures/1in/single_step/errors/single_step_errors.csv};
        \addplot table[
          col sep=comma,
          x=timestep,
          y=rel_Overall
        ] {Figures/meshgraphnet/1in/single_step_errors.csv};
        \legend{current,meshgraphnet}
      \end{axis}
    \end{tikzpicture}
    \caption{$L_2$ error using input sequence s=1.}
    \label{fig:sub-12a}
    \end{subfigure}%
    \begin{subfigure}[t]{0.48\textwidth}
    \centering
    \begin{tikzpicture}
      \begin{axis}[
        width=0.98\textwidth, 
        height=0.6\textwidth,
        xmin=0, xmax=60,
        ymin=0, ymax=1,
        xlabel={time},
        ylabel={$L_\infty$ Error},
        scaled y ticks  = true,
        scaled y ticks  = base 10:2,        
        legend pos=north east
      ]
        \addplot table[
          col sep=comma,
          x=timestep,
          y=linf_Overall
        ] {Figures/1in/single_step/errors/single_step_errors.csv};
        \addplot table[
          col sep=comma,
          x=timestep,
          y=linf_Overall
        ] {Figures/meshgraphnet/1in/single_step_errors.csv};
        \legend{current,meshgraphnet}
      \end{axis}
    \end{tikzpicture}
    \caption{$L_\infty$ error using input sequence s=1.}
    \label{fig:sub-12b}
    \end{subfigure}
    \hfill
    \begin{subfigure}[t]{0.48\textwidth}
    \centering
    \begin{tikzpicture}
      \begin{axis}[
        width=0.98\textwidth, 
        height=0.6\textwidth,
        xmin=3, xmax=60,
        ymin=0, ymax=0.1,
        xlabel={time},
        ylabel={Relative $L_2$ Error},
        scaled y ticks  = true,
        scaled y ticks  = base 10:2,        
        legend pos=north east
      ]
        \addplot table[
          col sep=comma,
          x=timestep,
          y=rel_Overall
        ] {Figures/4in/single_step/errors/single_step_errors.csv};
        \addplot table[
          col sep=comma,
          x=timestep,
          y=rel_Overall
        ] {Figures/meshgraphnet/4in/single_step_errors.csv};
        \legend{current,meshgraphnet}
      \end{axis}
    \end{tikzpicture}
    \caption{$L_2$ error using input sequence s=4.}
    \label{fig:sub-12c}
    \end{subfigure}
    \begin{subfigure}[t]{0.48\textwidth}
    \centering
    \begin{tikzpicture}
      \begin{axis}[
        width=0.98\textwidth, 
        height=0.6\textwidth,
        xmin=3, xmax=60,
        ymin=0, ymax=1,
        xlabel={time},
        ylabel={$L_\infty$ Error},
        scaled y ticks  = true,
        scaled y ticks  = base 10:2,        
        legend pos=north east
      ]
        \addplot table[
          col sep=comma,
          x=timestep,
          y=linf_Overall
        ] {Figures/4in/single_step/errors/single_step_errors.csv};
        \addplot table[
          col sep=comma,
          x=timestep,
          y=linf_Overall
        ] {Figures/meshgraphnet/4in/single_step_errors.csv};
        \legend{current,meshgraphnet}
      \end{axis}
    \end{tikzpicture}
    \caption{$L_\infty$ error using input sequence s=4.}
    \label{fig:sub-12d}
    \end{subfigure}%
    \caption{Single-step prediction error comparison between the proposed method and MeshGraphNet. Panels (a) and (b) show the time evolution of the relative $L_2$ and $L_\infty$ errors, respectively, when using an input sequence length of $s=1$. Panels (c) and (d) report the same error metrics for an input sequence length of $s=4$.}
    \label{Fig:errors-meshgraphnet}
\end{figure}

\paragraph{Sample Level Variability in Prediction Accuracy} \figref{fig:violin_errors} presents violin style density estimates of the relative \(L_{2}\) error across all test geometries at \(t=20\) and \(t=50\) for both single step and rollout evaluations. In the single step case (panel (a)), the error distribution at \(t=20\) is tightly concentrated around low values (peak near 2-3\%), and the error spread remains the same at \(t=50\), indicating that a minority of shapes -- particularly those with sharp features -- exhibit high instantaneous error. The rollout distributions (panel (b)) show substantially greater broadening: at \(t=20\), the median error is already higher than the single step case (peak near 15-20\%), and by \(t=50\) the density extends to over 40\% for most samples. The pronounced tails in both single step and rollout violins reveal that some geometries accumulate error much more rapidly, leading to a bimodal appearance in the density. This reflects that smoother shapes cluster at low error throughout, whereas irregular and high‐curvature geometries produce outliers with significantly degraded accuracy. Overall, these plots underscore that while the surrogate performs reliably on average, its worst case rollout performance can vary by an order of magnitude depending on sample shape.

\subsection{Benchmarking Against Graph Networks}

\begin{figure}[!t]
    \centering
    \begin{tikzpicture}
    \matrix [matrix of nodes,
           nodes={inner sep=0, anchor=south west},
           column sep=1pt, row sep=1pt]
    {
    \node{\subcaptionbox{$u$ GT, $t=5$}[0.30\textwidth]{%
      \includegraphics[width=0.3\textwidth]{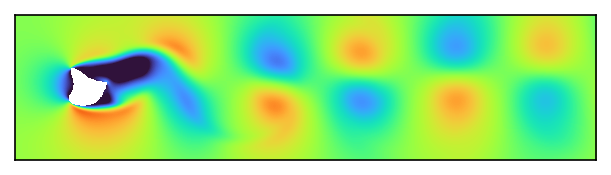}}}; &
    \node{\subcaptionbox{$u$ Predicted, $t=5$}[0.30\textwidth]{%
      \includegraphics[width=0.3\textwidth]{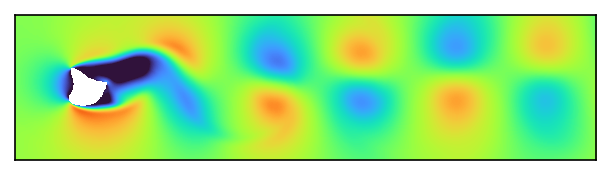}}}; &
    \node{\subcaptionbox*{}[0.037\textwidth]{\includegraphics[width=0.032\textwidth]{Figures/percentile/u_colorbar.png}}}; &
    \node{\subcaptionbox{$u$ Error, $t=5$}[0.30\textwidth]{%
      \includegraphics[width=0.3\textwidth]{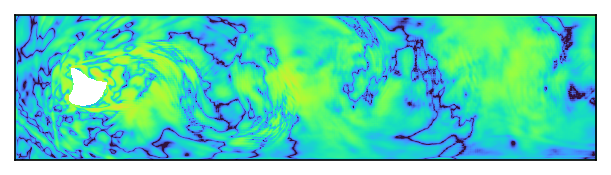}}}; &
    \node{\subcaptionbox*{}[0.037\textwidth]{\includegraphics[width=0.024\textwidth]{Figures/percentile/error_colorbar.png}}}; \\
    \node{\subcaptionbox{$v$ GT, $t=5$}[0.30\textwidth]{%
      \includegraphics[width=0.3\textwidth]{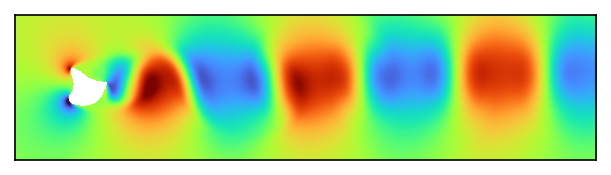}}}; &
    \node{\subcaptionbox{$v$ Predicted, $t=5$}[0.30\textwidth]{%
      \includegraphics[width=0.3\textwidth]{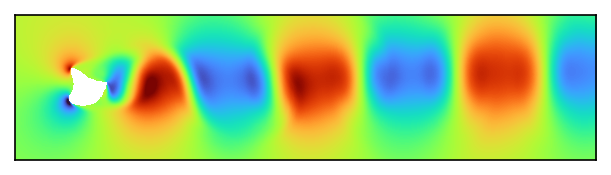}}}; &
    \node{\subcaptionbox*{}[0.037\textwidth]{\includegraphics[width=0.032\textwidth]{Figures/percentile/v_colorbar.png}}}; &
    \node{\subcaptionbox{$v$ Error, $t=5$}[0.30\textwidth]{%
      \includegraphics[width=0.3\textwidth]{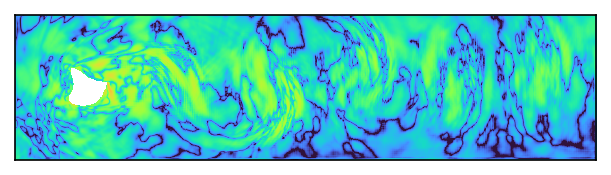}}}; &
    \node{\subcaptionbox*{}[0.037\textwidth]{\includegraphics[width=0.024\textwidth]{Figures/percentile/error_colorbar.png}}}; \\ 
    \node{\subcaptionbox{$u$ GT, $t=10$}[0.30\textwidth]{%
      \includegraphics[width=0.3\textwidth]{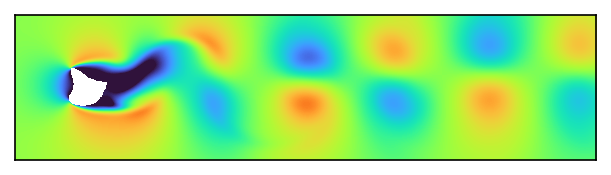}}}; &
    \node{\subcaptionbox{$u$ Predicted, $t=10$}[0.30\textwidth]{%
      \includegraphics[width=0.3\textwidth]{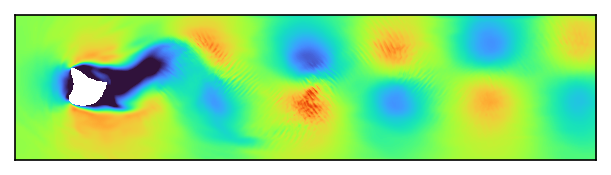}}}; &
    \node{\subcaptionbox*{}[0.037\textwidth]{\includegraphics[width=0.032\textwidth]{Figures/percentile/u_colorbar.png}}}; &
    \node{\subcaptionbox{$u$ Error, $t=10$}[0.30\textwidth]{%
      \includegraphics[width=0.3\textwidth]{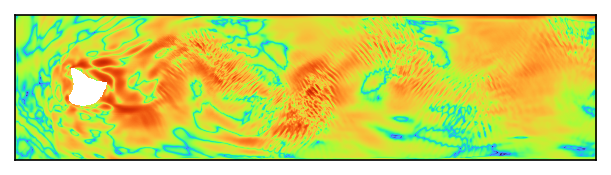}}}; &
    \node{\subcaptionbox*{}[0.037\textwidth]{\includegraphics[width=0.024\textwidth]{Figures/percentile/error_colorbar.png}}}; \\
    \node{\subcaptionbox{$v$ GT, $t=10$}[0.30\textwidth]{%
      \includegraphics[width=0.3\textwidth]{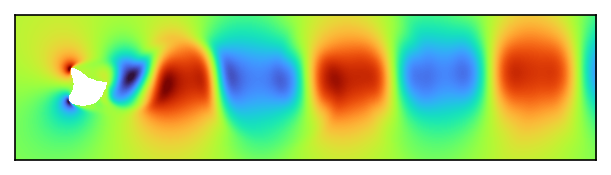}}}; &
    \node{\subcaptionbox{$v$ Predicted, $t=10$}[0.30\textwidth]{%
      \includegraphics[width=0.3\textwidth]{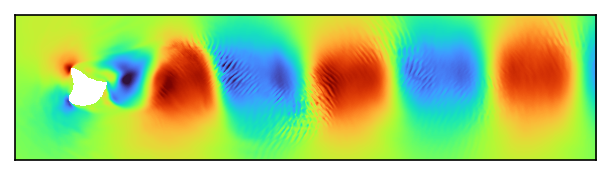}}}; &
    \node{\subcaptionbox*{}[0.037\textwidth]{\includegraphics[width=0.032\textwidth]{Figures/percentile/v_colorbar.png}}}; &
    \node{\subcaptionbox{$v$ Error, $t=10$}[0.30\textwidth]{%
      \includegraphics[width=0.3\textwidth]{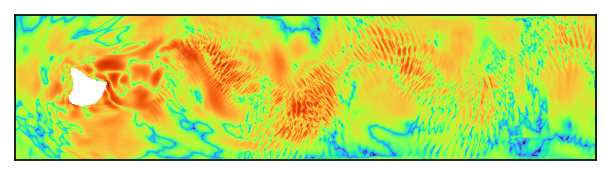}}}; &
    \node{\subcaptionbox*{}[0.037\textwidth]{\includegraphics[width=0.024\textwidth]{Figures/percentile/error_colorbar.png}}}; \\        
    \node{\subcaptionbox{$u$ GT, $t=20$}[0.30\textwidth]{%
      \includegraphics[width=0.3\textwidth]{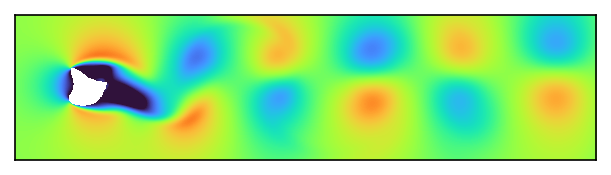}}}; &
    \node{\subcaptionbox{$u$ Predicted, $t=20$}[0.30\textwidth]{%
      \includegraphics[width=0.3\textwidth]{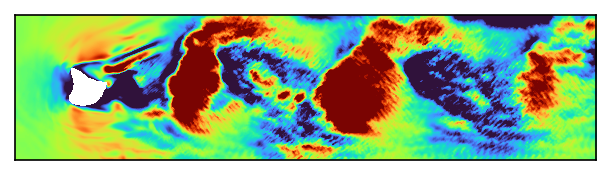}}}; &
    \node{\subcaptionbox*{}[0.037\textwidth]{\includegraphics[width=0.032\textwidth]{Figures/percentile/u_colorbar.png}}}; &
    \node{\subcaptionbox{$u$ Error, $t=20$}[0.30\textwidth]{%
      \includegraphics[width=0.3\textwidth]{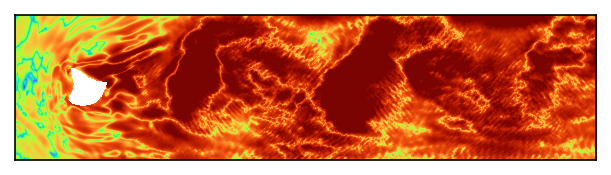}}}; &
    \node{\subcaptionbox*{}[0.037\textwidth]{\includegraphics[width=0.024\textwidth]{Figures/percentile/error_colorbar.png}}}; \\
    \node{\subcaptionbox{$v$ GT, $t=20$}[0.30\textwidth]{%
      \includegraphics[width=0.3\textwidth]{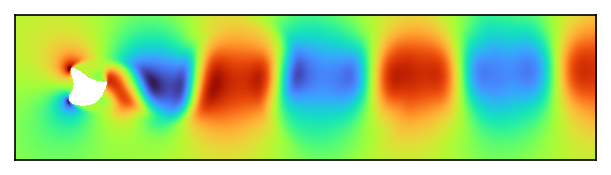}}}; &
    \node{\subcaptionbox{$v$ Predicted, $t=20$}[0.30\textwidth]{%
      \includegraphics[width=0.3\textwidth]{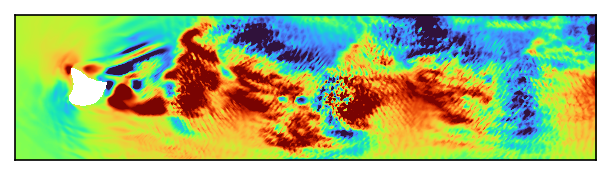}}}; &
    \node{\subcaptionbox*{}[0.037\textwidth]{\includegraphics[width=0.032\textwidth]{Figures/percentile/v_colorbar.png}}}; &
    \node{\subcaptionbox{$v$ Error, $t=20$}[0.30\textwidth]{%
      \includegraphics[width=0.3\textwidth]{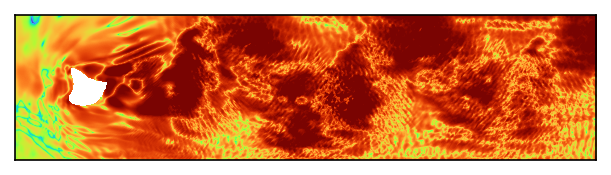}}}; &
    \node{\subcaptionbox*{}[0.037\textwidth]{\includegraphics[width=0.024\textwidth]{Figures/percentile/error_colorbar.png}}}; \\        
    };
    \end{tikzpicture}
    \caption{Rollout predictions of meshgraphnet for the geometry corresponding to $10$th percentile rollout error at $t=5$, $t=10$, and $t=20$. Shown are the ground-truth (GT), predicted, and absolute error fields for the velocity components $u$ and $v$. The absolute error fields are plotted on a logarithmic scale, showing increased error accumulation for higher-percentile geometries.}
    \label{Fig:rollout-meshgraphnet}
\end{figure}

\figref{Fig:errors-meshgraphnet} compares the single-step prediction accuracy between the proposed method and MeshGraphNet for two input sequence lengths, $s=1$ and $s=4$. Across all timesteps and both error metrics, MeshGraphNet consistently outperforms the proposed method in relative $L_2$ and $L_\infty$ errors, demonstrating its strong ability to capture instantaneous flow fields when ground-truth history is provided. However, as we show next, single-step prediction does not necessarily guarantee stable or accurate long-horizon predictions under autoregressive rollout. \figref{Fig:rollout-meshgraphnet} shows the autoregressive rollout behavior of MeshGraphNet using an input sequence length of $s=4$ for a representative geometry drawn from the $10$th percentile of rollout error. The first rollout snapshot at $t=5$ effectively corresponds to a single-step prediction, as the model is still conditioned on ground-truth history, and MeshGraphNet accurately reproduces the large-scale wake structure and velocity fields at this stage. As the rollout progresses, prediction errors begin to accumulate, and by $t=20$ the solution exhibits severe degradation, characterized by fragmented flow structures and widespread error growth. Beyond $t=20$, the prediction error diverges, indicating that the model is no longer producing stable predictions.

\section{Discussion and Conclusions} \label{sec:conclusion}

In this work, we introduce a Time-dependent Geometric Deep Operator Network that integrates an SDF based geometry encoding with a convolutional history encoder to predict unsteady, periodic flow around complex 2D shapes. Our extensive evaluation on the FlowBench flow past an object dataset demonstrates the following key findings:

\begin{itemize}
  \item \textbf{High Single‐Step Accuracy:} When provided with ground‐truth inputs, the surrogate achieves an average relative \(L_{2}\) error of approximately 5\% and stable RMSE values for both \(u\) and \(v\) components across 60 timesteps.
  \item \textbf{Error Accumulation in Rollouts:} Under autoregressive rollouts, prediction error grows monotonically to about 55\% relative \(L_{2}\) by \(t=60\), highlighting challenges in long horizon stability.
  \item \textbf{Geometric Generalization:} The model reliably captures vortex shedding patterns and wake structures across smooth shapes. Performance degrades the most for geometries with sharp corners, where rollouts exhibit smoothed peaks and displaced vortices. Predictions for smoother shapes maintain closer agreement with the ground truth data.
  \item \textbf{Point Wise Temporal Dynamics:} Time series at downstream probes (\(x=1D,2D\)) reveal accurate prediction of the time series for the single step predictions. The rollout prediction shows phase alignment and amplitude matching for the first 20-30 steps before error accumulates and degrades predictions.
  \item \textbf{Sample Level Variability:} Violin plot analysis shows pronounced tails and bimodality of the error distribution, indicating that complex geometries can incur an order-of-magnitude higher error than smoother counterparts.
\end{itemize}
These results underscore both the promise and limitations of purely data‐driven neural operators for unsteady flow: they offer dramatic speedups ($\times 1000$) and strong short term accuracy but require further developments to sustain long term autoregressive rollout on complex boundaries. While our trained MeshGraphNet exhibits limited extrapolation beyond approximately 20 rollout steps in our experiments, we believe that this is not a fundamental limitation of graph-based models. In particular, increasing model capacity (e.g., larger latent dimensions or deeper message-passing networks) and training for substantially longer durations could improve rollout stability over longer horizons. Our implementation prioritizes full-resolution training and comparable computational effort, and we therefore view the reported MeshGraphNet results as a conservative but fair baseline.

We focus on unsteady incompressible flows, where geometry-dependent effects such as vortex shedding and long-horizon autoregressive stability are central; extensions to other transient PDEs are left for future work. Future directions for this work span physics, generative refinement, temporal context, and generalization. First, one could incorporate explicit physical consistency during training by adding a Navier-Stokes residual regularization term (PINN style), which penalizes violations of incompressibility and momentum balance. This term could feature spatial weighting near the embedded boundary and time weighting across rollout steps, directly targeting long-horizon drift. Second, to improve robustness during rollout without sacrificing fast inference, one should investigate diffusion-based decoders (in the spirit of PDE--Refiner approaches) that take the operator’s coarse prediction and iteratively denoise/refine it, correcting accumulated phase and amplitude errors and sharpening near-boundary features when needed. Third, rather than relying on a fixed history length, we plan to explore adaptive history schemes that selectively incorporate longer temporal context only when the model detects increased uncertainty or onset of unstable dynamics, preserving efficiency in steady regimes while improving predictions during transients. Finally, it will be useful to rigorously evaluate out--of--distribution generalization by testing on geometries unseen during training, including sharper corners, different aspect ratios, and altered curvature/topology, using not only pointwise errors but also physics-centric rollout diagnostics (e.g., divergence, probe phase/Strouhal consistency) to identify failure modes and guide targeted data augmentation and model design. An additional important extension is to evaluate the proposed model on flows involving multiple interacting bodies, such as tandem or staggered geometries, where wake interference introduce additional complexity. A more challenging long-term goal is to develop models that generalize to flow configurations containing an arbitrary number $N$ of immersed geometries, requiring robustness to variable topology and interacting wakes.

Systematically addressing these challenges through physics-aware architectures and rigorous diagnostics will enable neural operators to transition from proof-of-concept demonstrations to reliable tools for design optimization and flow control in engineering practice.

\section*{Acknowledgements}
We gratefully acknowledge the NAIRR pilot program for enabling computational access, and we thank the ISU HPC cluster Nova and TACC’s Frontera for additional computing support. This research was funded by the AI Research Institutes program through NSF and USDA-NIFA under the AI Institute for Resilient Agriculture (Award No.\ 2021-67021-35329), with further support from NSF grants CMMI-2053760 and DMREF-2323716.

\section*{Data Availability}
This study utilizes the FlowBench Flow Past an Object (FPO) dataset, which is publicly accessible on HuggingFace at \url{https://huggingface.co/datasets/BGLab/FlowBench/tree/main/FPO_NS_2D_1024x256}. The dataset is licensed under a CC-BY-NC-4.0 license and serves as a benchmark for developing and evaluating scientific machine learning (SciML) models. The code used for training, to facilitate reproducibility or results, is available at \url{https://github.com/baskargroup/TimeDependent-DeepONet}.

\bibliographystyle{elsarticle-num-names}
\bibliography{Refs}

\newpage
\appendix

\section{Model Architecture Details}
\label{appendix:arch-details}

\tabref{tab:arch-details} gives a layer‐by‐layer specification of our time‐dependent Geometric DeepONet.  
\textbf{Notation:} $B$ batch size; $s=N_t$ number of input timesteps; $H,W$ spatial height and width; $P=H\times W$ total points; $m$ latent dimension; $c_3$ CNN branch channels; $fc_1,fc_2$ fusion channels; $C_{\rm out}$ output channels.

\begin{table}[h]
    \centering
    \caption{Architecture of the time‐dependent Geometric DeepONet.}
    \label{tab:arch-details}
    \footnotesize
    \begin{tabular}{p{0.25\linewidth} p{0.7\linewidth}}
    \toprule
    \textbf{Component} & \textbf{Configuration} \\
    \midrule
    \emph{Branch CNN Input} & $[B,\,2N_t,\,H,\,W]$ \\[\smallskipamount]
    \emph{CNN Encoder} & 3 Parallel conv streams ($1\times1$, $3\times3$, $5\times5$), each with $2\times2$ max pooling, producing $[B,c_3,H/8,W/8]$; concatenated to $[B,3c_3,H/8,W/8]$. \\[\smallskipamount]
    \emph{Encoder Fusion Stage 1} & $1\times1$ conv ($3c_3\!\to\!fc_1$), $2\times2$ pool $\to [B,fc_1,H/16,W/16]$. \\[\smallskipamount]
    \emph{Encoder Fusion Stage 2} & $1\times1$ conv ($fc_1\!\to\!fc_2$), $2\times2$ pool $\to [B,fc_2,H/32,W/32]$. \\[\smallskipamount]
    \emph{Branch MLP (Stage 1)} & $[\;fc_2\times\frac{H}{32}\times\frac{W}{32},\,256,\,128,\,m\;]$, ReLU \\[\smallskipamount]
    \emph{Trunk Input} & $[B,\,P,\,3]$ with channels corresponding to $(x,\,y,\,\mathrm{SDF})$ \\[\smallskipamount]
    \emph{Trunk MLP (Stage 1)} & $[3,\,128,\,128,\,m]$, ReLU \\[\smallskipamount]
    \emph{Stage 1 Fusion} & Element‐wise product of branch latent and trunk features $\to [B,P,m]$. \\[\smallskipamount]
    \emph{Branch MLP (Stage 2)} & $[\,m,\,128,\,128,\,m\times C_{\rm out}\;]$, ReLU \\[\smallskipamount]
    \emph{Trunk MLP (Stage 2)} & $[\,m,\,128,\,128,\,m\times C_{\rm out}\;]$, sine \\[\smallskipamount]
    \emph{Final Fusion} & Dot‐product over $m$ modes $\to [B,P,C_{\rm out}]$ \\
    \bottomrule
  \end{tabular}
\end{table}

\section{Training and Validation Loss} \label{subsec:loss-plots}
To further analyze training performance, we present the evolution of training and validation loss for our \emph{Time-Dependent Geometric-DeepONet} across four input sequence lengths: \(s=1\), \(s=4\), \(s=8\), and \(s=16\), as shown in \figref{fig:train-val-loss}.

\begin{figure}[!h]
    \centering
    \begin{subfigure}[b]{0.45\textwidth}
        \centering
        \begin{tikzpicture}
        \begin{semilogyaxis}[
            width=0.9\textwidth,
            height=0.55\textwidth,
            xlabel={Epoch},
            ylabel={Loss},
            title={Input sequence \(s=1\)},
            legend pos=north east,
            grid=major,
            xmin=0, xmax=1000,
            ymin=1e-2, ymax=10
        ]
        \addplot[blue, thick] table[x=epoch, y=1-train_loss, col sep=comma] {Figures/train_loss.csv};
        \addlegendentry{Training}
        \addplot[orange, thick] table[x=epoch, y=1-val_loss, col sep=comma] {Figures/train_loss.csv};
        \addlegendentry{Validation}
        \end{semilogyaxis}
        \end{tikzpicture}
    \end{subfigure}
    \begin{subfigure}[b]{0.45\textwidth}
        \centering
        \begin{tikzpicture}
        \begin{semilogyaxis}[
            width=0.9\textwidth,
            height=0.55\textwidth,
            xlabel={Epoch},
            ylabel={Loss},
            title={Input sequence \(s=4\)},
            legend pos=north east,
            grid=major,
            xmin=0, xmax=1000,
            ymin=1e-2, ymax=10
        ]
            \addplot[blue, thick] table[x=epoch, y=4-train_loss, col sep=comma] {Figures/train_loss.csv};
        \addlegendentry{Training}
        \addplot[orange, thick] table[x=epoch, y=4-val_loss, col sep=comma] {Figures/train_loss.csv};
        \addlegendentry{Validation}
        \end{semilogyaxis}
        \end{tikzpicture}
    \end{subfigure}
    \begin{subfigure}[b]{0.45\textwidth}
        \centering
        \begin{tikzpicture}
        \begin{semilogyaxis}[
            width=0.9\textwidth,
            height=0.55\textwidth,
            xlabel={Epoch},
            ylabel={Loss},
            title={Input sequence \(s=8\)},
            legend pos=north east,
            grid=major,
            xmin=0, xmax=1000,
            ymin=1e-2, ymax=10
        ]
        \addplot[blue, thick] table[x=epoch, y=8-train_loss, col sep=comma] {Figures/train_loss.csv};
        \addlegendentry{Training}
        \addplot[orange, thick] table[x=epoch, y=8-val_loss, col sep=comma] {Figures/train_loss.csv};
        \addlegendentry{Validation}
        \end{semilogyaxis}
        \end{tikzpicture}
    \end{subfigure}
    \begin{subfigure}[b]{0.45\textwidth}
        \centering
        \begin{tikzpicture}
        \begin{semilogyaxis}[
            width=0.9\textwidth,
            height=0.55\textwidth,
            xlabel={Epoch},
            ylabel={Loss},
            title={Input sequence \(s=16\)},
            legend pos=north east,
            grid=major,
            xmin=0, xmax=1000,
            ymin=1e-2, ymax=10
        ]
        \addplot[blue, thick] table[x=epoch, y=16-train_loss, col sep=comma] {Figures/train_loss.csv};
        \addlegendentry{Training}
        \addplot[orange, thick] table[x=epoch, y=16-val_loss, col sep=comma] {Figures/train_loss.csv};
        \addlegendentry{Validation}
        \end{semilogyaxis}
        \end{tikzpicture}
    \end{subfigure}
    \caption{Training and validation loss in semi-log scale for \emph{Time-Dependent Geometric-DeepONet} across 4 input sequence lengths \(s=1\)-\(s=16\). This figure presents the evolution of both training (blue) and validation (orange) losses over 1000 epochs for each sequence length.}
    \label{fig:train-val-loss}
\end{figure}

\section{Strouhal Number and Phase Lag with Increased Input Sequence Lengths}
\label{subsec:Str-Phaselag}

In the main text, we analyzed the Strouhal number and phase lag for a single-timestep input history (\(s = 1\)); see \figref{Fig:strouhal-phase-lag-1in}. Here, we repeat the same diagnostics for longer input sequences with \(s = 4\), \(s = 8\), and \(s = 16\). The corresponding results are shown in \figref{Fig:strouhal-phase-lag-4in}, \figref{Fig:strouhal-phase-lag-8in}, and \figref{Fig:strouhal-phase-lag-16in}, respectively. For each sequence length, the left column plots predicted versus ground-truth Strouhal numbers at four downstream probes (\(x/D = 1\)-\(4\)), while the right column shows the distribution of phase lag (prediction minus ground truth) across test samples at the same locations. Across all values of \(s\), the Strouhal scatter remains tightly clustered around the \(y = x\) line for every probe location, with a small number of outliers. As summarized in~\tabref{tab:strouhal_phase_all}, increasing the sequence length yields modest but consistent improvements in the Strouhal error: the relative \(L_2\) error decreases from \(0.23\) for \(s = 1\) to \(\approx 0.17\) for \(s = 16\). The phase-lag distributions are centered near zero for all \(s\), with a slightly reduced mean phase lag for \(s = 8\) and \(s = 16\). These diagnostics show that longer input histories provide small gains in both frequency and phase accuracy, while the overall temporal coherence of the wake is well captured with a single-timestep input (\(s = 1\)).

\begin{figure}[!t]
  \centering
  \begin{subfigure}[t]{0.48\textwidth}
    \centering
    \begin{tikzpicture}
      \begin{axis}[
        width=0.98\textwidth,
        height=0.6\textwidth,
        xmin=0, xmax=0.42,
        ymin=0, ymax=0.42,
        xlabel={GT $St$},
        ylabel={Pred.\ $St$},
        title={$x = 1D$},
      ]
        \addplot[
          only marks,
          mark=*,
          mark size=1pt
        ] table[
          col sep=comma,
          x=St_gt_x_2.5,
          y=St_pred_x_2.5
        ] {Figures/Strouhal/4in_rollout_strouhal.csv};

        \addplot[dashed] coordinates {(0,0) (0.4,0.4)};
      \end{axis}
    \end{tikzpicture}
  \end{subfigure}%
  \begin{subfigure}[t]{0.48\textwidth}
    \centering
    \begin{tikzpicture}
      \begin{axis}[
        width=0.98\textwidth,
        height=0.6\textwidth,
        xmin=0, xmax=265,
        ymin=-3.14, ymax=3.14,
        xlabel={Sample index},
        ylabel={Phase lag},
        title={$x = 1D$}
      ]
        \addplot[
          only marks,
          mark=*,
          mark size=1pt
        ] table[
          col sep=comma,
          x=sample_id,
          y=phase_lag_x_2.5
        ] {Figures/Strouhal/4in_rollout_phase_lag.csv};
      \end{axis}
    \end{tikzpicture}
  \end{subfigure}%
  \hfill

  \begin{subfigure}[t]{0.48\textwidth}
    \centering
    \begin{tikzpicture}
      \begin{axis}[
        width=0.98\textwidth,
        height=0.6\textwidth,
        xmin=0, xmax=0.42,
        ymin=0, ymax=0.42,
        xlabel={GT $St$},
        ylabel={Pred.\ $St$},
        title={$x = 2D$},
      ]
        \addplot[
          only marks,
          mark=*,
          mark size=1pt
        ] table[
          col sep=comma,
          x=St_gt_x_3.5,
          y=St_pred_x_3.5
        ] {Figures/Strouhal/4in_rollout_strouhal.csv};

        \addplot[dashed] coordinates {(0,0) (0.4,0.4)};
      \end{axis}
    \end{tikzpicture}
  \end{subfigure}%
  \begin{subfigure}[t]{0.48\textwidth}
    \centering
    \begin{tikzpicture}
      \begin{axis}[
        width=0.98\textwidth,
        height=0.6\textwidth,
        xmin=0, xmax=265,
        ymin=-3.14, ymax=3.14,
        xlabel={Sample index},
        ylabel={Phase lag},
        title={$x = 2D$},
      ]
        \addplot[
          only marks,
          mark=*,
          mark size=1pt
        ] table[
          col sep=comma,
          x=sample_id,
          y=phase_lag_x_3.5
        ] {Figures/Strouhal/4in_rollout_phase_lag.csv};
      \end{axis}
    \end{tikzpicture}
  \end{subfigure}%
  \hfill

  \begin{subfigure}[t]{0.48\textwidth}
    \centering
    \begin{tikzpicture}
      \begin{axis}[
        width=0.98\textwidth,
        height=0.6\textwidth,
        xmin=0, xmax=0.42,
        ymin=0, ymax=0.42,
        xlabel={GT $St$},
        ylabel={Pred.\ $St$},
        title={$x = 3D$},
      ]
        \addplot[
          only marks,
          mark=*,
          mark size=1pt
        ] table[
          col sep=comma,
          x=St_gt_x_4.5,
          y=St_pred_x_4.5
        ] {Figures/Strouhal/4in_rollout_strouhal.csv};

        \addplot[dashed] coordinates {(0,0) (0.4,0.4)};
      \end{axis}
    \end{tikzpicture}
  \end{subfigure}%
  \begin{subfigure}[t]{0.48\textwidth}
    \centering
    \begin{tikzpicture}
      \begin{axis}[
        width=0.98\textwidth,
        height=0.6\textwidth,
        xmin=0, xmax=265,
        ymin=-3.14, ymax=3.14,
        xlabel={Sample index},
        ylabel={Phase lag},
        title={$x = 3D$},
      ]
        \addplot[
          only marks,
          mark=*,
          mark size=1pt
        ] table[
          col sep=comma,
          x=sample_id,
          y=phase_lag_x_4.5
        ] {Figures/Strouhal/4in_rollout_phase_lag.csv};
      \end{axis}
    \end{tikzpicture}
  \end{subfigure}%
  \hfill

  \begin{subfigure}[t]{0.48\textwidth}
    \centering
    \begin{tikzpicture}
      \begin{axis}[
        width=0.98\textwidth,
        height=0.6\textwidth,
        xmin=0, xmax=0.42,
        ymin=0, ymax=0.42,
        xlabel={GT $St$},
        ylabel={Pred.\ $St$},
        title={$x = 4D$},
      ]
        \addplot[
          only marks,
          mark=*,
          mark size=1pt
        ] table[
          col sep=comma,
          x=St_gt_x_5.5,
          y=St_pred_x_5.5
        ] {Figures/Strouhal/4in_rollout_strouhal.csv};

        \addplot[dashed] coordinates {(0,0) (0.4,0.4)};
      \end{axis}
    \end{tikzpicture}
  \end{subfigure}%
  \begin{subfigure}[t]{0.48\textwidth}
    \centering
    \begin{tikzpicture}
      \begin{axis}[
        width=0.98\textwidth,
        height=0.6\textwidth,
        xmin=0, xmax=265,
        ymin=-3.14, ymax=3.14,
        xlabel={Sample index},
        ylabel={Phase lag},
        title={$x = 4D$},
      ]
        \addplot[
          only marks,
          mark=*,
          mark size=1pt
        ] table[
          col sep=comma,
          x=sample_id,
          y=phase_lag_x_5.5
        ] {Figures/Strouhal/4in_rollout_phase_lag.csv};
      \end{axis}
    \end{tikzpicture}
  \end{subfigure}%
  \hfill

  \caption{Strouhal number and phase‐lag (in radians) for the case of \(s=4\) at four downstream probe locations. Left column: predicted versus ground‐truth Strouhal number with a dashed $y=x$ reference line. Right column: phase lag (prediction minus ground truth) for test samples. Each row corresponds to a different streamwise position $x=1D$, $x=2D$, $x=3D$, and $x=4D$, where D is the geometry diameter.}
  \label{Fig:strouhal-phase-lag-4in}
\end{figure}

\begin{figure}[!htbp]
  \centering

  \begin{subfigure}[t]{0.48\textwidth}
    \centering
    \begin{tikzpicture}
      \begin{axis}[
        width=0.98\textwidth,
        height=0.6\textwidth,
        xmin=0, xmax=0.42,
        ymin=0, ymax=0.42,
        xlabel={GT $St$},
        ylabel={Pred.\ $St$},
        title={$x = 1D$},
      ]
        \addplot[
          only marks,
          mark=*,
          mark size=1pt
        ] table[
          col sep=comma,
          x=St_gt_x_2.5,
          y=St_pred_x_2.5
        ] {Figures/Strouhal/8in_rollout_strouhal.csv};

        \addplot[dashed] coordinates {(0,0) (0.4,0.4)};
      \end{axis}
    \end{tikzpicture}
  \end{subfigure}%
  \begin{subfigure}[t]{0.48\textwidth}
    \centering
    \begin{tikzpicture}
      \begin{axis}[
        width=0.98\textwidth,
        height=0.6\textwidth,
        xmin=0, xmax=265,
        ymin=-3.14, ymax=3.14,
        xlabel={Sample index},
        ylabel={Phase lag},
        title={$x = 1D$}
      ]
        \addplot[
          only marks,
          mark=*,
          mark size=1pt
        ] table[
          col sep=comma,
          x=sample_id,
          y=phase_lag_x_2.5
        ] {Figures/Strouhal/8in_rollout_phase_lag.csv};
      \end{axis}
    \end{tikzpicture}
  \end{subfigure}%
  \hfill

  \begin{subfigure}[t]{0.48\textwidth}
    \centering
    \begin{tikzpicture}
      \begin{axis}[
        width=0.98\textwidth,
        height=0.6\textwidth,
        xmin=0, xmax=0.42,
        ymin=0, ymax=0.42,
        xlabel={GT $St$},
        ylabel={Pred.\ $St$},
        title={$x = 2D$},
      ]
        \addplot[
          only marks,
          mark=*,
          mark size=1pt
        ] table[
          col sep=comma,
          x=St_gt_x_3.5,
          y=St_pred_x_3.5
        ] {Figures/Strouhal/8in_rollout_strouhal.csv};

        \addplot[dashed] coordinates {(0,0) (0.4,0.4)};
      \end{axis}
    \end{tikzpicture}
  \end{subfigure}%
  \begin{subfigure}[t]{0.48\textwidth}
    \centering
    \begin{tikzpicture}
      \begin{axis}[
        width=0.98\textwidth,
        height=0.6\textwidth,
        xmin=0, xmax=265,
        ymin=-3.14, ymax=3.14,
        xlabel={Sample index},
        ylabel={Phase lag},
        title={$x = 2D$},
      ]
        \addplot[
          only marks,
          mark=*,
          mark size=1pt
        ] table[
          col sep=comma,
          x=sample_id,
          y=phase_lag_x_3.5
        ] {Figures/Strouhal/8in_rollout_phase_lag.csv};
      \end{axis}
    \end{tikzpicture}
  \end{subfigure}%
  \hfill

  \begin{subfigure}[t]{0.48\textwidth}
    \centering
    \begin{tikzpicture}
      \begin{axis}[
        width=0.98\textwidth,
        height=0.6\textwidth,
        xmin=0, xmax=0.42,
        ymin=0, ymax=0.42,
        xlabel={GT $St$},
        ylabel={Pred.\ $St$},
        title={$x = 3D$},
      ]
        \addplot[
          only marks,
          mark=*,
          mark size=1pt
        ] table[
          col sep=comma,
          x=St_gt_x_4.5,
          y=St_pred_x_4.5
        ] {Figures/Strouhal/8in_rollout_strouhal.csv};

        \addplot[dashed] coordinates {(0,0) (0.4,0.4)};
      \end{axis}
    \end{tikzpicture}
  \end{subfigure}%
  \begin{subfigure}[t]{0.48\textwidth}
    \centering
    \begin{tikzpicture}
      \begin{axis}[
        width=0.98\textwidth,
        height=0.6\textwidth,
        xmin=0, xmax=265,
        ymin=-3.14, ymax=3.14,
        xlabel={Sample index},
        ylabel={Phase lag},
        title={$x = 3D$},
      ]
        \addplot[
          only marks,
          mark=*,
          mark size=1pt
        ] table[
          col sep=comma,
          x=sample_id,
          y=phase_lag_x_4.5
        ] {Figures/Strouhal/8in_rollout_phase_lag.csv};
      \end{axis}
    \end{tikzpicture}
  \end{subfigure}%
  \hfill

  \begin{subfigure}[t]{0.48\textwidth}
    \centering
    \begin{tikzpicture}
      \begin{axis}[
        width=0.98\textwidth,
        height=0.6\textwidth,
        xmin=0, xmax=0.42,
        ymin=0, ymax=0.42,
        xlabel={GT $St$},
        ylabel={Pred.\ $St$},
        title={$x = 4D$},
      ]
        \addplot[
          only marks,
          mark=*,
          mark size=1pt
        ] table[
          col sep=comma,
          x=St_gt_x_5.5,
          y=St_pred_x_5.5
        ] {Figures/Strouhal/8in_rollout_strouhal.csv};

        \addplot[dashed] coordinates {(0,0) (0.4,0.4)};
      \end{axis}
    \end{tikzpicture}
  \end{subfigure}%
  \begin{subfigure}[t]{0.48\textwidth}
    \centering
    \begin{tikzpicture}
      \begin{axis}[
        width=0.98\textwidth,
        height=0.6\textwidth,
        xmin=0, xmax=265,
        ymin=-3.14, ymax=3.14,
        xlabel={Sample index},
        ylabel={Phase lag},
        title={$x = 4D$},
      ]
        \addplot[
          only marks,
          mark=*,
          mark size=1pt
        ] table[
          col sep=comma,
          x=sample_id,
          y=phase_lag_x_5.5
        ] {Figures/Strouhal/8in_rollout_phase_lag.csv};
      \end{axis}
    \end{tikzpicture}
  \end{subfigure}%
  \hfill

  \caption{Strouhal number and phase‐lag (in radians) for the case of \(s=8\) at four downstream probe locations. Left column: predicted versus ground‐truth Strouhal number with a dashed $y=x$ reference line. Right column: phase lag (prediction minus ground truth) for test samples. Each row corresponds to a different streamwise position $x=1D$, $x=2D$, $x=3D$, and $x=4D$, where D is the geometry diameter.}
  \label{Fig:strouhal-phase-lag-8in}
\end{figure}

\begin{figure}[!htbp]
  \centering

  \begin{subfigure}[t]{0.48\textwidth}
    \centering
    \begin{tikzpicture}
      \begin{axis}[
        width=0.98\textwidth,
        height=0.6\textwidth,
        xmin=0, xmax=0.42,
        ymin=0, ymax=0.42,
        xlabel={GT $St$},
        ylabel={Pred.\ $St$},
        title={$x = 1D$},
      ]
        \addplot[
          only marks,
          mark=*,
          mark size=1pt
        ] table[
          col sep=comma,
          x=St_gt_x_2.5,
          y=St_pred_x_2.5
        ] {Figures/Strouhal/16in_rollout_strouhal.csv};

        \addplot[dashed] coordinates {(0,0) (0.4,0.4)};
      \end{axis}
    \end{tikzpicture}
  \end{subfigure}%
  \begin{subfigure}[t]{0.48\textwidth}
    \centering
    \begin{tikzpicture}
      \begin{axis}[
        width=0.98\textwidth,
        height=0.6\textwidth,
        xmin=0, xmax=265,
        ymin=-3.14, ymax=3.14,
        xlabel={Sample index},
        ylabel={Phase lag},
        title={$x = 1D$}
      ]
        \addplot[
          only marks,
          mark=*,
          mark size=1pt
        ] table[
          col sep=comma,
          x=sample_id,
          y=phase_lag_x_2.5
        ] {Figures/Strouhal/16in_rollout_phase_lag.csv};
      \end{axis}
    \end{tikzpicture}
  \end{subfigure}%
  \hfill

  \begin{subfigure}[t]{0.48\textwidth}
    \centering
    \begin{tikzpicture}
      \begin{axis}[
        width=0.98\textwidth,
        height=0.6\textwidth,
        xmin=0, xmax=0.42,
        ymin=0, ymax=0.42,
        xlabel={GT $St$},
        ylabel={Pred.\ $St$},
        title={$x = 2D$},
      ]
        \addplot[
          only marks,
          mark=*,
          mark size=1pt
        ] table[
          col sep=comma,
          x=St_gt_x_3.5,
          y=St_pred_x_3.5
        ] {Figures/Strouhal/16in_rollout_strouhal.csv};

        \addplot[dashed] coordinates {(0,0) (0.4,0.4)};
      \end{axis}
    \end{tikzpicture}
  \end{subfigure}%
  \begin{subfigure}[t]{0.48\textwidth}
    \centering
    \begin{tikzpicture}
      \begin{axis}[
        width=0.98\textwidth,
        height=0.6\textwidth,
        xmin=0, xmax=265,
        ymin=-3.14, ymax=3.14,
        xlabel={Sample index},
        ylabel={Phase lag},
        title={$x = 2D$},
      ]
        \addplot[
          only marks,
          mark=*,
          mark size=1pt
        ] table[
          col sep=comma,
          x=sample_id,
          y=phase_lag_x_3.5
        ] {Figures/Strouhal/16in_rollout_phase_lag.csv};
      \end{axis}
    \end{tikzpicture}
  \end{subfigure}%
  \hfill

  \begin{subfigure}[t]{0.48\textwidth}
    \centering
    \begin{tikzpicture}
      \begin{axis}[
        width=0.98\textwidth,
        height=0.6\textwidth,
        xmin=0, xmax=0.42,
        ymin=0, ymax=0.42,
        xlabel={GT $St$},
        ylabel={Pred.\ $St$},
        title={$x = 3D$},
      ]
        \addplot[
          only marks,
          mark=*,
          mark size=1pt
        ] table[
          col sep=comma,
          x=St_gt_x_4.5,
          y=St_pred_x_4.5
        ] {Figures/Strouhal/16in_rollout_strouhal.csv};

        \addplot[dashed] coordinates {(0,0) (0.4,0.4)};
      \end{axis}
    \end{tikzpicture}
  \end{subfigure}%
  \begin{subfigure}[t]{0.48\textwidth}
    \centering
    \begin{tikzpicture}
      \begin{axis}[
        width=0.98\textwidth,
        height=0.6\textwidth,
        xmin=0, xmax=265,
        ymin=-3.14, ymax=3.14,
        xlabel={Sample index},
        ylabel={Phase lag},
        title={$x = 3D$},
      ]
        \addplot[
          only marks,
          mark=*,
          mark size=1pt
        ] table[
          col sep=comma,
          x=sample_id,
          y=phase_lag_x_4.5
        ] {Figures/Strouhal/16in_rollout_phase_lag.csv};
      \end{axis}
    \end{tikzpicture}
  \end{subfigure}%
  \hfill

  \begin{subfigure}[t]{0.48\textwidth}
    \centering
    \begin{tikzpicture}
      \begin{axis}[
        width=0.98\textwidth,
        height=0.6\textwidth,
        xmin=0, xmax=0.42,
        ymin=0, ymax=0.42,
        xlabel={GT $St$},
        ylabel={Pred.\ $St$},
        title={$x = 4D$},
      ]
        \addplot[
          only marks,
          mark=*,
          mark size=1pt
        ] table[
          col sep=comma,
          x=St_gt_x_5.5,
          y=St_pred_x_5.5
        ] {Figures/Strouhal/16in_rollout_strouhal.csv};

        \addplot[dashed] coordinates {(0,0) (0.4,0.4)};
      \end{axis}
    \end{tikzpicture}
  \end{subfigure}%
  \begin{subfigure}[t]{0.48\textwidth}
    \centering
    \begin{tikzpicture}
      \begin{axis}[
        width=0.98\textwidth,
        height=0.6\textwidth,
        xmin=0, xmax=265,
        ymin=-3.14, ymax=3.14,
        xlabel={Sample index},
        ylabel={Phase lag},
        title={$x = 4D$},
      ]
        \addplot[
          only marks,
          mark=*,
          mark size=1pt
        ] table[
          col sep=comma,
          x=sample_id,
          y=phase_lag_x_5.5
        ] {Figures/Strouhal/16in_rollout_phase_lag.csv};
      \end{axis}
    \end{tikzpicture}
  \end{subfigure}%
  \hfill

  \caption{Strouhal number and phase‐lag (in radians) for the case of \(s=16\) at four downstream probe locations. Left column: predicted versus ground‐truth Strouhal number with a dashed $y=x$ reference line. Right column: phase lag (prediction minus ground truth) for test samples. Each row corresponds to a different streamwise position $x=1D$, $x=2D$, $x=3D$, and $x=4D$, where D is the geometry diameter.}
  \label{Fig:strouhal-phase-lag-16in}
\end{figure}

\section{MeshGraphNet Training Details and Loss Curve} \label{subsec:mgn-loss-plots}

MeshGraphNet was trained for 100 epochs on a single NVIDIA A100 GPU (80\,GB), with a total training time of 9.5 days. All experiments were performed at full spatial resolution ($256\times1024$ grid) without subsampling. The graph connectivity corresponds to a 4-neighborhood structured grid, and the model contains $8.4\times 10^{4}$ trainable parameters. Table~\ref{tab:mgn-hparams} summarizes the MeshGraphNet hyperparameters used for training using both input history lengths $s=1$ and $s=4$.

To further analyze training performance, we present the evolution of training and validation loss for our \emph{MeshGraphNet}, as shown in \figref{fig:mgn-train-val-loss}.

\begin{table}[!t]
\centering
\caption{MeshGraphNet (MGN) hyperparameters used in this work.}
\label{tab:mgn-hparams}
\renewcommand{\arraystretch}{1.1}
\setlength{\tabcolsep}{10pt}
\begin{tabular}{lc}
\toprule
\textbf{Hyperparameter} &  \\
\midrule
Latent dimension $D$ & 64 \\
Hidden dimension (MLPs) & 128 \\
MLP depth & 2 layers \\
Message-passing steps & 15 \\
Optimizer & Adam \\
Learning rate & $10^{-4}$ \\
Batch size & 4 \\
Epochs & 100 \\
\bottomrule
\end{tabular}
\end{table}

\begin{figure}[!t]
    \centering
    \begin{subfigure}[b]{0.48\textwidth}
        \centering
        \begin{tikzpicture}
        \begin{semilogyaxis}[
            width=0.9\textwidth,
            height=0.55\textwidth,
            xlabel={Epoch},
            ylabel={Loss},
            title={Input sequence \(s=1\)},
            legend pos=north east,
            grid=major,
            xmin=0, xmax=100,
            ymin=5e-5, ymax=1
        ]
        \addplot[blue, thick] table[x=epoch, y=1-train_loss, col sep=comma] {Figures/meshgraphnet/train_loss.csv};
        \addlegendentry{Training}
        \addplot[orange, thick] table[x=epoch, y=1-val_loss, col sep=comma] {Figures/meshgraphnet/train_loss.csv};
        \addlegendentry{Validation}
        \end{semilogyaxis}
        \end{tikzpicture}
    \end{subfigure}
    \hfill
    \begin{subfigure}[b]{0.48\textwidth}
        \centering
        \begin{tikzpicture}
        \begin{semilogyaxis}[
            width=0.9\textwidth,
            height=0.55\textwidth,
            xlabel={Epoch},
            ylabel={Loss},
            title={Input sequence \(s=4\)},
            legend pos=north east,
            grid=major,
            xmin=0, xmax=100,
            ymin=5e-5, ymax=1
        ]
            \addplot[blue, thick] table[x=epoch, y=4-train_loss, col sep=comma] {Figures/meshgraphnet/train_loss.csv};
        \addlegendentry{Training}
        \addplot[orange, thick] table[x=epoch, y=4-val_loss, col sep=comma] {Figures/meshgraphnet/train_loss.csv};
        \addlegendentry{Validation}
        \end{semilogyaxis}
        \end{tikzpicture}
    \end{subfigure}
    \caption{Training and validation loss in semi-log scale for \emph{MeshGraphNet} using two input sequence lengths \(s=1\), \(s=4\). This figure presents the evolution of both training (blue) and validation (orange) losses over 100 epochs for each sequence length.}
    \label{fig:mgn-train-val-loss}
\end{figure}

\end{document}